\documentclass[draftcls,12pt,onecolumn]{IEEEtran}
\usepackage{amsmath, amsfonts,amssymb, latexsym,epsfig,color,rotating,
times,float,subfigure,algorithm,verbatim}
\usepackage{epstopdf}

\newcommand{\T}{\mathbb{T}}
\renewcommand{\S}{\mathbb{S}}
\newcommand{\cop}{c}
\newcommand{\yi}{y^{(1)}}
\newcommand{\yii}{y^{(2)}}

\newcommand{\f}{\mathbf{f}}
\newcommand{\de}{\mathbf{d}}

\newcommand{\Io}{{\Pi^o(X)}}
\newcommand{\Ib}{{\Pi^b(X)}}
\renewcommand{\i}{\mathbf{i}}
\renewcommand{\j}{\mathbf{j}}

\newcommand{\pizero}{\pi_0}

\newcommand{\Ep}{\E^\mu_{\pi_0}}
\newcommand{\Epzero}{\E^\mu_{\pi_0}}

\newcommand{\Pone}{P^{(1)}}
\newcommand{\Ptwo}{P^{(2)}}

\newcommand{\Pp}{\P^\mu_{\pi_0}}
\newcommand{\Cb}{\bar{C}}
\newcommand{\Vb}{\bar{V}}

\renewcommand{\r}{\epsilon}

\newcommand{\F}{\mathcal{F}}

\newcommand{\Bs}{R^\pi} 
\newcommand{\Ts}{T}
\newcommand{\sigs}{\sigma}
\newcommand{\Tp}{T}
\newcommand{\sigp}{\sigma}
\newcommand{\priv}{\eta}

\newcommand{\R}{\mathcal{R}}

\renewcommand{\d}{\omega}

\newcommand{\discount}{\rho}

\newcommand{\Vl}{V^{LF}}

\newcommand{\bS}{X}
\newcommand{\Mu}{\boldsymbol{\mu}}
\newcommand{\U}{\mathcal{U}}
\newcommand{\A}{\mathbb{A}}
\newcommand{\X}{\mathbb{X}}
\newcommand{\Y}{\mathbb{Y}}
\newcommand{\ta}{\tilde{a}}

\newcommand{\ca}{c_a}

\newcommand{\hphi}{\hat{\phi}}

\newcommand{\ones}{\mathbf{1}}

\newcommand{\tpi}{\tilde{\pi}}
\newcommand{\nablat}{\widehat{\nabla}_{\phi}}

\renewcommand{\l}{\mathcal{L}}

\newcommand{\pie}{\pi^{\epsilon_1,\epsilon_2,\ldots,\epsilon_{S-1}}}

\newcommand{\pieone}{\pi^{\epsilon_1,\ldots\epsilon_j,\ldots, \epsilon_{S-1}}}
\newcommand{\pietwo}{\pi^{\epsilon_1,\ldots\bar{\epsilon}_j,\ldots, \epsilon_{S-1}}}

\newcommand{\p}{\prime}

\renewcommand{\H}{\mathcal{H}}
\renewcommand{\l}{\mathcal{L}}

\newcommand{\gtp}{\underset{\text{\tiny TP2}}{\geq}}

\newcommand{\gl}{\geq_{L_i}}
\newcommand{\glp}{\geq_{L_{i+1}}}

\newcommand{\glX}{\geq_{L_X}}
\newcommand{\glone}{\geq_{L_1}}

\newcommand{\E}                 {\Bbb{E}}
\renewcommand{\P}                 {\Bbb{P}}
\newcommand{\reals}{\mathbb{R}}
\newcommand{\I}{\Pi(X)}

\renewcommand{\u}{\{1,2\}}
\newcommand{\argmin}{\operatorname{argmin}}
\newcommand{\ole}{\stackrel{\triangle}{=}}

\newtheorem{theorem}{Theorem}
\newtheorem{corollary}{Corollary}

\newtheorem{result}{Result}
\newtheorem{lemma}{Lemma}
\newtheorem{definition}{Definition}
\newcommand{\nn}{\nonumber}
\newcommand{\gr}{\geq_r}
\newcommand{\lr}{\leq_r}
\newcommand{\gs}{\geq_s}
\newcommand{\ls}{\leq_s}

\newcommand{\bp}{{\bar{\pi}}}

\newcommand{\beq}{\begin{equation}}
\newcommand{\eeq}{\end{equation}}
\newcommand{\qed} {{$\hfill\blacksquare$}}

\newcommand{\thr}{\mathbf{\Gamma}}

\begin{document}

\title{Bayesian Sequential  Detection with Phase-Distributed Change Time and Nonlinear Penalty -- A POMDP Approach}
\author{Vikram Krishnamurthy  {\em Fellow, IEEE} 
\thanks{This work was partially supported by NSERC.}
\thanks{Vikram Krishnamurthy is
 with the Department of Electrical and Computer
Engineering, University of British Columbia, Vancouver, V6T 1Z4, Canada. 
(email:  vikramk@ece.ubc.ca).}}

\maketitle

\begin{abstract}  We show that the optimal decision policy
for several  types of Bayesian sequential
detection problems  has a threshold switching curve structure on the space
of posterior distributions. 
This   is established by using  lattice programming and stochastic orders  in a 
 partially observed Markov decision
process (POMDP) framework.
 A stochastic gradient algorithm is presented to estimate the optimal
linear approximation to this threshold curve.
We  illustrate these  results by  first considering quickest time detection with
phase-type distributed change time and a variance stopping penalty. Then it is proved that the threshold switching curve
 also arises in several other Bayesian decision problems
such as quickest transient detection, exponential delay (risk-sensitive) penalties, stopping time problems in
 social learning,  and  multi-agent scheduling in a changing world.  Using Blackwell dominance, it is shown that for dynamic decision making problems,  the optimal decision policy is lower bounded by a myopic policy.
 Finally,
it is shown how the achievable cost of the optimal decision policy varies with change time distribution
by imposing a partial order on transition matrices.
\end{abstract}


\section{Introduction} \label{sec:intro}

Quickest time change detection has applications in
biomedical signal processing, machine monitoring and finance   \cite{PH08,BN93}. 
There are two general formulations for quickest time
detection.  In the first
formulation, the change point $\tau^0$ is an unknown deterministic time,
and the goal is to determine a
stopping rule such that a certain worst case delay penalty is
minimized subject to a constraint on the false alarm frequency
(see, e.g., \cite{Mou86,Poo98,YKP99}).

The second formulation, which is the formulation we consider in this paper,  is  the  Bayesian approach.
The change time $\tau^0$ is a random variable specified by a prior distribution.
Consider a sequence of discrete time random measurements $\{y_k,k \geq 1\}$, such that 
 conditioned on the event $\{\tau^0 = t\}$, $y_k$, $k \leq t$  are i.i.d. random variables with distribution 
$B_1$ and $y_k, k >t$ are i.i.d. random variables with distribution $B_2$.
The quickest time detection problem involves detecting the change time $\tau^0$ with minimal cost. That is,
at each time $k=1,2,\ldots$, a decision $u_k \in \{\text{continue}, \text{stop and announce change}\}$ needs to be made to optimize a tradeoff
between false alarm frequency and linear delay penalty.  

In classical Bayesian  quickest time detection \cite{Shi78,Yak97,PH08},  the change time $\tau^0$ is  modelled by a geometric distribution.
A geometric distributed change time is realized by a two state discrete-time
Markov chain, which we denote as $x_k$.
Therefore, in  classical quickest time detection, the optimal decision policy at each time $k$ is a function of a two-dimensional
belief  state (posterior probability mass function)  $\pi_k(i) = P(x_k = i | y_1,\ldots,y_k,u_1,\ldots,u_{k-1})$,
$i=1,2$ with  $\pi_k(1)+\pi_k(2) =1$.
So  it suffices to consider one element, say $\pi_k(2)$,  of this
probability mass function. Classical quickest time change detection (see for example  \cite{PH08}) says that 
 there exists a threshold point $\pi^* \in [0,1]$ such that the optimal decision policy is  
\beq u_k = \begin{cases} \text{ continue } & \text{ if }
\pi_k(2) \geq \pi^* \\   \text{ stop and announce change  } &  \text{ if } \pi_k(2) < \pi^*
\end{cases} \label{eq:onedim}
\eeq  

 As a generalization of the  Bayesian framework,  \cite{Yak94,Yak97}  consider dependent observations from a finite state Markov chain with  transition probability matrix affected by the change point. 

\subsection*{Main Results and Organization of paper}

This paper considers  Bayesian quickest time detection
with  the following generalizations:  phase-type distributed change times,  a variance stopping penalty, optimal linear threshold policies, and  examples in  
transient detection, nonlinear delay penalty and 
stopping time problems in social  learning.  Our goal
is to exploit lattice programming techniques  to prove the existence of threshold
optimal decision policies 
 for a variety of quickest time detection problems.  Below is an overview of these results.

(i) {\em Phase-type Distributed Change Times}:  We consider
quickest time detection 
when  the change time $\tau^0$ has a phase-type (PH) distribution  \cite{Neu89}.
 PH-distributions  are used widely in modelling discrete event systems.
The optimal detection of a PH-distributed change point is useful  since the family of  PH-distributions forms a dense subset of the set of all distributions,
i.e., for any given distribution function $F$ such that $F(0) = 0$, one can find a sequence of PH-distributions	
$\{F_n , n	\geq	1\}$	 to		approximate	$F$	uniformly over $[0, \infty)$.
As described in \cite{Neu89}, a PH-distributed change time can be modelled
by a multi-state Markov chain with an absorbing state. (For a 2-state Markov chain, the PH-distribution
specializes to the geometric distribution).
So  for quickest time detection with PH-distributed change time, the
belief states  (Bayesian posterior)  lie in  a multidimensional simplex of probability
mass functions. 

(ii) {\em Variance Penalty}: The second generalization we consider is 
 a stopping penalty
comprising of the false alarm  and a  variance penalty. The variance
penalty is essential in stopping problems where one is interested in ultimately estimating the state $x$. It penalizes stopping too soon
if the uncertainty  of the state estimate is large.\footnote{In \cite{BB89}, a continuous time
stochastic control problem is formulated 
with a quadratic stopping cost, and the 
 existence of
the solution to the resulting quasi-variational inequality is proved.
 However, \cite{BB89} does not
deal with structural results.}
Since the variance  is quadratic in the belief state $\pi$, it is not possible to reformulate a variance penalty problem as a standard stopping time problem. 

{\em Under what conditions does there  exist   an optimal  threshold decision policy  for quickest  detection with PH-distributed change time
and variance penalty}?
How can the belief states (in a multi-dimensional simplex)
 be ordered and compared with a threshold? Sec.\ref{sec:qdbasic} formulates the quickest time detection problem as  a 
 partially observed Markov decision process (POMDP) and characterizes the optimal decision policy as the solution of
 a stochastic dynamic programming problem.
Using lattice programming   \cite{Top98}  
 our main result  (Theorems \ref{thm:1} and \ref{thm:modified} in Sec.\ref{sec:main}) shows that the optimal decision policy  is 
governed by a threshold switching curve on the space of Bayesian distributions (belief states). 
This result
 is  useful for several reasons: (a) It provides a multi-dimensional generalization of (\ref{eq:onedim}) to PH-distributed change times.
 (b) Efficient algorithms can be designed to estimate  
 optimal policies that satisfy this threshold structure.
 (c) The  result holds under set-valued constraints on the change time and observation
 distribution. 
 So there is an inherent robustness  since even if the underlying  model parameters are 
not exactly specified,  the threshold structure still holds. 

 Going from a 2 state Markov chain  (geometric distributed change time) to multiple states (PH-distributed change time) introduces substantial complications.
For 2 state Markov chains,  the posterior distribution can be parametrized by a scalar (as in (\ref{eq:onedim})) and therefore can be completely ordered.
However,  for more than 2 states,
 comparing posterior distributions requires  using stochastic orders  which are  {\em partial orders}. 
In this paper we use the 
{\em monotone likelihood ratio (MLR)} stochastic order \cite{Rie91,MS02,KD09,KW09} to prove our structural
results.
The MLR order is ideally suited for Bayesian problems since it is preserved under conditional expectations.
However, 
determining the optimal policy is non-trivial since the policy can only be characterized on a partially ordered set 
(more generally a lattice) within the unit simplex.
We modify the MLR stochastic order
to operate on line segments  within the  unit simplex of posterior distributions.
Such line segments form chains (totally ordered subsets of a partially ordered set) and permit us to prove that  the optimal decision policy has a  threshold structure.  Theorem \ref{cor:ph}  shows that for linear delay and false alarm penalties, the stopping region is convex.

(iii) {\em Optimal Linear Threshold}: Having  established the existence of a threshold curve, Theorem \ref{thm:dep} gives necessary and sufficient
conditions for the optimal linear hyperplane approximation to this  curve. Then a 
  simulation-based
stochastic 
 approximation algorithm (Algorithm \ref{alg1}) is presented to compute this optimal linear hyperplane
 approximation.

 The remainder of the paper illustrates the above structural results  in several examples.
 
 (iv) {\em Example 2: Quickest Transient Detection}: 
 Sec.\ref{sec:qtrans} considers quickest transient detection.
 We refer the reader to \cite{PKV10} for a nice description of the 
 quickest transient detection problem and 
 various
 cost functions. 
In quickest transient detection, a  Markov chain state jumps from a starting state to a
 transient state at a geometric distributed time, and then jumps out of the state to an absorbing state at another geometric distributed time. 
 We show in Theorem \ref{thm:qt} that a similar structural result to quickest time detection holds (i.e., 
 existence of a threshold switching curve and convexity of the stopping region).

 (v) {\em Example 3: Quickest Time Detection with Exponential Delay Penalty}: In Sec.\ref{sec:risk}, we generalize the results of Poor \cite{Poo98} to
 PH-distributed change times. 
 Poor \cite{Poo98} considers a novel variation of the quickest time detection problem where
 the time delay in the
detection is penalized  exponentially.  
By converting the resulting problem into a 
 standard stopping
problem \cite{CRS71}, it is shown in  \cite{Poo98}  that the optimal decision policy is a 
 threshold policy under mild conditions.
 
 The exponential delay penalty cost function in \cite{Poo98} is a special case of {\em risk sensitive stochastic
 control} with geometric change times.  Assuming more general PH-distributed change times,  Theorem \ref{thm:risk} 
 shows that the optimal detection policy is characterized by a  threshold switching curve and the stopping region is convex in the risk-sensitive belief state.
 
 Risk sensitive stochastic control is widely used in mathematical finance, see  \cite{Ben92,JBE94,EAM95} for  comprehensive treatments in
 discrete and continuous time.
 In simple terms,  quickest time detection seeks to optimize the  objective  $ \E\{J^0\}$
 where $J^0  $ is the accumulated  sample path cost until some stopping time $\tau$.
 In risk sensitive control, one seeks to optimize $J = \E\{\exp(\r J^0\}$. Note that $\r J$ can be written as
$
 \r J = \r + \r^2\E\{J^0\} + \text{ higher order terms}$. It therefore follows that for $\r > 0$, the scaled cost $\r J$ and hence $J$ is robust and penalizes
 heavily large sample path costs due to the presence of second order moments.  This is termed a risk-averse control and is of significant
 importance in mathematical finance, see \cite{Ben92}.
 Risk sensitive control provides a nice formalization of the exponential penalty delay cost  and allows us to generalize
 the results in Poor \cite{Poo98} to phase-distributed change times by applying lattice programming.

 (vi) {\em Example 4 and 5: Stopping Time Problems in Multi-agent Social Learning}:  Sec.\ref{sec:change} presents
 two examples  of stopping time problems involving   social learning
 amongst multiple agents. We consider:  How do local decisions in social learning affect the global decision in a stopping time problem?  Social learning has been used in economics \cite{Ban92,BHW92,Cha04}, for example to model behavior in financial markets; see also \cite{LADO07,SS00}.
In  social learning,  each agent  optimizes 
its local utility selfishly
and then broadcasts its action.  Subsequent agents then use their private observation together with the actions of previous agents
 to learn an underlying state.

Our first result (Example 4) deals with a multi-agent 
Bayesian stopping time problem where
 agents perform greedy social learning and reveal their  local actions to subsequent agents. 
 How can the multi-agent system  make a global decision when to stop?
 Such problems arise
in automated decision systems (e.g., sensor networks) where agents make local
decisions and reveal these local decisions to subsequent agents.
Theorem~\ref{thm:stopsocial} shows that 
the optimal decision policy of the stopping time problem has multiple thresholds.    This  is unusual: 
 if it is optimal to declare state 1 based on a Bayesian belief, it may not be optimal to declare state 1 when the belief about
state 1 is stronger. 
 We also give an explicit example of an optimal 
double threshold policy.  
The result  shows that making global decisions
based on local decisions involves non-monotone policies.

Our second result  (Example 5) deals with ``constrained optimal'' social learning.
A key result in social learning is that rational agents eventually herd, that is, they pick the same
 action irrespective of their private observation and social learning stops.  
  To enhance social learning,  
  Chamley \cite{Cha04} (see also \cite{SS97} for related work) formulated constrained social learning as a  stopping time
problem where agents either reveal their observations or they herd (which is equivalent to stopping
in a sequential decision problem). When should a multi-agent system make the global decision to stop  (herd)?
  Intuitively,
the decision to stop should be made when the state estimate is sufficiently accurate so that revealing private observations is no longer
required.
Theorem \ref{thm:socialopt} in Sec.\ref{sec:change}
shows that the  constrained optimal social learning proposed by Chamley \cite{Cha04}
  has a  threshold switching curve  in the space of public belief states.
Thus the global decision to stop  in \cite{Cha04} can be implemented efficiently in a multi-agent system.

 (vii) {\em Example 6: Multi-agent Scheduling in a  Changing World}:  
 In Sec.\ref{sec:fast} we examine: How can the optimal decision policy be bounded in terms of a myopic policy? How does the achievable 
 cost of the optimal policy vary with transition probabilities (and therefore change time distribution)?
We answer these two questions in a general setting where optimal decisions need to be made  when the 
underlying state $x$ evolves according to a finite state Markov chain without necessarily having an absorbing state.
 The problem is  no longer a stopping problem;  it is a more general
partially-observed stochastic control problem.

To formulate these results, Sec.\ref{sec:fast} considers a multi-agent scheduling problem.
Using Blackwell dominance, Theorem \ref{thm:compare2} shows that the optimal policy is lower bounded by a myopic policy.
The myopic policy can be computed efficiently and is a rigorous  lower bound to the computationally intractable
optimal policy.
Finally, Theorem \ref{thm:tmove} examines how the optimal expected  cost varies with transition matrix for
a  stopping time problem (e.g., quickest detection problem) and more general 
dynamic decision problem in a changing world.
The theorem shows that for the underlying Markovian state,
the larger the transition matrix (according to an order defined in Sec.\ref{sec:fast}), the cheaper 
the  optimal expected cost.


\section{Partially Observed Stochastic Control  Formulation} \label{sec:qdbasic}

In this section we present a partially observed  stochastic control formulation that allows us to tackle 
the various stopping time problems considered 
in subsequent  sections. 

\subsection{Stopping-Time Stochastic Control  Model} \label{sec:qdmodel}
The model comprises of the following ingredients

1. {\em Absorbing-state Markov chain and Phase-Type Distribution Change Time}:
We  model the change point  $\tau^0$ by a   
{\em  phase type (PH) distribution}. 
The family of all PH-distributions forms a dense subset for the set of all distributions
	\cite{Neu89} and hence can be used to  approximate   change points with an arbitrary distribution. This is done by
	constructing a multi-state  Markov chain as follows: Let $k=0,1,\ldots$ denote discrete time.
Assume  the state of nature $x_k$  evolves as a Markov chain on
the finite state space 
\beq  \{e_1,\ldots,e_X\} \text{  where $e_i$ is the $X$-dimensional unit vector with 1 in the $i$-th position}. \label{eq:ei}\eeq
 Here  state `1'  (corresponding to $e_1$) is an absorbing state
and denotes the state after the jump change.  The states $2,\ldots,X$ (corresponding to $e_2,\ldots,e_X$) can be viewed as  a single composite state that $x$ resides in before the jump. Denote 
\beq \X = \{1,2,\ldots,X\}. \eeq
 We assume that  the change occurs after at least one measurement. So the initial distribution 
\beq \pi_0 = (\pi_0(i), i \in \X),\;  \pi_0(i) = 
P(x_0 = e_i) \;
\text{ satisfies } \pi_0(1) = 0. \label{eq:init} \eeq
The $X\times X$ 
transition probability matrix $P$ with elements $P_{ij} = P(x_{k+1}=e_j|x_k=e_i)$ is 
\beq \label{eq:phmatrix}
P = \begin{bmatrix}  1 & 0 \\ \underline{P}_{(X-1)\times 1} & \bar{P}_{(X-1)\times (X-1)} \end{bmatrix}.
\eeq
Let the ``change time" $\tau^0$ denote the time at which $x_k$ enters the absorbing state 1,
i.e., \beq \tau^0 = \inf\{k: x_k = 1\} . \label{eq:tau}\eeq
The distribution of $\tau^0$ is determined by choosing the transition probabilities $\underline{P}, \bar{P}$ in 
 (\ref{eq:phmatrix}).  To ensure that $\tau^0$ is finite, assume states $2,3,\ldots X$ are transient.
This  is equivalent to $\bar{P}$ satisfying  $\sum_{n=1}^\infty \bar{P}^{n}_{ii} < \infty$ for $i=1,\ldots,X-1$ (where $\bar{P}^{n}_{ii}$ denotes the $(i,i)$ element of the $n$-th power
of matrix $\bar{P}$).
 The distribution of $\tau^0$ (which is equivalent to the  
distribution of the absorption time to state 1)   is given by
\beq \label{eq:nu}
 \nu_0 = \pi_0(1), \quad \nu_k = \bar{\pi}_0^\p \bar{P}^{k-1} \underline{P}, \quad k\geq 1, \eeq
 where $\bar{\pi}_0 = [\pi_0(2),\ldots,\pi_0(X)]^\p$.
The key idea  is that by appropriately choosing the pair $(\pi_0,P)$ 
and the associated state space dimension $X$,
 one can approximate any given discrete distribution on $[0, \infty)$ by the distribution $\{\nu_k, k \geq 0\}$; see 
\cite[pp.240-243]{Neu89}.
 The event $\{x_k = 1\}$ means the change point has occurred at time $k$ according
 to PH-distribution (\ref{eq:nu}). Of course, in the special case when $x$ is a 2-state Markov chain,
 the change time $\tau^0$  is geometrically distributed.

2. {\em Observation}:
At time $k$, the noisy observation $y_k  \in \Y$ given state $x_k$ has
conditional probability distribution
 \beq \label{eq:obs}
 P(y_k \leq \bar{y}|x_k = e_i) = \sum_{y\leq \bar{y}} B_{iy},\quad i \in  \X. \eeq
Here $\sum_y$ denotes integration with respect to the Lebesgue measure
(in which case $\Y\subset \reals$ and $B_{iy}$ is the conditional probability density function)
or counting measure (in which case $\Y$ is a subset of the integers and $B_{iy}$ is the conditional probability
mass function $B_{iy}= P(y_k=y|x_k=e_i)$).

3. {\em  Belief State}: At time $k$, the  belief state is the posterior probability mass function of $x_k$ given  the observation history $y_1,\ldots,y_k$ and 
past decisions $u_1,\ldots, u_{k-1}$. That is 
\beq \pi_k = (\pi_k(i), \, i \in \X) , \quad \pi_k(i) = P(x_k = e_i|y_1,\ldots,y_{k},u_1,\ldots,u_{k-1}), \text{ initialized by } \pi_0.
\eeq
Equivalently, denote the filtration 
  \beq
  \label{eq:sigf} \F_k = \text{ $\sigma$-algebra generated by } (y_1,\ldots,y_k,u_1,\ldots,u_{k-1}). \eeq
  Then $\pi_k = \E\{x_k|\F_k\}$. (The notational advantage of choosing unit vectors (\ref{eq:ei})
  for the state space is that conditional probabilities and conditional expectations coincide).

The belief state
 is updated via the Bayesian (Hidden Markov Model) filter 
\begin{align}  \label{eq:hmm} \pi_k &= \Tp(\pi_{k-1},y_k),  \text{ where }
\Tp(\pi,y) = 
\frac{B_y P^\p\pi}{\sigp(\pi,y)},
\; \sigp(\pi,y) = \mathbf{1}_X^\p B_y  P^\p\pi 
\\
B_y &= \text{diag}(P(y|x=e_i),i\in \X) . \nonumber
 \end{align}
 Here $\ones_X$ denotes the $X$ dimensional vector of ones.
The belief state   $\pi$ in (\ref{eq:hmm})  is an $X$-dimensional probability vector. It belongs to the
 $X-1$ dimensional unit-simplex denoted as
\begin{align}
\I \ole \left\{\pi \in \reals^{X}: \mathbf{1}_{X}^{\p} \pi = 1,
\quad 
0 \leq \pi(i) \leq 1 \text{ for all } i \in \X \right\}. \label{eq:Pi}
\end{align}
For example, $\Pi(2)$ is a one dimensional simplex (unit line segment),
$\Pi(3)$ is a two-dimensional  simplex (equilateral triangle); $\Pi(4)$
is a tetrahedron, etc.
The states  $e_1,e_2,\ldots, e_X$ of the Markov chain $x$ are the vertices of $\I$.

4. {\em Sequential Decision and Costs}:  At each time  $k$, a decision $u_k$ is taken
where 
\beq \label{eq:actionpolicy}
 u_k = \mu(\pi_k)  \in \U = \{1 \text{ (announce change and stop)} ,2 \text{ (continue) }\}. \eeq
In (\ref{eq:actionpolicy}),  the policy $\mu$  belongs to the class of stationary decision
policies denoted $\Mu$.

(i) {\em  Cost of announcing change and stopping}:  
If decision $u_k=1$  is chosen, then the problem terminates. If  $u_k=1$ is chosen before  the change point $\tau^0$, then a false
alarm and variance penalty is paid.  If $u_k=1$ is chosen at or after the change point $\tau^0$, then only a variance penalty is paid.
Below we formulate these costs.

Let $g = (g_1,\ldots,g_X)^\p$ specify the  physical state levels  associated with states $1,2,\ldots, X$ of the Markov chain $x$.
 The {\em variance  penalty}   is 
\beq \E\{ \|(x_k -  \pi_k)^\p g \|^2 \mid \F_k \} = G^\p \pi_k(i) - (g^\p \pi_k)^2,  \text { where }
G_i = g_i^2 \text{ and } G=(G_1,G_2,\ldots,G_X). 
\label{eq:varder}
\eeq
This  conditional variance   penalizes choosing the stop action if the uncertainty in the state estimate is large, see also
\cite{BB89}.

Next, the false alarm event  $\cup_{i\geq 2}   \{x_k =  e_i\} \cap \{u_k = 1\} = \{x_k\neq e_1\} \cap \{u_k = 1\}$ represents the event
that a change is announced before the change happens at time $\tau^0$.
To evaluate the {\em false alarm penalty}, 
 let  $f_i I(x_k=e_i,u_k=1)$ denote the cost of a false alarm in state $e_i$, $i \in \X$, where
$f_i \geq 0$. Of course,
$f_1 = 0$ since a false alarm is only incurred if the stop action is picked in states $2,\ldots, X$.  The expected false alarm penalty is 
\beq  \sum_{i \in \X} f_i \E\{I(x_k=e_i,u_k=1)|\F_k\} = \f^\p \pi_k  I(u_k=1), \quad
\text{ where }
  \f = (f_1,\ldots,f_X)^\p, \; f_1 = 0. \label{eq:falsev}\eeq
The false alarm  vector $\f$ is chosen with increasing elements so that states  further
from  state~1 incur larger  penalties.

 Then with $\alpha, \beta$ denoting non-negative  constants that weight the relative importance of these costs, the 
 expected stopping cost at time $k$ is
 \beq \label{eq:cp1}
\Cb(\pi_k,u_k=1) = \alpha (G^\p \pi_k - (g^\p \pi_k)^2) + \beta \, \f^\p \pi_k . \eeq
 One can also view $\alpha$ informally as a Lagrange multiplier in a stopping time problem
 that seeks to minimize a cumulative cost (as in (\ref{eq:csdef}) below) subject to a variance stopping constraint.

(ii) {\em Delay cost of continuing}: 
We allow two possible choices for the delay costs: \\
(a) If decision $u_k=2$ is taken then 
 $ \{x_{k+1} = e_1, u_k = 2\}$ is the event that no change
is declared at time $k$ even though the state has changed at time $k+1$.
So with  $d$ denoting a non-negative constant,
 $d \,I(x_{k+1} = e_1, u_k=2)$ depicts a {\em delay cost}.
The expected delay cost  for decision $u_k=2$ is 
\beq \Cb(\pi_{k},u_k=2) = d \,\E\{I(x_{k+1}= e_1,u_k=2)|\F_{k}\} 
= d e_1^\p P^\p \pi_k .  \label{eq:exd} \eeq
The above cost is motivated by  applications (e.g., sensor networks) where  if the decision maker chooses $u_k=2$, then
it needs to gather observation $y_{k+1}$  thereby
incurring an additional operational cost denoted as $\cop$. Strictly speaking, $\Cb(\pi,2) = d e_1^\p P^\p \pi + \cop$.
Without loss of generality
 set the constant $\cop$ to zero, as it does not affect our structural results.
The penalty   $d \,I(x_{k+1} = e_1, u_k=2)$ gives incentive for
 the decision maker to predict the state $x_{k+1}$. \\
(b) 
Instead of the above, the more `classical'  formulation is that a delay cost is incurred when the event  $ \{x_{k} = e_1, u_k = 2\}$ occurs.
Then the expected delay cost is \beq  \Cb(\pi_{k},u_k=2) = d\,\E\{I(x_{k} = e_1, u_k=2) | \F_k\}
= d e_1^\p \pi_k  \label{eq:exd2}. \eeq

\noindent  {\em Remark}:  Due to the variance penalty, the   cost $\Cb(\pi,1)$ in (\ref{eq:cp1}) is quadratic in the belief state $\pi$. Therefore,
the formulation cannot be reduced to a standard stopping problem
with linear costs in the belief state. 

\subsection{Quickest Time Detection Objective} \label{sec:qdobjective}
Let $(\Omega,\mathcal{F})$ be the underlying measurable space where $\Omega = (\X \times \U \times \Y)^\infty$ is the product space, which is endowed with the product topology and $\mathcal{F}$  is the corresponding product sigma-algebra. For any $\pi_0\in \I$,  and policy $\mu \in \Mu$,
 there exists a (unique) probability measure $\P^\mu_{\pi_0}$ on  $(\Omega, \mathcal{F})$, 
   see 
 \cite{HL96} for details. Let $\Epzero$  denote the expectation with respect to the measure  $\P^\mu_{\pi_0}$.
  \footnote{The formulation on $\Omega = (\X \times \U \times \Y)^\infty$ is as follows, see \cite{HL96}. Augment $\I$ to include the  fictitious stopping state $e_{X+1}$ which is cost free, i.e.,
  $C(e_{X+1},u) = 0$ for all $u\in \U$. When decision $u_k=1$ is chosen, the belief state $\pi_{k+1}$ transitions to $e_{X+1}$ and remains there indefinitely.
   Then (\ref{eq:csdef}) is equivalent to
$ J_\mu(\pizero) = \Ep\{\sum_{k=1}^{\tau-1} \discount^{k-1} \Cb(\pi_{k},u_k=2)
+  \rho^{\tau-1} \Cb(\pi_{\tau},u_\tau = 1) + \sum_{k=\tau+1}^\infty \rho^{k-1} \Cb(e_{X+1},u_{k})
 \} $, where the last summation is zero.}
 
 Let $\tau$ denote  a stopping time adapted to the sequence of 
 $\sigma$-algebras $\F_k,k\geq 1$, defined in  (\ref{eq:sigf}).
 That is, with $u_k$ determined by decision policy (\ref{eq:actionpolicy}),  
\beq
\tau = \{ \inf k:  u_k = 1\} .   \label{eq:tauu}\eeq
 For each initial distribution $\pi_0 \in \I$, and policy
 $\mu$, the following cost  is associated:
\beq \label{eq:csdef}
J_\mu(\pizero) = \Ep\{\sum_{k=1}^{\tau-1} \discount^{k-1} \Cb(\pi_{k},u_k=2)
+  \rho^{\tau-1} \Cb(\pi_{\tau},u_\tau = 1)
 \}.
\eeq
Here $\discount \in [0,1]$ denotes an economic discount factor.
Since  $\Cb(\pi,1)$, $\Cb(\pi,2)$ are non-negative and bounded for all $\pi \in \I$, stopping is guaranteed
in finite time, i.e., $\tau$ is finite with probability 1 for any $\discount  \in [0,1]$ (including $\rho = 1$). 

\noindent {\em Remark}:  For the special case $\alpha =0 $, $\X = \{1,2\}$ (i.e., geometric distributed change time), $\f=e_2$,
and delay cost (\ref{eq:exd}),  it is easily shown that
\beq  \label{eq:altcost}J_\mu(\pizero) =   d \,\Ep\{(\tau - \tau^0)^+\} + \beta \,\P^\mu_{\pi_0}(\tau < \tau^0) +  d P_{21} \Ep\{ (\tau^0 - 1) I(\tau^0 < \tau) \}\eeq
 (where $\tau^0$ is defined in (\ref{eq:tau}) and $\tau$ is defined
in (\ref{eq:tauu})). For the delay cost (\ref{eq:exd2}),
 the cost function assumes the classical Kolmogorov--Shiryayev
criterion for detection of disorder \cite{Shi63},
 namely
 \beq J_\mu(\pizero) =   d \Ep\{(\tau - \tau^0)^+\} + \beta \P^\mu_{\pi_0}(\tau < \tau^0) .
\label{eq:ksd} \eeq
    \qed


The goal is to determine the change time $\tau^0$ defined in (\ref{eq:tau})  with minimal cost, that is, compute the optimal policy $\mu^* \in \Mu$ to minimize (\ref{eq:csdef}),
i.e., $J_{\mu^*}(\pizero) = \inf_{\mu \in \Mu} J_\mu(\pizero)$. The existence of an optimal stationary policy $\mu^*$ follows from \cite[Prop.1.3, Chapter 3]{Ber00b}.
Considering  the above  cost (\ref{eq:csdef}),
the optimal stationary policy $\mu^*: \I \rightarrow \u$ and associated value function
 $\Vb(\pi)$
are the solution of the following
 ``Bellman's dynamic programming  equation'' 
\begin{align} \label{eq:dp_initial}
\mu^*(\pi)&= \arg\min\{ \Cb(\pi,1), \;\Cb(\pi,2)
+ \discount \sum_{y \in \Y}  \Vb\left( \Tp(\pi ,y) \right) \sigp(\pi,y)\} \\
J_{\mu^*}(\pi) &= \Vb(\pi) = \min \{ \Cb(\pi,1),\; \Cb(\pi,2)
+ \discount \sum_{y \in \Y}  \Vb\left( \Tp(\pi ,y) \right) \sigp(\pi,y)\}
  \nonumber
\end{align}
Before proceeding, we rewrite the above in a form that is more amenable\footnote{The  reason for changing coordinates from $\Cb(\pi,1)$,  $\Cb(\pi,2)$ 
to $C(\pi,1)$, $C(\pi,2)$ is to make our analysis compatible with
existing results in quickest time detection. To ensure this compatibility, we  need $C(\pi,1)$ to be decreasing with respect to the MLR order (see Appendix)
when $\alpha = 0$. As shown in the proof of Theorem \ref{thm:1} in the Appendix, $C(\pi,1)$ is MLR decreasing.
In  comparison  $\Cb(\pi,1)$ is not MLR deceasing. Of course, the stopping set $\mathcal{R}_1$ (see (\ref{eq:stopset})) and optimal policy $\mu^*$ are invariant to the choice of coordinates. This idea of changing coordinates is  described in \cite{HS84}, albeit for the  simpler  fully observed 
Markov decision process case.}
 for analysis.
Define 
\begin{align}
V(\pi) &= \Vb(\pi) -  (\alpha+\beta)  \f^\p \pi,
\quad C(\pi,1) = \alpha(G^\p \pi - (g^\p \pi)^2 ) - \alpha \f^\p \pi  \nonumber \\
C(\pi,2) &= \Cb(\pi,2) - (\alpha+\beta) \f^\p \pi + \rho(\alpha+\beta) \f^\p P^\p \pi.
%
 \label{eq:costdef} \end{align}
Then clearly $V(\pi)$ satisfies Bellman's dynamic programming  equation 
\begin{align} \label{eq:dp_alg}
\mu^*(\pi)&= \arg\min_{u \in \U} Q(\pi,u) , \;J_{\mu^*}(\pi) = V(\pi) = \min_{u \in \{1,2\}} Q(\pi,u),\\
 \text{ where }  Q(\pi,2) &=  C(\pi,2)
+ \discount \sum_{y \in \Y}  V\left( \Tp(\pi ,y) \right) \sigp(\pi,y),\quad
Q(\pi,1) =   C(\pi,1)
  \nonumber
\end{align}
Thus the goal is to determine the optimal stopping set 
\begin{align} 
\R_1 &=  \{\pi \in \I : \mu^*(\pi) = 1\} = \{\pi \in \I:  C(\pi,1) <  C(\pi,2)
+ \discount \sum_{y \in \Y}  V\left( \Tp(\pi ,y) \right) \sigp(\pi,y) \}  \nonumber \\
&=\{\pi \in \I:  \Cb(\pi,1) <  \Cb(\pi,2)
+ \discount \sum_{y \in \Y}  \Vb\left( \Tp(\pi ,y) \right) \sigp(\pi,y) \}.
 \label{eq:stopset}
 \end{align}
In Sec.\ref{sec:discussion}, sufficient conditions are given to ensure that $\R_1$ is non-empty.

\noindent {\em Value Iteration Algorithm and methodology}:
We comment briefly here on our analysis methodology which is detailed
in Sec.\ref{sec:main}.
Let $k=1,2,\ldots,$ denote iteration number (the fact that we used
$k$ previously to denote time  should not result in confusion).
The   value iteration
algorithm is  a fixed point iteration of 
Bellman's equation and proceeds as follows:
\begin{align}\nonumber
V_0(\pi) &= -(\alpha+\beta) \f^\p  \pi, \quad V_{k+1}(\pi)= \min_{u \in \u} Q_{k+1}(\pi,u), \quad
\mu^*_{k+1}(\pi)= \argmin_{u \in \u} Q_{k+1}(\pi,u) \\
\text{ where } & Q_{k+1}(\pi,2) =  C(\pi,2) 
+ \discount \sum_{y \in \Y}  V_k\left( T(\pi,y) \right) \sigma(\pi,y),
\;
Q_{k+1}(\pi,1) = C(\pi,1). 
\label{eq:vi}
\end{align}
Let $\mathcal{B}(X)$ denote the set of bounded real-valued functions on $\I$.
Then for any $V$ and  $\tilde{V} \in \mathcal{B}(X)$, define the sup-norm metric
$\sup\|V(\pi) - \tilde{V}(\pi)\|$, $\pi \in \I$. Then $\mathcal{B}(X)$ is a Banach
space. The value iteration algorithm (\ref{eq:vi}) will generate a sequence of value functions
$\{V_k\} \subset \mathcal{B}(X)$ that will converge uniformly (sup-norm metric) as $k\rightarrow \infty$ to $V(\pi) \in \mathcal{B}(X)$, the optimal value
 function of Bellman's equation.
However, since the belief state space $\I$ is an uncountable  set, the
value iteration algorithm (\ref{eq:vi}) do not 
translate into practical solution methodologies as
$V_k(\pi) $ needs to be evaluated at each $\pi \in \I$, an
uncountable set.
Indeed, due to the nonlinearity in the belief states,
the formulation is more complex than a partially observed Markov decision process
which is known to be 
PSPACE hard \cite{PT87}. Although value iteration is not useful from a computational point of view, in Sec.\ref{sec:main}, we exploit the
structure of the value iteration 
recursion (\ref{eq:dp_alg}), (\ref{eq:vi}) to prove that $\R_1$ is characterized by a threshold switching curve.
We then exploit this structure to devise polynomial complexity algorithms
for approximating the optimal  policy $\mu^*$ and thus determining the stopping set $\R_1$.

{\em Remark}:
Computational algorithms based on value iteration 
such as Sondik's algorithm, Monahans's algorithm, Cheng's algorithm, Witness
algorithm (see \cite{Cas98,POMDP} for a tutorial description) and Lovejoy's suboptimal algorithms \cite{Lov91b}
solve POMDPs with linear costs (i.e.,  $\alpha = 0$) over finite horizons.  
These algorithms
require finite observation spaces and are computationally intractable except for small $\X$ and $\Y$.
They are not applicable directly to stopping problems considered in this paper since
we consider nonlinear penalty costs, possibly continuous observation
space $\Y$ (Examples 1,  2 and 3), and problems where the observation probabilities depend on the belief state (social learning
in Examples 4 and 5).

\section{Example 1:  Quickest Time Detection with PH-distributed Change Time and Variance Penalty} \label{sec:main}

This section  considers  quickest time detection with PH-distributed change time and variance penalty.
 Sec.\ref{subsec:main} below gives the main results of this paper, namely the optimal decision policy
is characterized by  a threshold curve. Sec.\ref{sec:discussion} and \ref{sec:assd}
discuss the implications
and main assumptions.
 Sec.\ref{sec:linear} then parametrizes the optimal linear approximation
 to this threshold curve. Finally, Sec.\ref{sec:spsa}  gives a stochastic
 optimization algorithm (Algorithm \ref{alg1}) to compute this optimal linear approximation.

The quickest time detection problem is a special case of the
  stochastic control problem formulated  in Sec.\ref{sec:qdbasic}. 
 The states $2, 3,.\ldots, X$ are fictitious and are  defined to generate  the change
 time $\tau^0$ with PH-distribution (\ref{eq:nu}).
So   states  $2, 3,.\ldots, X$ are indistinguishable in terms of the observation $y$.
That is, the observation probabilities $B$ in (\ref{eq:obs}) and Markov chain state levels $g$ in (\ref{eq:varder}) satisfy
 \beq B_{2y} =B_{3y} = \cdots = B_{Xy} \text{ for all } y\in \Y, \quad g_1 = 0,\;  g_2 = g_3= \cdots = g_X = 1.
\label{eq:obsin}
\eeq
The above choice of $g= \ones_X-e_1$ is without loss of generality since  the variance penalty
(\ref{eq:varder})  is translation invariant with respect to $c \ones_X$ for any $c$.

{\em Notation}: 
Notation and definitions regarding stochastic orders, lattice
programming, 
the poset $[\I,\gr]$ and submodularity
 are given in Appendix \ref{sec:mlrdef}. Below $\gr$ denotes the monotone likelihood ratio order, $\glX$ denotes the likelihood ratio order on lines $\l(e_X,\bp)$,
$\glone$  denotes the likelihood ratio order on lines $\l(e_1,\bp)$,  and
$\gs$ denotes first order stochastic dominance.

\subsection{Main Result: Existence of Decision Curve Policy for
Quickest Time Detection} \label{subsec:main}

This  section  gives three main results:
The
 optimal policy  for quickest detection with PH-distributed change time and variance penalty
 is characterized by  a threshold curve (Theorems \ref{thm:1} and \ref{thm:modified}). Also for $\alpha = 0$, it is  shown that the stopping set $\R_1$ is convex
 (Theorem \ref{cor:ph}). 

\subsubsection{Quickest Detection with Delay Penalty (\ref{eq:exd})}
For the stopping cost $\Cb(\pi,1)$ in  (\ref{eq:cp1}), choose    $\f = [0, 1, \cdots, 1]^\p = \ones_X - e_1$.
This  weighs the states $2,\ldots,X$ equally in the false alarm penalty. 
With assumption (\ref{eq:obsin}), the variance penalty  (\ref{eq:varder}) becomes $\alpha(e_1^\p  \pi - (e_1^\p \pi)^2)$.
The delay cost $\Cb(\pi,2)$ is chosen as (\ref{eq:exd}).
   To summarize (\ref{eq:costdef}), (\ref{eq:dp_alg}), (\ref{eq:stopset}) hold with
\beq  \label{eq:stylized}
\Cb(\pi,1) = \alpha\left(e_1^\p \pi - (e_1^\p \pi)^2\right) + \beta(1-e_1^\p \pi), \quad
\Cb(\pi,2)  = d e_1^\p P^\p \pi. \eeq

Theorem \ref{thm:1} below is our main  result on the structure of the optimal decision  policy $\mu^*(\pi)$.
It is based on the following assumptions (discussed in Sec.\ref{sec:assd}).
\begin{itemize}
\item[(A1-Ex1)] 
$ d \geq \rho(\alpha+\beta)$
\item[(A2)] The observation distribution $B_{xy}$ in (\ref{eq:obs}) is TP2 in $(x,y)$ (see Defn.\ref{def:tp2}(iii) in Appendix \ref{sec:mlrdef}). Equivalently, from (\ref{eq:obsin}), 
 $B_{2y} \gr B_{1y}$.

\item[(A3)] The transition matrix $P$ in (\ref{eq:phmatrix}) is TP2, i.e. all its second order minors
are non-negative.

\item[(S-Ex1)] 
$(d-\rho(\alpha+\beta))(1-P_{21}) \geq  \alpha -\beta $
 \end{itemize}

 (A1-Ex1) and (S-Ex1) are constraints on the delay and stopping cost functions (that the decision maker can design), while (A2) and (A3) are assumptions
on the underlying observation (\ref{eq:obs}) and PH-distribution (\ref{eq:tau}).

\begin{theorem}[Switching Curve Optimal Policy] \label{thm:1} 
Consider the  quickest time detection problem (\ref{eq:csdef}) with costs defined in (\ref{eq:stylized})
 and PH-distributed change time $\tau^0$ defined in (\ref{eq:tau}).
Then for $\pi \in \I$,  under (A1-Ex1), (A2), (A3), (S-Ex1), there
  exists an optimal policy $\mu^*(\pi)$ that is $\glX$ increasing
on lines $\l(e_X,\bp)$ and $\glone$ increasing on lines $\l(e_1,\bp)$.
As a consequence:\\
(i) The stopping set $\R_1$ defined in (\ref{eq:stopset}) has the following structure:
There exists a 
 threshold switching curve $\Gamma$ 
that partitions 
 belief state space $\I$ into two individually connected
regions $\R_1$, $\R_2$, such
that
the optimal   policy is
\beq  \mu^*(\pi) = \begin{cases} \text{continue} = 2 & \text{ if } \pi \in \R_2 \\
                                  \text{stop} =1 & \text{ if } \pi \in \R_1 \end{cases}
                    \label{eq:mustar}              \eeq
                    (A set  is connected if it cannot be expressed as the union of two
                    disjoint nonempty closed sets \cite{Rud76}). 
                  The threshold curve $\Gamma$ intersects each line $\l(e_X,\bp)$ and $\l(e_1,\bp)$ at most once.
                     \\
                  (ii)   
                  There exists
an $i^* \in \{0,\ldots,X\}$, such that  $e_1,e_2,\ldots,e_{i^*} \in \R_1$ and $e_{i^*+1},\ldots,e_X \in \R_2$. 
\\
                 (iii) For geometric distributed change time $\tau^0$, there exists a unique threshold point $\pi^*(2)$  such that 
(\ref{eq:onedim}) holds.
(Note (A3) holds trivially in this case).
  \qed            
\end{theorem}

Theorem \ref{thm:1}  is proved in Appendix \ref{sec:appth1} and uses meta Theorem \ref{thm:key} in Appendix \ref{sec:applp} as a key step. The intuition behind Theorem \ref{thm:1} is discussed in Sec.\ref{sec:discussion} and \ref{sec:assd} below. Fig.\ref{fig:structure} gives a pictorial illustration.
Note that if $\alpha=0$, then (S-Ex1) holds trivially if (A1-Ex1) holds.

\subsubsection{Quickest Detection with Delay Penalty (\ref{eq:exd2})}
Next  consider the `classical' delay cost $\Cb(\pi,2)$ in  (\ref{eq:exd2}) and stopping cost $\Cb(\pi,1)$ in (\ref{eq:cp1}) with $g$
in (\ref{eq:obsin}).
 Then (\ref{eq:costdef}), (\ref{eq:dp_alg}), (\ref{eq:stopset}) hold with
\beq  \Cb(\pi,1) = \alpha\left(e_1^\p \pi - (e_1^\p \pi)^2\right)  + \beta \, \f^\p \pi_k , \quad
\Cb(\pi,2) =  d e_1^\p\pi . \label{eq:modified2}\eeq
Below we show that Theorem \ref{thm:1} continues to hold, if the decision maker designs
the false alarm vector $\f$ to satisfy the following 
 linear constraints:
\begin{itemize}
\item[(AS-Ex1)]
 (i) $f_i \geq \max\{1, \rho \frac{\alpha+\beta}{\beta} \f^\p P^\p e_i + \frac{\alpha-d}{\beta}\}$, $i\geq 2$.\\
(ii) $f_j - f_i  \geq \rho \f^\p P^\p( e_j -e _i) $, $j \geq i, i \in \{2,\ldots,X-2\}$ \\
(iii) $f_X - f_i \geq \frac{\rho(\alpha+\beta)}{\beta} \f^\p P^\p (e_X-e_i)$, $i \in \{2,\ldots,X-1\}$.
\end{itemize}
Feasible choices of $\f$ are easily obtained by a linear programming solver.

\begin{theorem} \label{thm:modified}
Consider the quickest detection problem with delay and stopping costs in  (\ref{eq:modified2}).
 Then under (AS-Ex1), (A2), (A3), Theorem 1 holds.  \qed
\end{theorem}

\subsubsection{Convexity of Stopping Region when $\alpha = 0$}
Finally, we present the following result for the case $\alpha= 0$,
 i.e., no variance penalty.

\begin{theorem} \label{cor:ph}
For arbitrary PH-distributed change time $\tau^0$, and no variance penalty ($\alpha = 0$),  the stopping region    $\R_1   $ is a convex subset of $\I$.
\qed
  \end{theorem}

The proof of Theorem \ref{cor:ph} is in Appendix \ref{app:thm2}; it was proved in \cite{Lov87a} in a POMDP setting.
Theorem~\ref{cor:ph} says that  as long as costs $\Cb(\pi,1)$, $\Cb(\pi,2)$  are linear in $\pi$
 (i.e., no variance penalty), then the  stopping set is convex for any size $X$ (i.e., arbitrary PH-distribution);  no assumptions are required
 on the transition matrix $P$ or observation likelihood matrix $B$.
However, even though $\R_1$ is convex (and therefore connected), Theorem \ref{cor:ph} does not guarantee that $\R_2$ is connected.
As described in Sec.\ref{sec:assd}, Theorem \ref{thm:1} and Theorem \ref{thm:modified} go much further than Theorem \ref{cor:ph} in characterizing $\R_1$ and $\R_2$,
even for the case $\alpha = 0$.

\subsection{Discussion of Theorems \ref{thm:1}, \ref{thm:modified}} \label{sec:discussion}

Theorems \ref{thm:1} and \ref{thm:modified} imply that  since the optimal decision policy
is characterized 
by a threshold curve,  quickest time detection for PH-distributed change times and variance penalty can be implemented 
efficiently, see Sec.\ref{sec:linear}.
 Without this result, the stopping set $\R_1$
 is not necessarily a connected region as will be shown
in Sec.\ref{sec:numerical}.

\subsubsection{Geometric distributed change time}  When the  change time $\tau^0$ is geometrically distributed,
since the state space $\X=\{1,2\}$, $\I$ is a one dimensional simplex.  Then the stochastic orders $\gr$, $\gl$ and $\gs$ defined in Appendix \ref{sec:mlrdef} coincide, and become  total orders. 
Also (A3) holds automatically for this case. 
Below we discuss the cases of $\alpha\neq 0$ and $\alpha  = 0$.\\
(i) $\alpha\neq 0$:
For  geometric distributed change time, Theorems \ref{thm:1} and 
\ref{thm:modified} say that the
classical threshold policy depicted in  (\ref{eq:onedim}) continues to hold  when a  nonlinear variance penalty
is considered.
For example,
consider Theorem \ref{thm:modified} with non-zero $\alpha$,  delay in (\ref{eq:modified2}) and
false alarm vector $\f^\p=e_2$. So the
 false alarm cost is $\f^\p \pi = 1 - e_1^\p \pi$ (which
is identical to (\ref{eq:stylized})). One can view this as  the  Kolmogorov-Shiryayev criterion (\ref{eq:ksd})  with an additional variance penalty.  Theorem~\ref{thm:modified} holds under
the conditions (AS-Ex1) and (A2). Here (AS-Ex1) equivalent to the constraint that
$\alpha \leq \frac{d + \beta(1-\rho P_{22})}{1+ \rho P_{22}}$.
 (Choose
$f_1=0$, $f_2=1$
in (AS-Ex1)(i)).  So for $\alpha \leq d/2$, (AS-Ex1) always holds. 
\\
(ii) $\alpha = 0$: For  quickest time detection with geometric distributed change time and no variance penalty ($\alpha = 0$), the 
well known existence of a threshold point (e.g., for the   Kolmogorov--Shiryayev
criterion~(\ref{eq:ksd}))
follows trivially from Theorem~\ref{cor:ph}.
  Since
$\I$ is a one dimensional simplex, convexity of 
stopping set  $\R_1$ (Theorem \ref{cor:ph}) implies that there is a threshold point $\pi^*$ that satisfies (\ref{eq:onedim}).

\subsubsection{Avoiding trivial cases}

To ensure the stopping set $\R_1$ contains state $e_1$,  assume $\Cb(e_1,1)<\Cb(e_1,2)$.  From (\ref{eq:stylized}) or (\ref{eq:modified2}) this is equivalent
to $d>0$.
The strict inequality also implies that  $\R_1 \cap \Io$ is non-empty, where
 $\Io$ denote the interior of the simplex $\I$.
 
For the detection problem to be non-trivial, we want  $\Cb(e_i,1) > \Cb(e_i,2)$ for $i\geq  2$, otherwise  it is always optimal to stop at time 1. 
For the case of Theorem \ref{thm:1}, from (\ref{eq:stylized}), a sufficient  condition  is
that 
$ \beta > d P_{i1} $, $i=2,\ldots,X$.
Since the transition matrix  $P$ is TP2, it follows\footnote{Proof: We prove the 
contrapositive, that is, $P_{i1}<P_{i+1,1}$ implies $P$ is not TP2.
Recall from (A3-Ex1), TP2 means that $P_{i1} P_{i+1,j} \geq P_{i+1,1} P_{ij}$ for all $j$. So assuming  $P_{i1}<P_{i+1,1}$, to show
that $P$ is not TP2, we need to show that there is at least one $j$ such that
$P_{i+1,j} < P_{ij}$. But $P_{i1}<P_{i+1,1}$ implies $\sum_{k\neq 1} P_{i+1,k} < \sum_{k\neq 1} P_{ik}$, which in turn implies that 
at least for one $j$, $P_{i+1,j} < P_{ij}$. \label{foot}}
 that $P_{21} \geq P_{31} \geq \cdots \geq P_{X1}$. Therefore, 
it is sufficient for  the decision maker to choose constants $\beta$ and $d$ such
that $  \beta > d  P_{21}$. For  Theorem \ref{thm:modified}, 
from (\ref{eq:modified2}), $\Cb(e_i,1) > \Cb(e_i,2)$ always
holds for $\beta>0$ and $f_i>0$, $i\geq 2$.

\subsubsection{Non-degenerate Threshold Curve} \label{sec:nondeg}
Let $\Io$ and $\Ib$, respectively, denote the interior and boundary of the simplex $\I$.
Determining the threshold switching curve $\Gamma$ in Theorem \ref{thm:1}  requires determining $\Gamma \cap \Io$ (portion
of  curve
that lies in the interior of the simplex) and 
$\Gamma \cap \Ib$ (portion of  curve that lies on the boundary of the simplex).
Since $\Ib$ comprises of sub-simplices, to determine
$\Gamma \cap \Ib$
one would need to search for the threshold curve $\Gamma$ within these sub-simplices.
While
conceptually straightforward,
we can eliminate this search  by  ensuring that the belief state $\pi$ always lives in the interior $\Io$ of the simplex. The following lemma gives sufficient conditions for the sequence of belief states $\pi_k$ over time to lie
in $\Io$.

\begin{lemma}\label{lem:nondeg}
Suppose each column of transition matrix $P$ has at least one non-zero element, and the observation likelihoods satisfy $B_{1y} \neq 0$ for $y \in \Y$. Then
for initial belief state $\pi_0$ satisfying (\ref{eq:init}) with $\pi_0(i) \neq 0$, $i >1$,  subsequent belief states $\pi_k$ lie in $ \Io \cup \{e_1\}$ for all time $k\geq 1$.
\end{lemma}

The proof follows straightforwardly from the belief state update (\ref{eq:hmm}).
 Since the sequence of belief states $\pi_k$, $k\geq 1$, lives in  $\Io$, 
 one only needs to compute the threshold curve inside the simplex, i..e.,  $\Gamma \cap \Io$.
 Recall from the previous remark that $\R_1 \cap \Io$ is non-empty and consequently $\Gamma \cap \Io$ is non-empty.

\subsection{Assumptions and Proofs of Theorem \ref{thm:1} and Theorem \ref{thm:modified}} \label{sec:assd}
Below we discuss the  main assumptions of Theorem \ref{thm:1} and Theorem \ref{thm:modified} in Sec.\ref{subsec:main}, then outline the structure 
of the proof, and finally give  intuitive examples that illustrate the structure of  stopping set
$\R_1$.

\subsubsection{Discussion of Assumptions}
Recall (A1-Ex1) and (S-Ex1) are design constraints the decision maker uses to choose
the stopping and delay costs.
In contrast, (A2) and (A3) are assumptions on the underlying stochastic model.

As described in the Appendix \ref{sec:applp}, {\bf (A1-Ex1)} is sufficient for $C(\pi,2)$ 
to be $\gr$ decreasing. We also require $C(\pi,1)$ in (\ref{eq:stylized}) to be $\gr$ decreasing,  but this  holds trivially
in our setup.

({\bf S-Ex1)} is a submodularity condition, see Defn.\ref{def:supermod} in Appendix. We refer to 
\cite{Ami05,Top98} for extensive treatments of lattice programming and submodularity. 
The key idea  is that if $Q(\pi,u)$ is submodular on the partially ordered
set $[\I,\gr]$, then the optimal policy $\mu^*(\pi) = \arg\min_u Q(\pi,u)$ is monotone increasing
with respect to $\gr$.

In our setting, submodularity of $Q(\pi,u)$  in  (\ref{eq:dp_alg})  is equivalent to showing that $Q(\pi,2) - Q(\pi,1)$ is decreasing with respect to $\pi$
(in terms of the MLR order $\gr$, see discussion of structure of proof given below). 
Since by (A1-Ex1), $C(\pi,1)$ and $C(\pi,2)$ are MLR decreasing in $\pi$, it will be proved
 that $V(\pi)$ is MLR decreasing
in $\pi$ providing (A2) and (A3) hold. Since the sum of submodular functions is submodular,
 establishing submodularity of $Q(\pi,u)$ in  (\ref{eq:dp_alg}) is equivalent to establishing submodularity of $C(\pi,u)$.
Clearly if $\alpha = 0$, then from (\ref{eq:stylized}), $C(\pi,1) = 0$ is independent of
$\pi$. So if $\alpha = 0$ then the submodular condition (S-Ex1) holds trivially since  $C(\pi,2)$ is decreasing in $\pi$ via (A1-Ex1). So for quickest time detection with PH-distributed change time
and no variance penalty, submodularity holds by construction.
Note that when the variance penalty is included,
(S-Ex1) always holds if $\alpha < \beta$.

{\bf AS-Ex1} in Theorem \ref{thm:modified} ensures that $C(\pi,1)$ and $C(\pi,2)$ 
in (\ref{eq:costdef}) for the modified delay cost in  (\ref{eq:modified2})
are monotone decreasing and that $C(\pi,u)$ is submodular. It  is analogous to (A-Ex1) and (S-Ex1).

{\bf (A2)} is required
for preserving the MLR ordering with respect to observation $y$ of the Bayesian filter update
$T(\pi,y)$ - this is a key step in showing $V(\pi)$ is MLR decreasing in $\pi$. Theorem~\ref{thm:key}(ii) in the Appendix states that $T(\pi,y)$ is MLR increasing in $y$, iff
(A2) holds. 

(A2) 
is satisfied by numerous continuous and discrete distributions, see a classical
detection theory book such as \cite{Poo93}. Examples include Gaussians, Exponential, Binomial, Poisson, etc.

{\bf (A3)} is essential for  the Bayesian update $T(\pi,y)$ 
 preserving monotonicity with respect to $\pi$. Theorem \ref{thm:key}(1) in the appendix shows that 
 $T(\pi,y)$ is MLR increasing in $\pi$ iff $P^\p \pi$ is MLR increasing in $\pi$, and
 (A3) is a sufficient condition for the latter. TP2 stochastic orders and kernels  have been
 studied in great detail in \cite{KR80}.
 
 (A3) is
 satisfied by several classes of  transition matrices; see  \cite{Kij97,KK77}.
Consider, for example, a  tridiagonal transition probability matrix $P$ with
 $P_{ij} = 0$ for $ j\geq i+2$ and $j \leq i-2$. As shown in
 \cite[pp.99--100]{Gan60}, a necessary and sufficient condition for
 tridiagonal $P$ to be TP2 is that $P_{ii} P_{i+1,i+1} \geq P_{i,i+1}
 P_{i+1,i} $. 

\subsubsection{Structure of Proof of Theorem \ref{thm:1}}
The proof  in the appendix
comprises of three steps. Steps 1 and 2 below are proved in meta Theorem \ref{thm:key} in Appendix \ref{sec:applp} under general
conditions (A1), (A2), (A3) and (S). \\
{\em Step 1}: We first show  that
the value function $V(\pi)$ is MLR decreasing (see Appendix \ref{sec:mlrdef} for definition). As shown in the proof, the general conditions
(A1), (A2), (A3) are sufficient  for $V(\pi)$ to be $\gr$ decreasing on $\I$.
This involves showing that $C(\pi,1)$, $C(\pi,2)$ are MLR decreasing.  (A1)(i) and (A1)(ii) in the appendix are sufficient conditions for this.
The proof that (A1)(i)
is sufficient  for $C(\pi,1)$ to be MLR decreasing is similar in spirit to the
Schur-convexity proof (Theorem A.3 of \cite{MO79}) with the 
difference that in Schur convexity the vectors $\pi$ have elements in
ascending order while in our case the elements of $\pi$ can be in any
order.  

 Conditions (A2) and (A3) are  required for MLR monotone updates $T(\pi,y)$ of the belief state, and also
first order stochastic dominance monotonicity of $\sigma(\pi,y)$, see Appendix for definition.

\begin{figure} \centering
\includegraphics[scale=0.4]{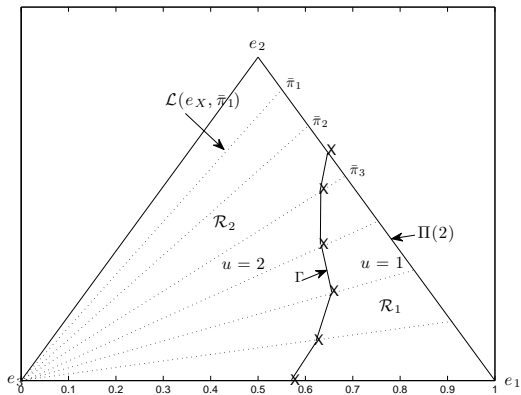}
\caption{Illustration of threshold switching decision curve $
\Gamma$. Here $X=3$ and hence  $\I$ is an equilateral
triangle. Theorem
\ref{thm:1} shows that the stopping region $\R_1$  is a connected and $e_1\in \R_1$. Also  
$\R_2$ is connected.
 The lines
segments $\l(e_X,\bp_1) $ connecting the sub-simplex $\Pi(2)$ to $e_3$ are
defined in (\ref{eq:lines}). Theorem \ref{thm:1} says that the threshold curve $\Gamma$ can intersect
each line $\l(e_X,\bp) $ only once (and similarly intersect each line $\l(e_1,\bp)$ only once).
In the special case $\alpha =0$, Theorem \ref{cor:ph} says that $\R_1$ is a  convex set.
} \label{fig:structure}
\end{figure}

\noindent {\em Step 2}: We then prove that  $Q(\pi,u)$ is submodular on $[\l(e_X,\bp),\glX]$ and $[\l(e_1,\bp),\glone]$.
 (S) is sufficient for $C(\pi,u)$ to be submodular on lines $\l(e_X,\bp)$ and $\l(e_1,\bp)$.
 Since we only require submodularity on lines $\l(e_X,\bp)$  and $\l(e_1,\bp)$,  and these are chains (i.e., totally ordered subsets
 of a partially ordered set), the condition (S) is less restrictive that requiring submodularity
 on the entire
simplex $\I$. Finally (A1),(A2),(A3),(S) are sufficient for  $Q(\pi,u)$ to be  submodular
 on lines $\l(e_X,\bp)$ and $\l(e_1,\bp)$. 
 So Theorem \ref{thm:key} in Appendix \ref{sec:applp}
implies a monotone policy on each chain  $[\l(e_{X},\bp),\glX]$. 
So there exists
a threshold belief state on each line where the optimal policy switches
from 1 to 2.  (A similar argument holds for lines  $[\l(e_{1},\bp),\glone]$). \\
{\em Step 3}: Step 3 is proved in Appendix \ref{sec:appth1}. The entire simplex $\I$ can
be covered by the union of lines $\l(e_X,\bp)$. The union of the resulting
threshold belief states yields the threshold curve $\Gamma(\pi)$. This is illustrated in
Fig.\ref{fig:structure}.

\subsubsection{Some Intuition}  
\begin{figure} \centering
\mbox{\subfigure[Example 1]
{\epsfig{figure=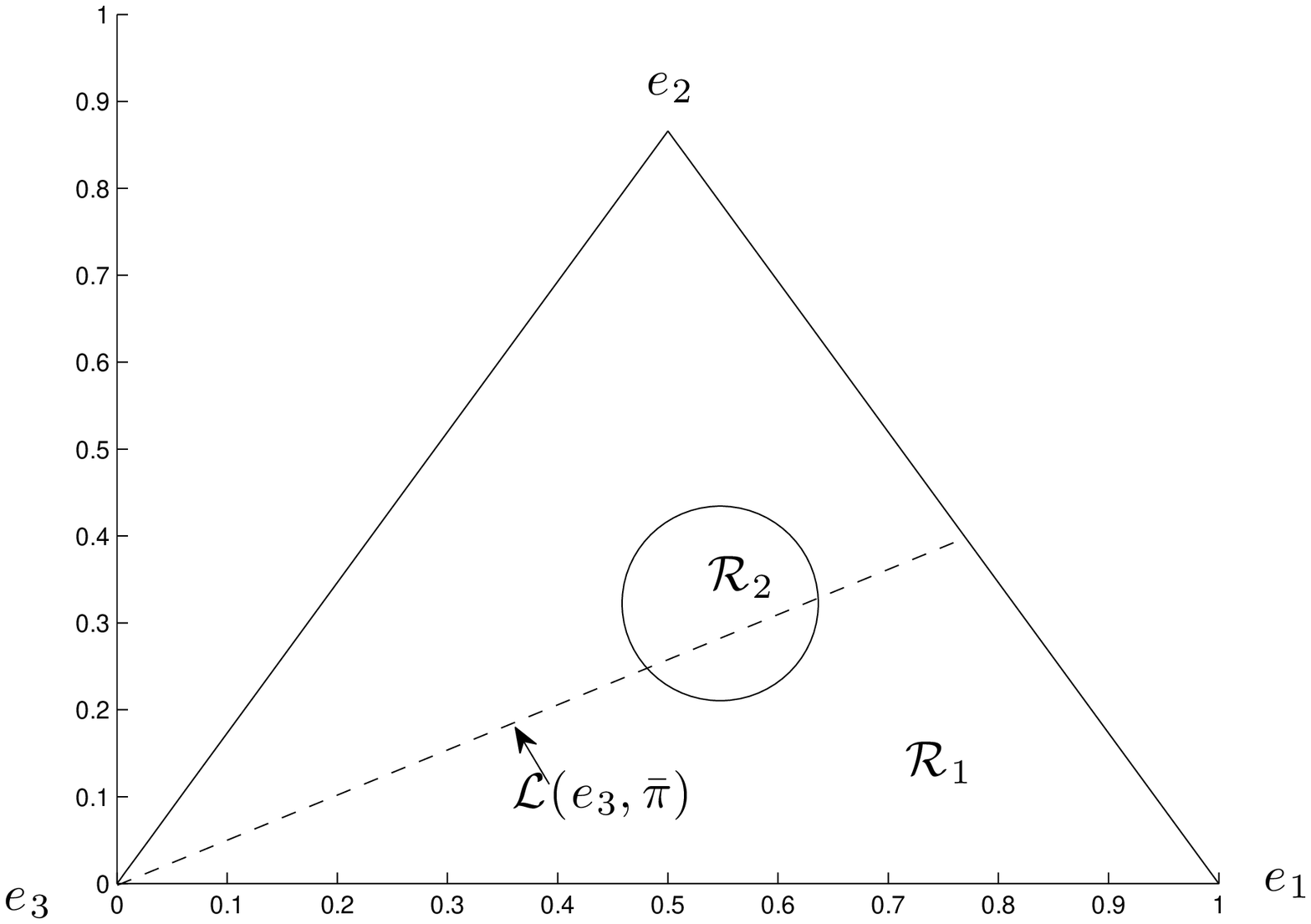,width=0.31\linewidth}} \quad
\subfigure[Example 2]{\epsfig{figure=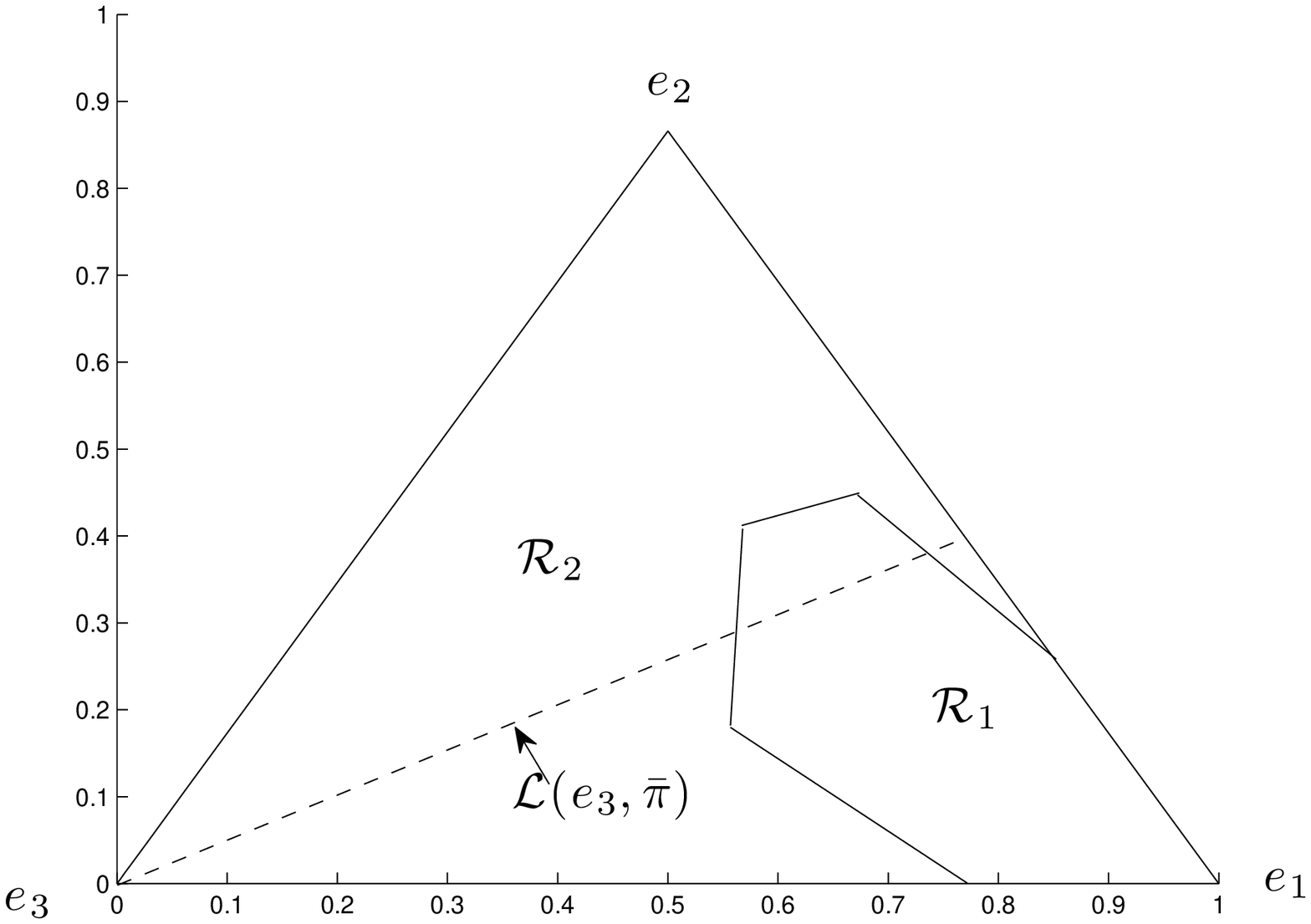,width=0.31\linewidth}} \quad
\subfigure[Example 3]{\epsfig{figure=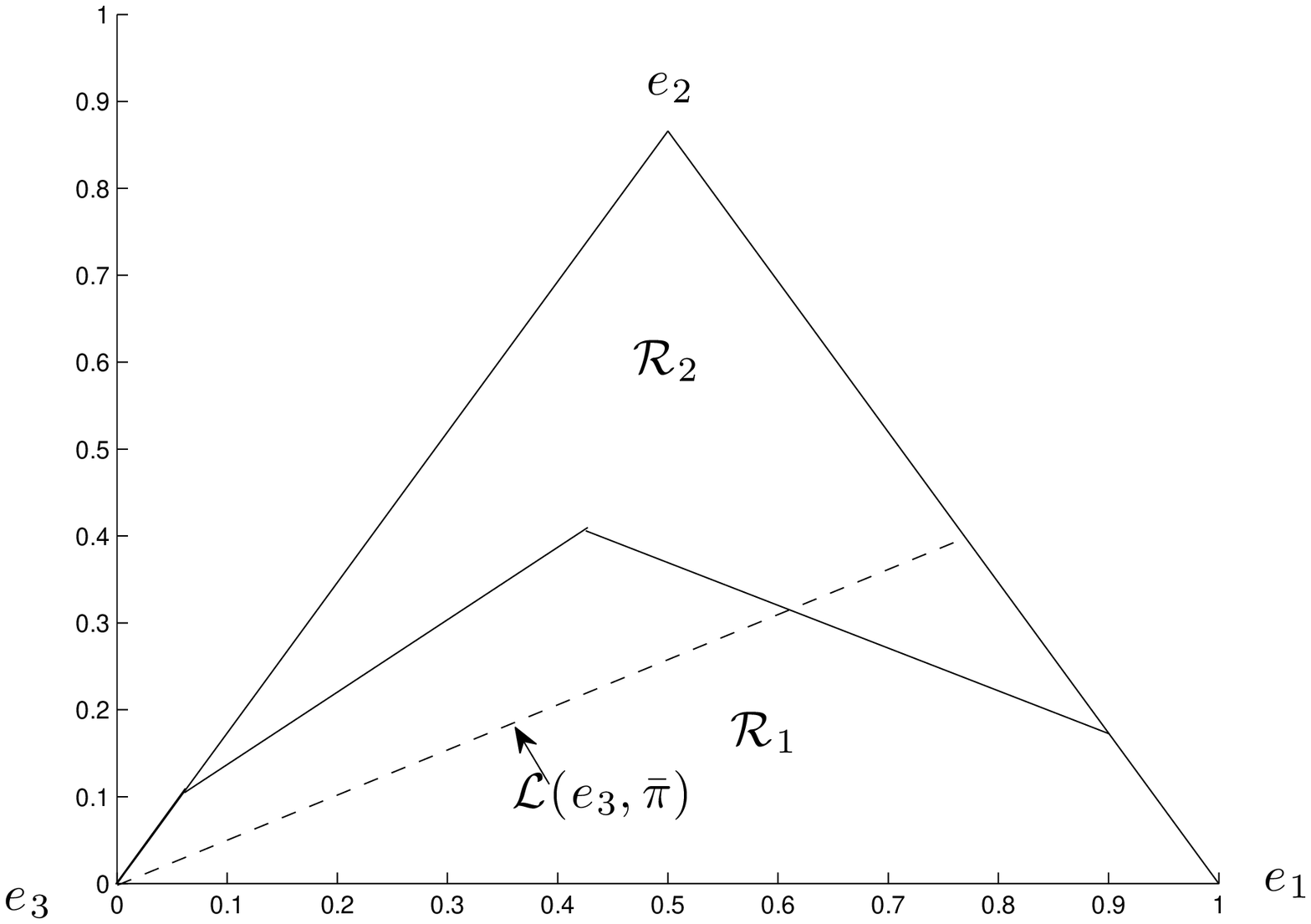,width=0.31\linewidth}}}
\mbox{\subfigure[Example 4]{\epsfig{figure=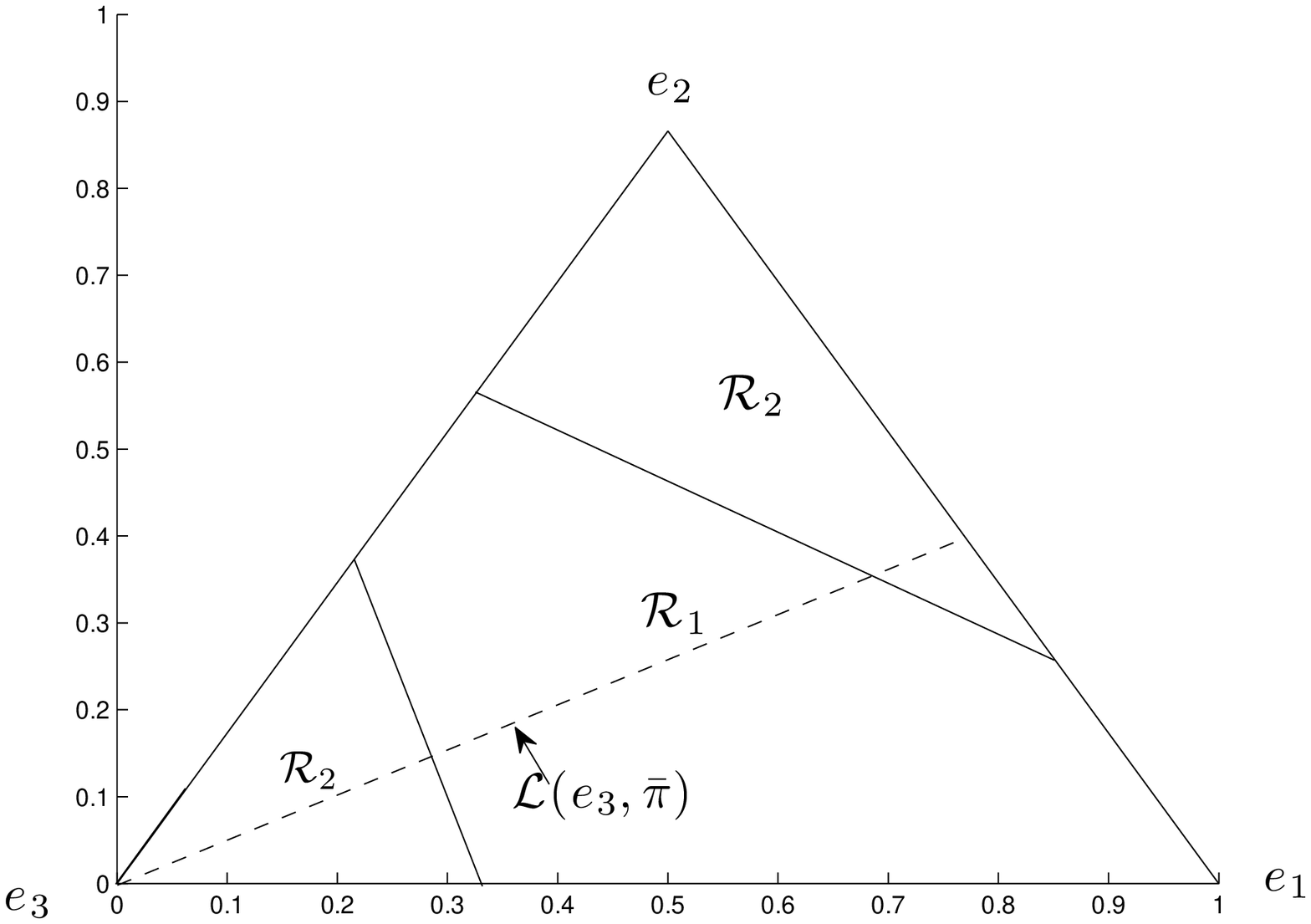,width=0.31\linewidth}} \quad
\subfigure[Example 5]{\epsfig{figure=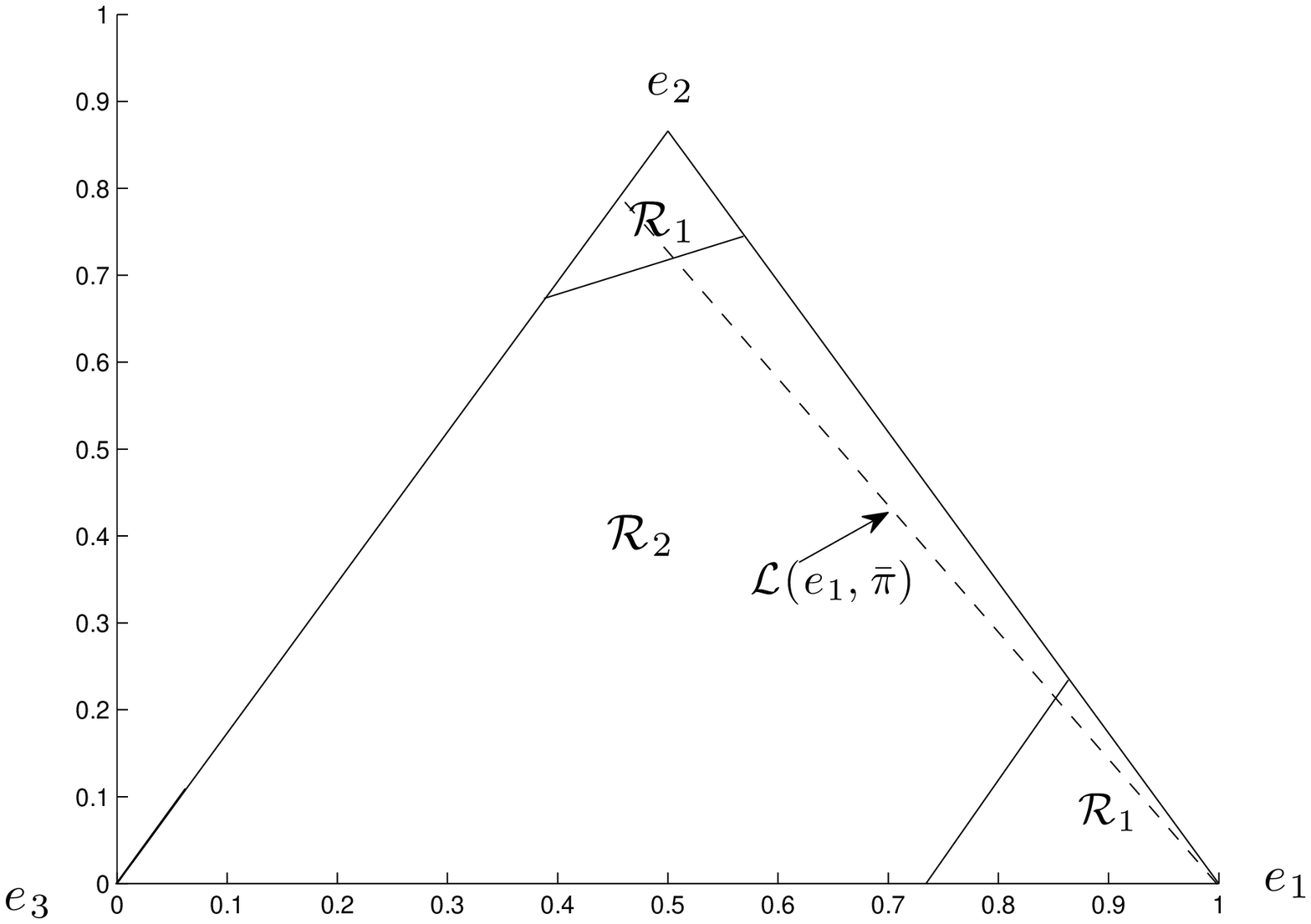,width=0.31\linewidth}}\quad
\subfigure[Example 6]{\epsfig{figure=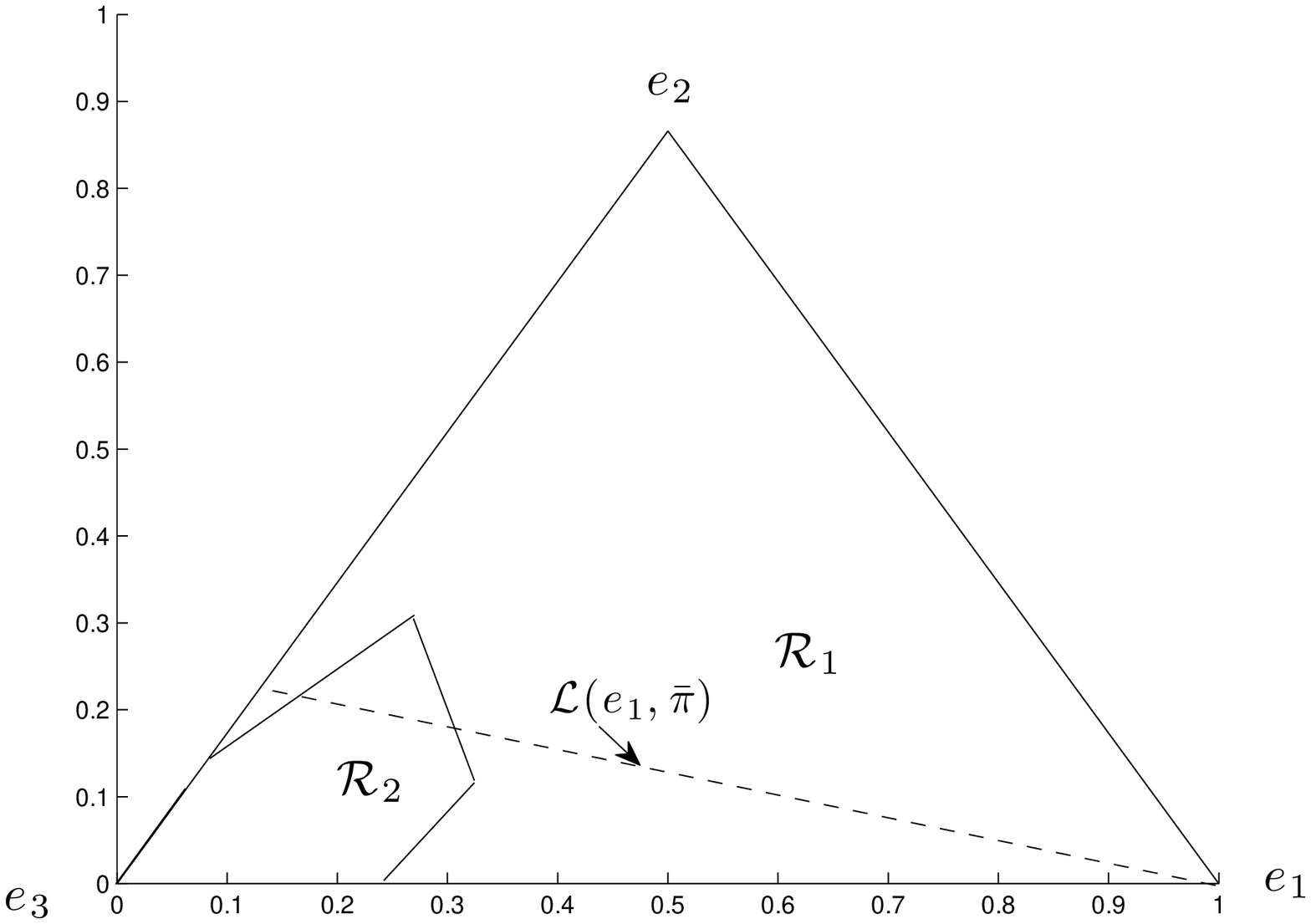,width=0.31\linewidth}}
}
\caption{Examples of decision regions that violate Theorem \ref{thm:1} on belief space $\Pi(3)$.} \label{fig:invalid}
\end{figure}

Recall for $\X=\{1,2,3\}$, the belief state space $\Pi(3)$ is
 an equilateral triangle.
So on $\Pi(3)$, more insight can be given to visualize what the above theorem says.\footnote{The threshold curve 
$\Gamma$ intersects each line segment from vertex $e_3$  and each line segment from vertex $e_1$ at most once. This
implies $\Gamma$ can be parametrized by a pair of monotonically decreasing angles with respect to vertices $e_1$ and $e_3$.
By Lebesgue theorem \cite{Rud76},  a monotone function is differentiable almost everywhere. So for $\X=\{1,2,3\}$, $\Gamma$
is differentiable almost everywhere.}
 In Fig.\ref{fig:invalid},
six examples are given of decision regions that violate the theorem. To make these examples non-trivial, we have included $e_1\in \R_1$ in all cases.

The decision regions in
Fig.\ref{fig:invalid}(a) violate the condition that $\mu^*(\pi)$ is  increasing on lines
towards $e_3$. Even though $\R_1$ and $\R_2$ are  individually connected regions, 
the depicted line $\l(e_3,\bp)$ intersects the boundary of $\R_2$ more than once (and so violates
Theorem \ref{thm:1}). 

The decision regions in Fig.\ref{fig:invalid}(b)  satisfy Theorem \ref{cor:ph} since the stopping set $\R_1$ is convex. As mentioned  in Sec.\ref{subsec:main},
  Theorem \ref{thm:1} gives more structure to $\R_1$ and $\R_2$.
Indeed,  the decision regions in Fig.\ref{fig:invalid}(b)  violate Theorem \ref{thm:1}. They violate the statement that the policy is increasing on lines
towards $e_3$ since the boundary of $\R_1$ (i.e., threshold curve $\Gamma$) cannot intersect a line from $e_3$ more than
once. 
Therefore, Theorem \ref{thm:1} says a lot more about the structure of the boundary than convexity does.
In particular, for the PH-distributed change time without variance penalty, Theorems \ref{thm:1} and
\ref{cor:ph} together say that the threshold curve $\Gamma$ is convex and cannot intersect a line
$\l(e_3,\bp)$ or a line $\l(e_1,\bp)$ more than once.

Fig.\ref{fig:invalid}(c) also satisfies Theorem \ref{cor:ph} since $\R_1$ is convex.
But the decision regions in  Fig.\ref{fig:invalid}(c) violate Statement (ii) of Theorem \ref{thm:1}.
 In particular, if $e_1$ and $e_3$ lie in $\R_1$, then $e_2$ should also
 lie in $\R_1$. Again this reveals that Theorem \ref{thm:1} says a lot more about the structure
 of the stopping region even for the case of zero variance penalty ($\alpha = 0$).

Fig.\ref{fig:invalid}(d) also satisfies Theorem  \ref{cor:ph} since $\R_1$ is convex; but does not
satisfy Theorem \ref{thm:1} since $\R_2$ is not a connected set. Indeed when $\R_2$ is not connected
as shown in the proof of Theorem \ref{thm:1}, the policy $\mu^*(\pi)$ is not monotone on the line
$\l(e_3,\bp)$ since it goes from 2 to 1 to 2.

 Fig.\ref{fig:invalid}(e) and (f) violate   Theorem \ref{thm:1} since the optimal policy
$\mu^*(\pi)$ is not monotone on 
 line  $\l(e_1,\bp)$; it goes from 1 to 2 to 1. For the case $\alpha = 0$,  Fig.\ref{fig:invalid}(e) and (f) violate Theorem \ref{cor:ph} since the stopping region
$\R_1$ is non-convex. 

Since the conditions of Theorem \ref{thm:1} are sufficient conditions, what happens when they do not hold?
In Sec.\ref{sec:numerical}, we will give a numerical example  where (S-Ex1) is violated and 
 $\R_1$ is no longer a connected set (Fig.\ref{fig:qd1}(d)). It is straightforward to construct other examples where
 both  $\R_1$ and $\R_2$ are disconnected regions when the assumptions of Theorem \ref{thm:1} are violated.

\subsection{Characterization of Optimal Linear Decision Threshold} \label{sec:linear}
This subsection assumes that  (A1-Ex1), (A2), (A3), (S-Ex1) of Sec.\ref{subsec:main} hold. So
 Theorem~\ref{thm:1} applies and  computing
the optimal policy $\mu^*$ reduces to estimating the threshold  curve $\Gamma$. In general, any user-defined basis function approximation can be used to parametrize
this curve. However, any such approximation needs to capture the essential feature
of Theorem \ref{thm:1}: the parametrized optimal policy needs to be MLR increasing on lines.
(An identical discussion applies to Theorem \ref{thm:modified} with assumptions (AS-Ex1), (A2), (A3)).

Below, we derive  the optimal {\em linear} approximation
to the threshold curve $\Gamma$
on 
simplex $\I$.
 Such a linear decision threshold has two attractive properties:
(i) Estimating it is  computationally  efficient.
(ii) We give  conditions on the  coefficients of the linear threshold that are necessary and
sufficient for the resulting  policy to be MLR increasing
on lines. Due to the necessity and sufficiency of the condition, optimizing
over the space of linear thresholds on $\I$   yields the optimal linear
approximation to  threshold curve $\Gamma$.

On  $\I$, define the linear threshold policy $\mu_{\theta}(\pi)$ as 
\beq \label{eq:linear}
\mu_{\theta}(\pi)  = \begin{cases}
\text{stop} = 1 & \text{ if }  \begin{bmatrix} 0 & 1 & \theta^\p\end{bmatrix}^\p \begin{bmatrix}\pi \\ -1 \end{bmatrix} < 0
\\
\text{continue} = 2 & \text{ otherwise }  
  \end{cases}, \quad \pi \in \I.
 \eeq
 Here $\theta = (\theta(1),\ldots,\theta(X-1))^\p \in \reals^{X-1}$ denotes the parameter vector of the linear threshold policy.
(Since $\I \subset \reals^{X-1}$, a linear hyperplane on $\I$ is parametrized by $X-1$ coefficients).
 
  Theorem \ref{thm:dep} below characterizes the optimal  linear decision threshold approximation  to the threshold  curve on $\I$.
 Assume conditions (A-Ex1), (A2), (A3),  (S-Ex1) 
hold  for the quickest detection  problem (\ref{eq:csdef}) so that from Theorem \ref{thm:1},
the optimal policy $\mu^*(\pi)$ is 
MLR increasing on lines $\l(e_X,\bp)$ and $\l(e_1,\bp)$. Assume the conditions of Lemma \ref{lem:nondeg} hold, so that
one only needs to search for the optimal linear threshold in the interior of $\I$.  
Finally, the 
requirement that $e_1$ lies in the stopping set, means $\mu_\theta(e_1) < 0$ which 
 implies
$\theta(X-1) >0$.

\begin{theorem}[Optimal Linear Threshold Policy] 
\label{thm:dep}
For belief states $\pi \in \I$,
the   
linear threshold policy $\mu_{\theta}(\pi)$ defined
in (\ref{eq:linear}) is  \\
(i)  MLR increasing
on lines $\l(e_X,\bp)$  iff   $\theta(X-2) \geq 1 $ and $\theta(i) \leq \theta(X-2)$ for $i< X-2$. \\
(ii) MLR increasing
on lines  $\l(e_1,\bp)$  iff $\theta(i)\geq 0$,
for $i<X-2$.\qed
\end{theorem}

The proof of Theorem \ref{thm:dep} is in Appendix \ref{app:dep}.
As a consequence of Theorem \ref{thm:dep}, the optimal
 linear threshold
approximation to  threshold curve $\Gamma$ of Theorem \ref{thm:1} is 
the solution of the following constrained optimization problem:
\beq
\theta^* = \arg \min_{{\theta} \in \reals^{X}} J_{\mu_{\theta}}(\pizero), \; \text{ subject to  $0 \leq \theta(i)\leq \theta(X-2)$, $\theta(X-2) \geq 1$ and $\theta(X-1)>0$
}
\label{eq:tc}\eeq
 where the cost $J_{\mu_{\theta}}(\pizero)$ is obtained as in  (\ref{eq:csdef})
by applying threshold policy
$\mu_{\theta}$ in (\ref{eq:linear}). 
 
 \noindent {\em Remark}:
The constraints in (\ref{eq:tc}) are necessary {\em and} sufficient for the linear threshold policy
(\ref{eq:linear})
to be MLR increasing on lines $\l(e_X,\bp)$ an $\l(e_1,\bp)$. That is, (\ref{eq:tc}) 
defines the  set of all
MLR increasing linear threshold policies -- it does not leave out any MLR
increasing polices; nor does it  include
any non MLR increasing policies.
 Therefore optimizing  over the space of MLR increasing linear 
threshold policies yields the optimal linear  approximation to
 threshold curve $\Gamma$.

\begin{figure} \centering
\mbox{\subfigure[Case 1]
{\epsfig{figure=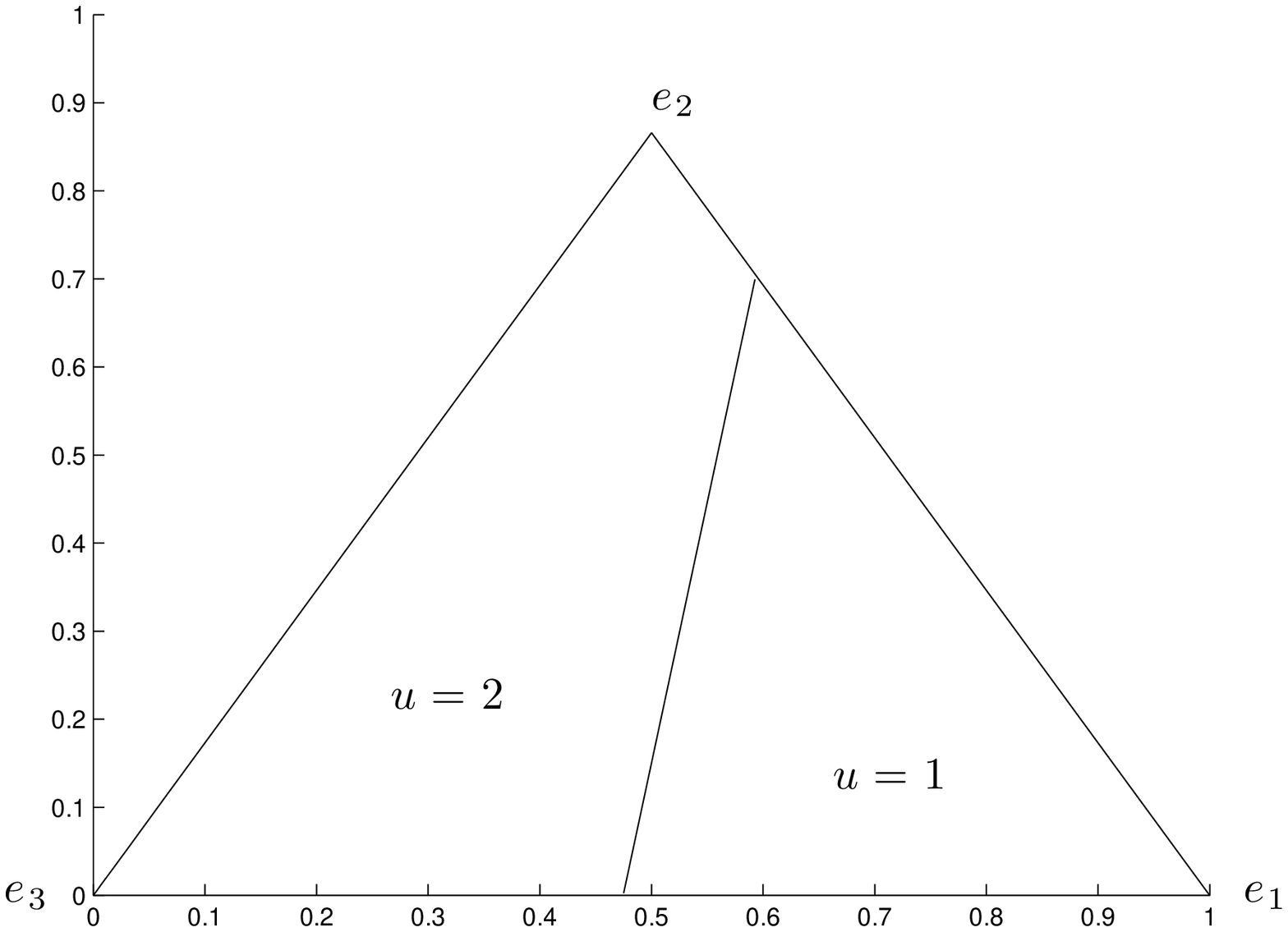,width=0.31\linewidth}} \quad
\subfigure[Case 2 ]{\epsfig{figure=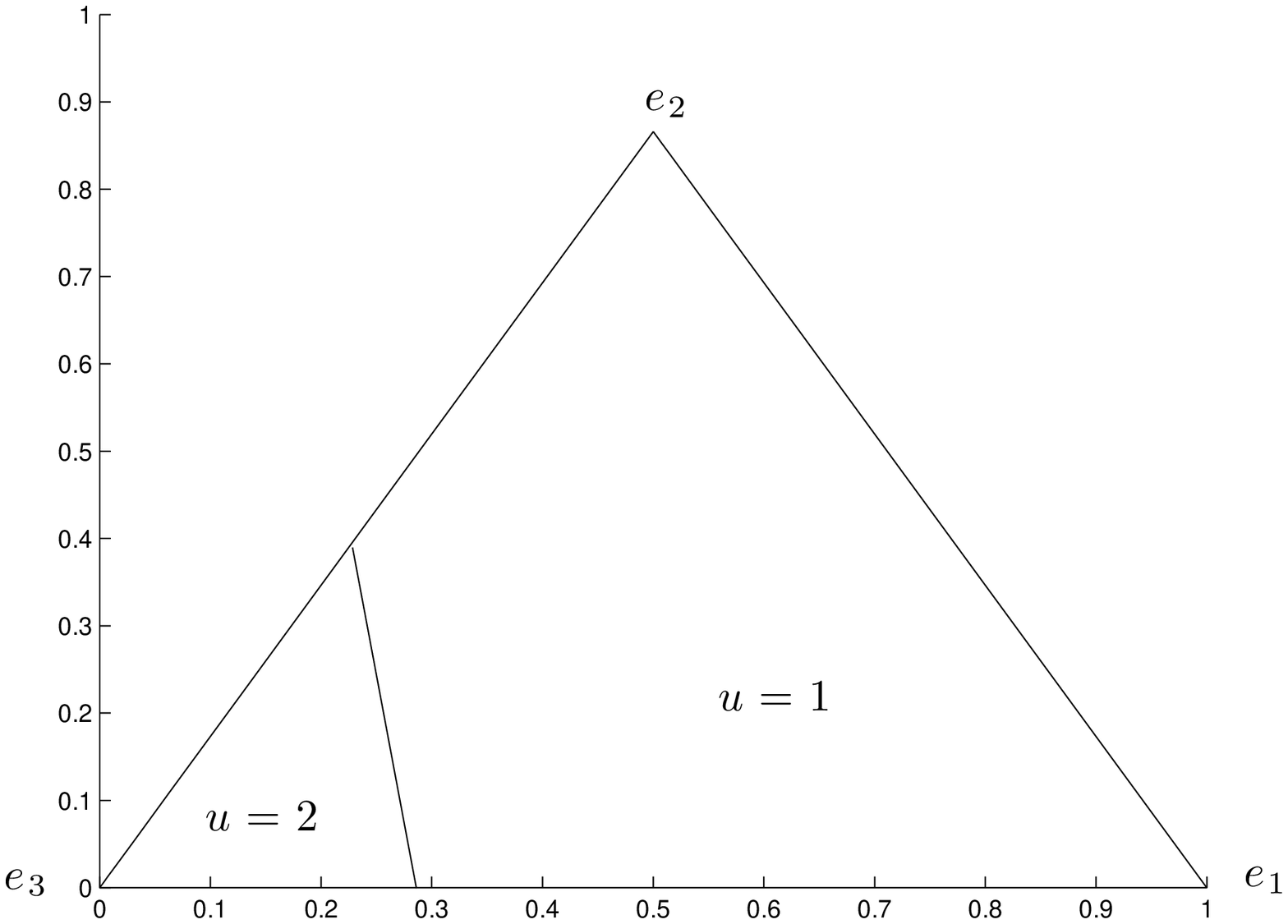,width=0.31\linewidth}} \quad
\subfigure[Case 3 (invalid)]{\epsfig{figure=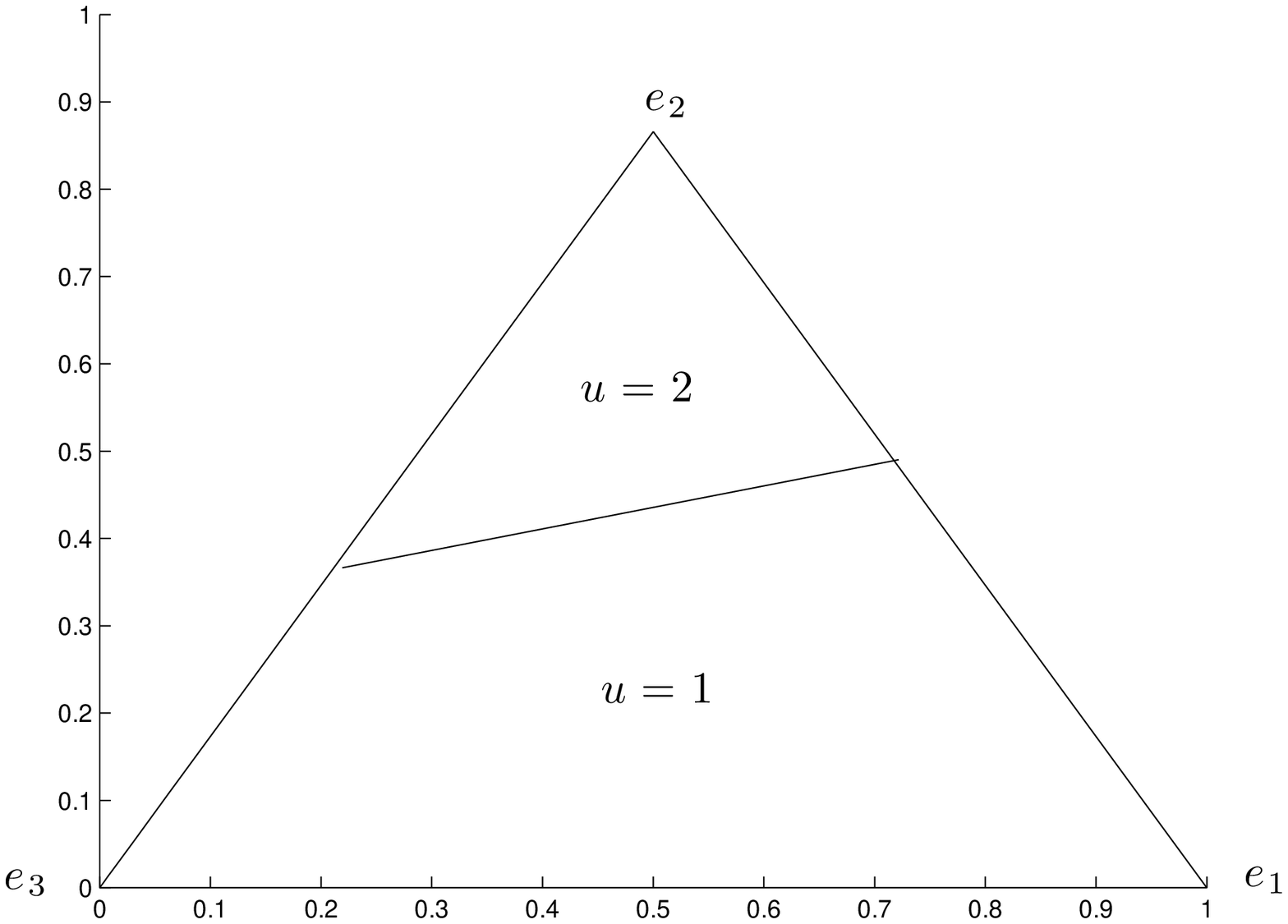,width=0.31\linewidth}}
}
\caption{Examples of Valid Linear Threshold Policies on belief space $\I$ for $X=3$ (Case 1 and Case 2). 
Case 3 is   invalid.} \label{fig:ex1}
\end{figure}

{\em Intuition}: Consider $X=3$, $\X = \{1,2,3\}$ so that the belief space $\I$ is an equilateral triangle.
 Then with $(\omega(1),\omega(2))$ denoting Cartesian coordinates in the equilateral triangle, clearly $\pi(2)=2\omega(2)/\sqrt{3}$, $\pi(1) = 
 \omega(1) - \omega(2)/\sqrt{3}$ and the linear threshold satisfies
\beq 
 \omega(2) = \frac{\sqrt{3}\theta(1)}{2-\theta(1)} \omega(1) + \bigl(\theta(2)-\theta(1)\bigr) \frac{\sqrt{3}}{2-\theta(1)}  \label{eq:cartesian}\eeq
So the conclusion of  Theorem \ref{thm:dep} that $\theta(1)\geq 1$ implies that the  linear MLR increasing threshold has slope of $60^o$ or larger. For $\theta(1)>2$, it follows from (\ref{eq:cartesian}) that the slope of the linear threshold
becomes negative, i.e., more than $90^o$. For a non-degenerate threshold, the $\omega(1)$ intercept of the line should lie in $[0,1]$ implying
$\theta(1)>\theta(2) $ and $\theta(2) > 0$.  Fig.\ref{fig:ex1} illustrates these results.
  Fig.\ref{fig:ex1}(a) and (b)  illustrate  valid linear thresholds. In Fig.\ref{fig:ex1}(a) and (b), the conditions of Theorem \ref{thm:dep} hold (the slope is larger than
  $60^o$ and $\omega(1)$ intercept is in $[0,1]$).
  Fig.\ref{fig:ex1}(c) shows an invalid threshold (since the slope is smaller than $60^o$ and $\omega(1)$ intercept lies outside $[0,1]$). 
  In other words,  Fig.\ref{fig:ex1}(c)  shows an  invalid threshold
since it violates the requirement that $\mu_\theta(\pi)$ is decreasing
on lines towards $e_3$ on $\Pi(3)$. (A line segment $\l(e_3,\bp)$ starting from some point $\bp $ on facet $(e_2,e_1)$ and connected to $e_3$ would
start in the region $u=2$ and then go to region $u=1$. This  violates  the requirement that $\mu_\theta(\pi)$ is increasing on lines towards $e_3$).

\subsection{Algorithm to compute the optimal linear decision curve} \label{sec:spsa}
In this section a stochastic approximation algorithm is presented to estimate the optimal threshold vector $\theta^*$ in (\ref{eq:tc}).
  Because the cost $J_{\mu_\theta}(\pizero)$  in (\ref{eq:tc})  cannot be computed in closed form, we resort to simulation
 based stochastic optimization.  Let 
$n=1,2\ldots,$ denote iterations of the algorithm.  The aim is to  solve the following linearly constrained stochastic optimization problem:
\beq \text{  Compute } 
\theta^* = \arg\min_{\theta \in \Theta} \E\{ {J}_n(\mu_{\theta})  \} 
\;
\text{ subject to  $0 \leq \theta(i)\leq \theta(X-2)$, $\theta(X-2) \geq 1$ and $\theta(X-1)>0$.
}
 \label{eq:stochopt2}\eeq
Here, for each initial condition $\pi_0$, the 
sample path cost   $ {J}_n(\mu_{\theta},\pi_0) $ is evaluated as
\begin{align}
J_n(\mu_\theta,\pi_0)&=  \sum_{k=1}^\infty \discount^{k-1} C(\pi_k,u_k)  \quad
\text{ where } u_k = \mu_\theta(\pi_k) \text{ is computed via (\ref{eq:linear}) } \label{eq:jnmu}\\
J_n(\mu_\theta) &= \frac{1}{L} \sum_{l=1}^L J_n(\mu_\theta,\pi_0^{(l)}) 
\text { where prior $\pi_0^{(l)}$ is sampled uniformly from simplex $\I$.} \nonumber
\end{align}
A convenient way of sampling uniformly from $\I$ is to use 
the Dirichlet distribution (i.e., $\pi_0(i) = x_i/\sum_i x_i$, where $x_i \sim $ unit exponential distribution).

The above constrained stochastic optimization problem can be solved by a variety of methods. One method is to convert  it into an equivalent unconstrained
problem via the following parametrization: Let  $\phi = (\phi(1),\ldots\phi(X-1))^\p \in \reals^{X-1}$
and parametrize $\theta$ as
\beq \theta^\phi = \begin{bmatrix} \theta^\phi(1),\ldots,\theta^\phi(X-1)
\end{bmatrix}^\p, \;\text{ where }
\theta^\phi(i) =
\begin{cases}
\phi^2(X-1) & i =X-1 \\
 1 + \phi^2(X-2) & i = X-2\\
(1+ \phi^2(X-2)) \sin^2(\phi(i)) & i =1,\ldots,X-3
 \end{cases}
\label{eq:transformation}
\eeq
Then 
$\theta^\phi$ trivially  satisfies constraints in (\ref{eq:stochopt2}).
So
(\ref{eq:stochopt2}) is equivalent to the following unconstrained stochastic
optimization problem: 
\begin{align}  \text{  Compute } & \mu_{{\phi^*}}(\pi) \text{ where }
\phi^* = \arg\min_{\phi \in \reals^{X-1}}
 \E\{ {J}_n(\phi)  \} \text{ and }   \nonumber \\
  {J}_n(\phi)  & \text{ is computed using (\ref{eq:jnmu}) with policy 
  $\mu_{\theta^\phi}(\pi)$ evaluated
  according to (\ref{eq:transformation}). }
\label{eq:phi}
\end{align}

Algorithm \ref{alg:spsa} below, 
uses the Simultaneous Perturbation Stochastic Approximation (SPSA) algorithm
\cite{Spa03}
to
generate a sequence of estimates $\hphi_{n}$,
 $n=1,2,\ldots,$ that
converges to a local minimum of  the optimal
linear threshold $\phi^*$ 
 with
policy $ \mu_{\phi^*}(\pi) $.


\begin{algorithm}[h]
\caption{Policy Gradient Algorithm for computing optimal linear threshold policy} \label{alg1}
Assume (A1-Ex1), (A2),  (A3), (S-Ex1) hold so that the optimal social policy is
characterized by a threshold switching curve in Theorem \ref{thm:1}.\\
Step 1: Choose initial threshold coefficients  $\hphi_{0}$ and linear threshold policy
$\mu_{\hphi_{0}}$.
\\
Step 2: For iterations $n=0,1,2,\ldots$ 
\begin{itemize}
\item Evaluate sample cost 
$
{J}_n({\hphi_n})$ using (\ref{eq:phi}). 
Compute gradient estimate   
$$ \nablat {J}_n({\hphi_n}) = \frac{\displaystyle
J_n({\hphi_n+\Delta_n \d_n}) - J_n({\hphi_n-\Delta_n \d_n})}{\displaystyle 2
\Delta_n } \d_n, \quad 
\d_n(i) = \begin{cases}
-1 & \text{ with probability } 0.5 \\
+1 & \text{with probability } 0.5.\end{cases}.
 $$ 
 Here $\Delta_n= \Delta/(n+1)^\gamma$ denotes the gradient step size with $0.5 \leq \gamma \leq 1$ and $\Delta > 0$.

\item Update threshold coefficients  $\hphi_n$  via 
stochastic approximation algorithm
\beq \label{eq:sa}
\hphi_{n+1} = \hphi_n - \epsilon_{n+1} \nablat  {J}_n({\hphi}_n), \quad
\epsilon_n = \epsilon/(n+1+s)^\zeta, \quad 0.5 < \zeta \leq  1, \;
\text{ and }\epsilon , s > 0. \eeq
\end{itemize} \label{alg:spsa}
\end{algorithm}

The above SPSA algorithm \cite{Spa03} 
picks a single
random direction $\d_n$ along which direction the derivative is
evaluated at each batch $n$.  Unlike the Kiefer-Wolfowitz
finite difference algorithm 
to evaluate the gradient estimate $\nablat J_n$ in (\ref{eq:sa}), SPSA
requires only 2 batch simulations, i.e., the number of evaluations is
independent of dimension of  parameter  $\phi$.
Because  the stochastic gradient algorithm
(\ref{eq:sa}) converges to  local optima, it is necessary
to try several initial conditions $\hphi_0$. The computational cost  at each
iteration is linear in the dimension of $\theta$ and is independent of 
 the observation alphabet size.

For fixed $\theta$, the samples $J_n(\mu_\theta)$
in  (\ref{eq:jnmu}) are simulated independently and 
have identical distribution.
Thus the proof that $\theta_n= \theta^{\hphi_n}$ 
 generated by 
Algorithm \ref{alg1} converges
to a local optimum of $\E\{J_n(\mu_\theta)\}$
(defined in (\ref{eq:stochopt2})) with probability one, is a straightforward application of techniques in  \cite{KY03}  (which gives general convergence methods
for Markovian dependencies).

\noindent{\em Remark}: More sophisticated gradient estimation methods can be used instead of the SPSA finite difference algorithm given here. For example,
\cite{VKM03,Pfl96} present  score function and weak derivative approaches for
estimating the gradient of a Markov process with respect to a policy. In 
\cite{BB02} the score function method is used to perform gradient-based reinforcement learning.
These algorithms are  applicable  to solve the constrained stochastic optimization
problem (\ref{eq:stochopt2}) thereby yielding
the optimal linear threshold policy. If the change time distribution (specified by $P$)
and the observation likelihoods (specified by  $B$) are not completely specified,
but (A2) and (A3) hold,  Theorem \ref{thm:1} applies and  reinforcement learning
algorithms \cite{BB02} can be used to  solve (\ref{eq:stochopt2}). Moreover \cite{KY02,YKI04}
analyze the tracking properties of stochastic approximation algorithms when the 
 transition and observation matrices and time varying.

\section{Example 2:  Quickest Transient Detection with variance penalty}
\label{sec:qtrans}
Our second example deals with Bayesian quickest transient detection.\footnote{The author gratefully acknowledges Dr. Venugopal Veeravalli at U.\ Illinois for describing quickest transient detection
and giving access to the preprint \cite{PKV10}.} 
We show below that under similar  assumptions to quickest time detection,
the threshold switching curve of Theorem~\ref{thm:1} holds. Therefore
the linear threshold results and Algorithm \ref{alg:spsa}  hold.

The set up is identical to Sec.\ref{sec:qdbasic} with state space $\X=\{1,2,3\}$. 
The 
 transition probability matrix and initial distribution are
\beq \label{eq:transientp}
 P = \begin{bmatrix}      1 & 0 & 0 \\ p_{21}  &  p_{22} & 0 \\ 
0 & p_{32} & p_{33}  
\end{bmatrix} , \quad \pi_0 = e_3.\eeq
So the Markov chain starts in state 3. After some geometrically distributed time  it  jumps to the transient  state 2.
Finally after residing in state 2 for some geometrically distributed time, it then 
 jumps to the absorbing state 1.  
 
 In quickest transient detection, we are interested in detecting transition to state 2 with minimum cost.
  The action space is $\U  = \{1 \text{ (stop)}, 2 \text{ (continue)} \}$. The stop action $u=1$
 declares that transient state 2 was visited.
 
 We choose
the following costs (see \cite{PKV10} for other choices). Similar to (\ref{eq:exd2}),
let $d_i I(x_{k}=e_i,u_k=2)$ denote the delay cost in state $e_i$, $i\in \X$. Of course 
$d_3= 0$
since $x_k=e_3$ implies that the transient state has not yet been visited. So the expected delay cost
is 
\beq  \sum_{i\in \X} d_i \E\{x_{k}=e_i,u_k=2|\F_k\} = \de^\p \pi_k \text{  where }
\de = (d_1,d_2,d_3)^\p,\; d_3 = 0. \label{eq:delayt} \eeq
 Typically the elements of the delay  vector $\de$ are chosen as $d_1\geq d_2 > 0 $
so that  state 1 (final state) accrues a larger delay than the  transient state. This gives 
incentive to declare that transient state 2 was visited when the current state is 2, rather than wait until the process reaches
state 1.

The false alarm cost for declaring $u=1$ (transient state 2 was visited) when  $x= 1$ is zero  
since the final state 1 could only have been reached
after visiting transient state 2. So the false alarm penalty is $
\E\{ I (x_k=e_3,u_k=1)|\F_k\} = 1 - (e_1+e_2)^\p \pi_k$ for action $u_k=1$. For convenience,
in the  variance penalty (\ref{eq:varder}), we  choose $g= [0,0, 1]^\p$.
So from (\ref{eq:cp1}), (\ref{eq:exd2}), the expected stopping cost and continuing costs are 
\beq
\Cb(\pi,1) = \alpha \bigl( g^\p \pi - \bigl( g^\p \pi \bigr)^2 \bigr)  + \beta ( 1 - (e_1+e_2)^\p \pi) ,\quad
\Cb(\pi,2)=\de  ^\p \pi.
\label{eq:transcost}
\eeq
The optimal decision policy $\mu^*(\pi)$ and stopping set $\R_1$ are as in (\ref{eq:costdef}), (\ref{eq:dp_alg}) and (\ref{eq:stopset}).

{\em Main Result}: The following assumptions are similar to those in quickest time detection.
Note that due to its structure, $P$ in (\ref{eq:transientp}) is always TP2
(i.e., (A3) in Sec.\ref{subsec:main} holds).

\begin{itemize}
\item[(S-Ex2)] The scaling factor for the variance penalty satisfies
 $\alpha \leq  \frac{d_2 + \beta - \rho\beta P_{33}}{1 + \rho P_{33}} $.
\end{itemize}

For $\alpha = 0$
(zero variance penalty)
(S-Ex2)  holds trivially.

\begin{theorem} \label{thm:qt}Consider the quickest transient detection problem  with delay and stopping
 costs in  (\ref{eq:transcost}).
Then  under  (A2), (\ref{eq:transientp}), (S-Ex2),  the conclusions of Theorem \ref{thm:1} hold. Also if the observation likelihoods are non-zero, then for $k \geq 2$, $\pi_k \in \Io$ and the threshold is non-degenerate, see
Sec.\ref{sec:nondeg} and  Lemma \ref{lem:nondeg}.
Thus  Algorithm \ref{alg1}  estimates
the optimal linear threshold.  (The proof follows from meta Theorem \ref{thm:key} in Appendix \ref{sec:applp} and  Theorem \ref{thm:1}). \qed
\end{theorem}

\subsubsection*{PH-distributed change times}
More generally, suppose the process $x$ jumps after a PH-distributed time to transient state.
Then after another PH-distributed time period, it jumps to the absorbing state. We show that 
Theorem \ref{thm:qt} continues to hold.

To model the two PH-distributed change times, let 1 denote the absorbing state, $\T = \{2,\ldots,X_1+1\}$ denote the set of
 transient states and $\S
= \{X_1+2,\ldots,X_{1}+X_2+1\}$ denote set of starting states.
Define the $(X_1+X_2+1) \times (X_1+X_2+1)$ transition matrix
\beq 
P =  \begin{bmatrix} e_{X_1}^\p  & \mathbf{0} _{X_2} \\
     \underline{P}_{X_1 \times X_1} & \mathbf{0}_{X_1 \times X_2} \\ \mathbf{0}_{X_2} &
     \bar{P}_{(X_1+X_2-1)\times X_2}
      \end{bmatrix}  .\eeq
Suppose the Markov chain starts in $\S$. Then after a PH-distributed time it jumps to
$\T$ and finally after another PH-distributed time, jumps to state 1.
Just as in  Sec.\ref{sec:qdbasic}, $\bar{P}$  and $\underline{P}$  determine the PH-distribution in the 
start and transient states, respectively. Let $\de$ and $\f$ denote the delay and false alarm
vectors. $\de$ is a vector with decreasing elements with  $d_i$ are $d_i = 0$, $i\in \S$.
$\f$ is a vector with increasing elements with $f_i=0$, $i\neq \S$.
The delay and stopping costs are
$\Cb(\pi,1) = \beta \f^\p \pi$, $\Cb(\pi,2) = \de^\p \pi $.
Then the conclusions of Theorem \ref{thm:qt} hold under (A2), (A3) if the decision
maker designs $\f$ and $\de$ to satisfy the following linear constraints:
\beq
f_{X_1+2} \geq 1, \quad \left(\de+\beta (\rho P - I) \f\right)^\p (e_i-e_{i+1} ) \geq 0,
\quad i = 1,\ldots, X_1 + X_2. \label{eq:ddm} \eeq
The decision maker can design suitable  $\f$ and $\de$ satisfying (\ref{eq:ddm})  using a linear programming solver.

{\em Remark}:
 In the formulation of \cite{PKV10}, it is assumed that states 1 and 3 are indistinguishable in terms of observations, i.e., $B_{1y} = B_{3y}$ for all $y\in \Y$.
In this case, obviously (A2) does not hold. At this stage, we are unable to prove the structural result of Theorem \ref{thm:qt} when (A2) does not hold.
(Relabelling state 1 as 2 and state 2 as 1 does not work. Then $B$ satisfies (A2) but the transition matrix $P$ is longer TP2).
Nevertheless, we have the following result which follows from Theorem \ref{cor:ph}.
\begin{corollary}
For  $\alpha=0$, (no variance penalty), then 
the stopping region $\R_1$ in quickest transient detection is a convex subset of $\I$. \qed\end{corollary}

\section{Example 3: Quickest Detection with Exponential Penalty for Delay} \label{sec:risk}
In this example, we generalize the results of Poor \cite{Poo98}, which deals with exponential delay penalty
and geometric change times. We consider  exponential delay penalty with PH-distributed change time. Our formulation
involves risk sensitive partially observed stochastic control, see Sec.\ref{sec:intro} for motivation.
We first show  that the exponential penalty cost function 
in \cite{Poo98} is a special case of risk-sensitive stochastic control cost function when the state space dimension $X=2$.  We then use the
risk-sensitive stochastic control formulation to derive structural results for 
PH-distributed change time. In particular, the main result below (Theorem~\ref{thm:risk}) shows that 
the threshold switching curve still characterizes the optimal stopping region $\R_1$. The assumptions and main results are conceptually similar to Theorem \ref{thm:1}.

Since our aim is to interpret and extend the results of \cite{Poo98} using risk sensitive control, we consider the same costs as in \cite{Poo98}, so  $\alpha = 0$ (no variance penalty).
 Below, we will use $c(e_i,u=1)$ to denote  false alarm costs and
$c(e_i,u=2)$ to denote delay costs, where $i \in \X$.

Risk sensitive control \cite{Ben92} considers the exponential cost function
\beq \label{eq:risk}
J_\mu(\pizero) = \Ep\biggl\{ \exp \biggl( \r \sum_{k=1}^{\tau-1} c(x_k,u_k=2) + \r c(x_\tau, u_\tau = 1) \biggr)\biggr\} \eeq
where $\r >  0$ is the risk sensitive parameter.

Let us first
show that the exponential penalty cost in \cite{Poo98} is  a special case of (\ref{eq:risk}) for consider the case $X=2$ (geometric distributed change time). For the state $x \in \{e_1,e_2\}$, choose
$c(x,u=1) = \beta I(x\neq e_1,u=1) = \beta (1 -e_1^\p x)$ (false alarm cost) , $c(x,u=2) = d I(x=e_1,u=2)
= d e_1^\p x
$ (delay cost).
Then it is easily seen that
$\sum_{k=1}^{\tau-1} c(x_k,u_k=2) +  c(x_\tau, u_\tau = 1) =  d \, |\tau - \tau^0|^+ + \beta I(\tau < \tau^0)$. Therefore   (recall $\tau^0$ is defined in (\ref{eq:tau}) and $\tau$ is defined
in (\ref{eq:tauu})),
\begin{align}
J_\mu(\pizero) &= \Ep\bigl\{ \exp \bigl(\r d \, |\tau - \tau^0|^+ + \r \beta I(\tau < \tau^0) \bigr)\bigr\} \left[I(\tau <\tau^0) + I(\tau = \tau^0) + I(\tau > \tau^0) 
\right] \nonumber\\
&= \Ep\bigl\{\exp(\r \beta) I(\tau < \tau^0) + \exp(\r d\,  |\tau-\tau^0|^+) I(\tau > \tau^0) + 1\bigr\}\nonumber\\
&=\Ep\bigl\{ (e^{\r \beta} - 1) I(\tau < \tau^0) + e^{\r d |\tau-\tau^0|^+ }\bigr\} \nonumber\\
&= (e^{\r \beta} - 1) \Pp(\tau < \tau^0) + \Ep \{e^{\r d|\tau-\tau^0|^+ }\} \label{eq:poorexp}
\end{align}
which is identical to Poor's exponential delay cost function \cite[Eq.40]{Poo98}. Thus the Bayesian quickest time detection
with exponential delay penalty in  \cite{Poo98} is a special case of a risk sensitive stochastic control problem.

We consider the delay cost as in (\ref{eq:exd}); so
for state $x\in \{e_1,\ldots,e_X\}$,  $c(x,u_k=2) = d e_1^\p P^\p  x$. To get an intuitive
feel for this modified delay cost function, for the case  $X=2$,
$$\sum_{k=1}^{\tau-1}c(x_k,u_k=2) + c(x_\tau,u_\tau=1) = d|\tau - \tau^0|^+ + \beta I(\tau < \tau^0) 
+ dP_{21} (\tau^0-1) I(\tau^0 < \tau)$$
Therefore,
  by using (\ref{eq:altcost}),  for $X=2$, the exponential delay cost function is
\beq
J_\mu(\pizero)= (e^{\r \beta} - 1) \Pp(\tau < \tau^0) +  \Ep\bigl\{ e^{\r d \left[ |\tau-\tau^0|^+ + P_{21}(\tau_0-1)I(\tau_0<\tau) \right]}\bigr\}.
\eeq
This is similar to (\ref{eq:poorexp}) except for the additional term $P_{21}(\tau_0-1)I(\tau_0<\tau)$ in the exponential.

With the above motivation, in the rest of this section we consider risk sensitive quickest time detection
for PH-distributed change time, i.e. $X \geq 2$.
Let $\pi$ denote the risk sensitive belief state, see \cite{EAM95,JBE94} for  extensive descriptions of the risk sensitive belief state and verification theorems 
for dynamic programming in  risk sensitive control.
It can be shown \cite{EAM95} that the value function $\Vb(\pi)$ satisfies
\beq \label{eq:dprisk}
\Vb(\pi) = \min\{ \Cb(\pi,1) ,\sum_{y\in \Y} \Vb(T(\pi,Y)) \sigma(\pi,y) \}
\eeq
where with $R_1 = (1, e^{\r \beta},\ldots, e^{\r \beta})^\p$, $R_2 = (e^{\r d}, e^{\r d P_{21}}, \ldots, e^{\r d P_{X1}})^\p$, $B_y$ defined in (\ref{eq:hmm})
 \beq \Cb(\pi,1) = R_1^\p  \pi, \;
T(\pi,y) = \frac{B_y P^\p \text{diag}(R_2) \pi}{\sigma(\pi,y)}, \;
\sigma(\pi,y) = \ones^\p B_y P^\p \text{diag}(R_2) \pi \label{eq:riskpi}\eeq

As in Sec.\ref{sec:qdobjective}, define
$V(\pi) = \Vb(\pi) - \Cb(\pi,1)$. Then $V(\pi)$ satisfies Bellman's equation (\ref{eq:dp_alg}) with
\beq C(\pi,1) = 0,\quad 
C(\pi,2) = R_1^\p (P^\p  \text{diag}(R_2) - I)  \pi. 
\label{eq:riskcosts}
\eeq

Assume the following condition holds
\begin{itemize}
\item[(A1-Ex3)] The elements of $R_1^\p (P^\p  \text{diag}(R_2) - I)$ are decreasing wrt $i=1,2,\ldots,X$.
\end{itemize}
Evaluating $ C(\pi,2) = R_1^\p (P^\p  \text{diag}(R_2) - I)  \pi$, then (A1-Ex3) 
 is equivalent to
$$ e^{\r d}-1 \geq e^{\r d P_{21}}(P_{21} + e ^{\r \beta}(1 - P_{21}))-e ^{\r \beta} \text{ and }
e^{\r d P_{i1}}(P_{i1}+ e^{\r \beta}(1 - P_{i1})) \text{ decreasing in $i\in \{2,\ldots,X\}$ } $$
For example, if $d=\epsilon=1$, then for $\beta \geq 1$, the following are verified by elementary 
calculus:\\
(i)  (A1-Ex3) always holds
for $\beta \geq 1$ when $X=2$ (geometric distributed change time). \\
(ii) For PH-distributed change time,
if (A3) holds, then (A1-Ex3) always holds
providing $P_{21} < 1/(e^\beta-1)$.

\begin{theorem}\label{thm:risk}
The stopping region $\R_1$ is a convex subset of $\I$.
Under (A1-Ex3), (A2), (A3), Theorem \ref{thm:1} holds. Thus  Algorithm \ref{alg1}  estimates
the optimal linear threshold. \qed
\end{theorem}

The proof is in Appendix \ref{app:risk}.

\noindent{\em Remarks: (i) Delay Formulation in \cite{Poo98}}: Consider the formulation in Poor \cite{Poo98} which is equivalent to  (\ref{eq:poorexp}). Then for 
the geometric distributed case $X=2$, the convexity of $\R_1$ holds using a similar proof to above. Since $\I$ is a 1-dimensional simplex
and $e_1 \in \R_1$, convexity implies there exists (a possible degenerate) threshold point $\pi^*$ that characterizes $\R_1$ such
that the optimal policy is of the form (\ref{eq:onedim}).
As a sanity check,  the 
analogous condition
to (A1-Ex3) reads $e^{\r d} -1 > P_{21}(1 - e^{\r \beta})$. This always holds for $\r \geq 0$. Therefore, assuming (A2) holds,
the above theorem holds
for Poor's \cite{Poo98} exponential delay penalty case under (A2).  (Recall (A3) holds trivially when $X=2$).
Finally, for $X>2$, using a similar proof (see  Theorem \ref{thm:modified}), one can again show that the
conclusions of Theorem \ref{thm:risk} hold.
\\
{\em (ii) Other Examples}:  With the above risk sensitive formulation, the dynamic programming equation (\ref{eq:dprisk})  for the exponential delay case  is very 
similar to the other examples in this paper. Therefore, it is straightforward to generalize the  above exponential penalty result  to  quickest transient  detection (of Sec.\ref{sec:qtrans}),
and  social learning stopping time problems considered below. 

\section{Example 4 \& 5: Stopping Time Problems in  Multi-agent Social Learning} \label{sec:change}

Here we consider stopping time problems in multi-agent social learning. We present two results:\\
(i) Sec.\ref{sec:social1} (Example 4) considers a multi-agent system seeking to solve a Bayesian stopping time
problem.\\
(ii) Sec.\ref{sec:cso} (Example 5) deals with {\em constrained optimal} social learning  which is formulated
in  Chamley \cite{Cha04} as a sequential stopping time problem.
 We show   that  the optimal policy 
 has a  threshold switching curve similar to Theorem \ref{thm:1}. 


\subsection{Motivation: Social Learning amongst myopic agents}\label{sec:herd}
Since social learning only serves as a motivation for subsequent subsections, our description is brief;
see \cite{Cha04}.
Consider a countable infinite number of agents performing social learning  to estimate an underlying random state $x$.
Each agent acts once  in a predetermined sequential order indexed by $k=1,2,\ldots$. One can also view $k$ as the discrete time instant when agent $k$ acts. A key difference between social learning compared to
the formulation in previous sections is that  agent $k$ does not have access to the belief state or private
observations of previous agents. Instead each agent  $k$ only has access to the actions taken by previous agents together with its
own current private observation $y_k$.

Throughout this section, we assume that the observation space $\Y$ is finite.
Let $y_k \in \Y =  \{1,2,\ldots,Y\}$ denote the private observation of agent $k$ and $a_k \in \A =  \{1,2,,\ldots, A\}$ denote the action agent $k$ takes. Define the sigma algebras:
\begin{align}  \mathcal{H}_k & \quad 
\sigma\text{-algebra generated by } (a_1,\ldots,a_{k-1},y_k),  \nonumber\\
\mathcal{G}_k & \quad
\sigma\text{-algebra generated by } (a_1,\ldots,a_{k-1},a_k).   \label{eq:sigg} \end{align}

The social learning model \cite{BHW92,Cha04} comprises of the following  ingredients:
\\
(i) The state of nature $x$ as in Sec.\ref{sec:qdmodel}  except that the transition matrix is $P=I$. That is, the state of nature is a random variable
with distribution $\pi_0$ (see (\ref{eq:init})) instead of a random process.
\\ (ii) At time $k$,
agent $k$  records a private observation $y_k\in \Y $ 
from the observation distribution $B_{iy} = P(y|x=e_i)$, $i \in \X$.
\\
(iii) 
{\em Private belief}:  Using the public belief $\pi_{k-1} $ available at time $k-1$ (defined in Step (v) below), agent $k$  then updates its Bayesian private belief  $\priv_k$ as in (\ref{eq:hmm}) with $P = I$.
Here 
\beq \priv_k = \E\{x|\mathcal{H}_k\} =  (\priv_k(i), \;i \in \X), \quad\priv_k(i) = P(x = e_i| a_1,\ldots,a_{k-1},y_k),\quad \text{initialized with  $\pi_0$. } \label{eq:mudef}
\eeq
 (iv)   {\em Myopic Action}: Agent  $k$ then takes  action $a_k\in \A = \{1,2,,\ldots, A\}$ to  minimize myopically its expected cost $a_k = \arg\min_{a\in \A} \{c_a^\p\priv_k\}$.
 Here $\ca = (c(e_i,a), i \in \X)$ denotes an $X$ dimensional cost vector, and $c(e_i,a)$ denotes the cost  incurred when the underlying state is $e_i$ and the  agent picks action $a$.
Thus agent $k$ chooses  action 
\beq 
a_k = a(\pi_{k-1},y_k) = \arg\min_{a \in \A} \E\{c(x,a)|\mathcal{H}_k\}  =\arg\min_{a\in \A} \{c_a^\p\priv_k\}
\label{eq:step2} \eeq
(v) {\em Social learning and Public belief}:   
Finally agent $k$ broadcasts this action $a_k$ to subsequent agents.
Define   the public belief  $\pi_k$  as the posterior distribution
 of the state $x$ given all actions taken up to time $k$. 
\beq \label{eq:pidef}
\pi_k =  \E\{x| \mathcal{G}_{k}\} = (\pi_k(i), \;i \in \X), \quad
\pi_k(i)  = P(x = e_i|a_1,\ldots a_k), \quad \text{initialized with   $\pi_0$ }
\eeq
Based on the action $a_k$  every  agent (apart from $k$) perform social learning to
 update their public belief according to the following ``social learning Bayesian filter":
\beq \pi_k = \Ts(\pi_{k-1},a_k), \text{ where } \Ts(\pi,a) = 
 \frac{\Bs_a \pi}{\sigs(\pi,a)},
\; \sigs(\pi,a) = \mathbf{1}_X^\p \Bs_a \pi
\label{eq:piupdate} \eeq
In (\ref{eq:piupdate}), $\Bs_a  = \text{diag}(P(a|x=e_i,\pi),i\in \X ) $ with elements
\begin{align} \label{eq:aprob}
 P(a_k = a|x=e_i,\pi_{k-1}=\pi) = \sum_{y\in \Y} P(a_k=a|y,\pi)P(y|x=e_i) \\
= \sum_{y\in \Y} \prod _{\ta \in \A - \{a\}}I(c_a^\p B_{y} \pi < c_{\ta}^\p B_{y} \pi) P(y|x=e_i) \nn \end{align}
where $I(\cdot)$ denotes the indicator function and $B_y$ is defined in (\ref{eq:hmm}).

The following well known result \cite{BHW92,Cha04}  states that eventually  after some finite time $\bar{k}$, all agents  pick the same action and the private belief freezes.  This is termed 
 an information cascade. The proof follows via an elementary application of the martingale convergence
 theorem.

\begin{theorem}[\cite{BHW92}] 
\label{thm:herd} The above social learning model leads to an information cascade (i.e., all agents herd)  in finite time
with probability~1. That is there
exists a finite time $\bar{k}$ after which social learning ceases, i..e, public belief $\pi_{k+1} = \pi_k$, $k \geq \bar{k}$, and all agents pick the same action, i.e., $a_{k+1} = a_k$, $k\geq \bar{k}$.
  \qed\end{theorem}

\subsection{Example 4: Sequential Detection with Social Learning}  \label{sec:social1}
Suppose a multi-agent system makes local decisions and performs social learning as above. 
 Given such a protocol, how can the multi-agent system  make a global decision when to stop?
As mentioned in Sec.\ref{sec:intro}, such problems are motivated
in decision systems where a global decision needs to be made based on local
decisions of agents.

 We consider a Bayesian sequential detection problem
for state $x=e_1$.  The main result
below (Theorem \ref{thm:stopsocial}) is that the global decision of when to stop is a multi-threshold function of the belief state. 
This unusual behavior is because in social learning,  the action likelihood probabilities $\Bs_a$ in (\ref{eq:aprob}) 
depend on the belief state $\pi$. 

Consider  $\X = \Y=\{1,2\}$ and $\A = \{1,2\}$ and
 the social learning model of Sec.\ref{sec:herd}, where  the costs $c(e_i,a)$ satisfy
\beq c(e_1,1) < c(e_1,2) ,  \quad  c(e_2,2) < c(e_2,1) . \label{eq:costdom}\eeq
Otherwise one action will always dominate the other action and the problem is un-interesting.

Redefine the sigma algebras in (\ref{eq:sigg}) to include the action history:
\begin{align}  \mathcal{H}_k & \quad 
\sigma\text{-algebra generated by } (a_1,\ldots,a_{k-1},y_k,u_1,\ldots,u_k),  \nonumber\\
\mathcal{G}_k & \quad
\sigma\text{-algebra generated by } (a_1,\ldots,a_{k-1},a_k,u_1,\ldots,u_k).   \label{eq:siggn} \end{align}
Let $\tau$ denote a stopping time adapted to the sequence of sigma-algebras $\mathcal{G}_{k}$, $k \geq 1$  (see  (\ref{eq:siggn})). 
In words, each agent has only the public belief obtained via social learning to make the global
decision of whether to continue or stop. The goal is to solve the following sequential
detection problem to detect state $e_1$: Pick the stopping time $\tau$ to minimize
\beq
  J_\mu(\pizero)  =
   \Ep\left\{\sum_{k=1}^{\tau-1} \discount^{k-1}  \E\left\{ \left. 
  d I(x = e_1) \right\vert \mathcal{G}_{k-1}\right\}  + 
  \rho^{\tau-1}
  \beta \E\{ I(x \neq e_1) | \mathcal{G}_{\tau-1}\} \right\}.  \label{eq:costsocial1}
\eeq
As in previous sections, the first term is the delay cost and penalizes the decision of choosing
$u_k=2$ (continue) when the state is $e_1$ by the non-negative constant $d$.   The second term is the
stopping cost incurred by choosing $u_\tau = 1$ (stop and declare state 1) at time  $k =\tau$. It is 
the error probability of  declaring state $e_1$ when the actual state is $e_2$.
  In terms of the public belief, 
\begin{align}  \label{eq:stopsocial}
J_\mu(\pizero) &= 
 \Ep \{\sum_{k=1}^{\tau-1} \discount^{k-1} \Cb(\pi_{k-1},u_k=2) + \rho^{\tau-1} \Cb(\pi_{\tau-1} , u_\tau=1) \}  \\
\Cb(\pi,2) &=  
   d e_1^\p \pi,  
   \quad 
   \Cb(\pi,1) = \beta
e_2^\p \pi.
 \nonumber  \end{align}

The global  decision $u_k = \mu(\pi_{k-1}) \in \{1 \text{ (stop) } ,2 \text{ (continue)}\} $ is a function of  the  public belief $\pi_{k-1}$ updated according to the social learning protocol (\ref{eq:mudef}), (\ref{eq:piupdate}).
The optimal policy $\mu^*(\pi)$ and value function $V(\pi)$ satisfy Bellman's equation (\ref{eq:dp_alg}) with
\begin{align} \label{eq:dpsocialstop}
Q(\pi,2) &= C(\pi,2) + \discount \sum_{a \in \A}  V\left( \Tp(\pi ,a) \right) \sigp(\pi,a) \text{ where } \\
C(\pi,2) &= \Cb(\pi,2) -  (1-\discount) \Cb(\pi,1), \quad Q(\pi,1) = C(\pi,1) = 0. \nonumber
\end{align}
Here $T(\pi,a)$ and $\sigp(\pi,a)$ are obtained from the social learning Bayesian filter (\ref{eq:piupdate}).

Since  $\X=\{1,2\}$, the public belief state $\pi = [1-\pi(2), \;\pi(2)]^\p$ is parametrized by the scalar $\pi(2) \in [0,1]$, i.e.,
$\I$ is the interval $[0,1]$. In order to state the main result,
define  the following
 four intervals which form a partition of the interval  [0,1]:
\begin{align*}
\mathcal{P}_l &= \{\pi(2):  \eta_l <\pi(2) \leq \eta_{l-1} \},  \quad l=1,\ldots, 4  \;\text { where } \\ \eta_0 &= 1, \;
\eta_1  = \frac{(c(e_1,2) - c(e_1,1)) B_{11}}{(c(e_1,2) - c(e_1,1)) B_{11} + (c(e_2,1) - c(e_2,2)) B_{21} } \\
\eta_2 &=  \frac{(c(e_1,2) - c(e_1,1))} {(c(e_1,2) - c(e_1,1))  + (c(e_2,1) - c(e_2,2))  } \\
\eta_3 & = \frac{(c(e_1,2) - c(e_1,1)) B_{12}}{(c(e_1,2) - c(e_1,1)) B_{12} + (c(e_2,1) - c(e_2,2)) B_{22} } , \quad \eta_4 = 0.\end{align*}Note that $\eta_0$ corresponds to belief state $e_2$, and $\eta_4$ corresponds to belief state $e_1$.
 (See discussion
at the end of this section for more intuition about the intervals $\mathcal{P}_i$).

It is readily verified that if the observation matrix $B$ is TP2, then  $\eta_3 \leq \eta_2 \leq \eta_1$.
The following is the main result. 

\begin{theorem} \label{thm:stopsocial} Consider the stopping time problem (\ref{eq:stopsocial}) where agents
perform social learning using the social learning Bayesian filter (\ref{eq:piupdate}). Assume (\ref{eq:costdom}), $ d \geq \rho \beta$,  and $B$ is TP2 symmetric.
Then
the optimal stopping policy $\mu^*(\pi)$  has the following structure: The stopping set $\R_1$ is the union of  at most three
intervals. That is  $\R_1 = \R_1^a \cup \R_1^b  \cup \R_1^c$ where $\R_1^a$, $\R_1^b$, $\R_1^c$ are possibly empty intervals. Here \\
(i)  The stopping interval $\R_1^a \subseteq  \mathcal{P}_1 \cup  \mathcal{P}_4$ and is characterized by a threshold point. That is, 
	if $ \mathcal{P}_1$ has a threshold point $\pi^*$,  then $\mu^*(\pi) = 1$ for all $\pi(2) \in \mathcal{P}_4$ and 
	\beq
\mu^*(\pi) = \begin{cases} 2  & \text{ if } \pi(2) \geq \pi^* \\
	1 & \text{ otherwise }  \end{cases}, \quad \pi(2) \in \mathcal{P}_1 .\eeq
	Similarly, if $\mathcal{P}_4$ has a threshold point $\pi_4^*$, then $\mu^*(\pi) =2 $ for all $\pi(2) \in \mathcal{P}_1$.
\\	
(ii)  The stopping intervals $\R_1^b \subseteq \mathcal{P}_2$ and     $\R_1^c \subseteq \mathcal{P}_3$ 
		\\
	(iii) The intervals $\mathcal{P}_1$ and $\mathcal{P}_4$ are regions of information cascades.  That is, if $\pi_k \in \mathcal{P}_1
\cup \mathcal{P}_4$, then social learning ceases and $\pi_{k+1} = \pi_k$ (see Theorem \ref{thm:herd} for definition
of information cascade). \qed
\end{theorem}

The proof of Theorem \ref{thm:stopsocial}  is in Appendix \ref{app:stopsocial}. The proof depends on properties of the social learning
filter and these are summarized in Lemma \ref{lem:social} in Appendix \ref{app:stopsocial}.

\begin{figure} \centering
\mbox{\subfigure[Policy $\mu^*(\pi)$]
{\epsfig{figure=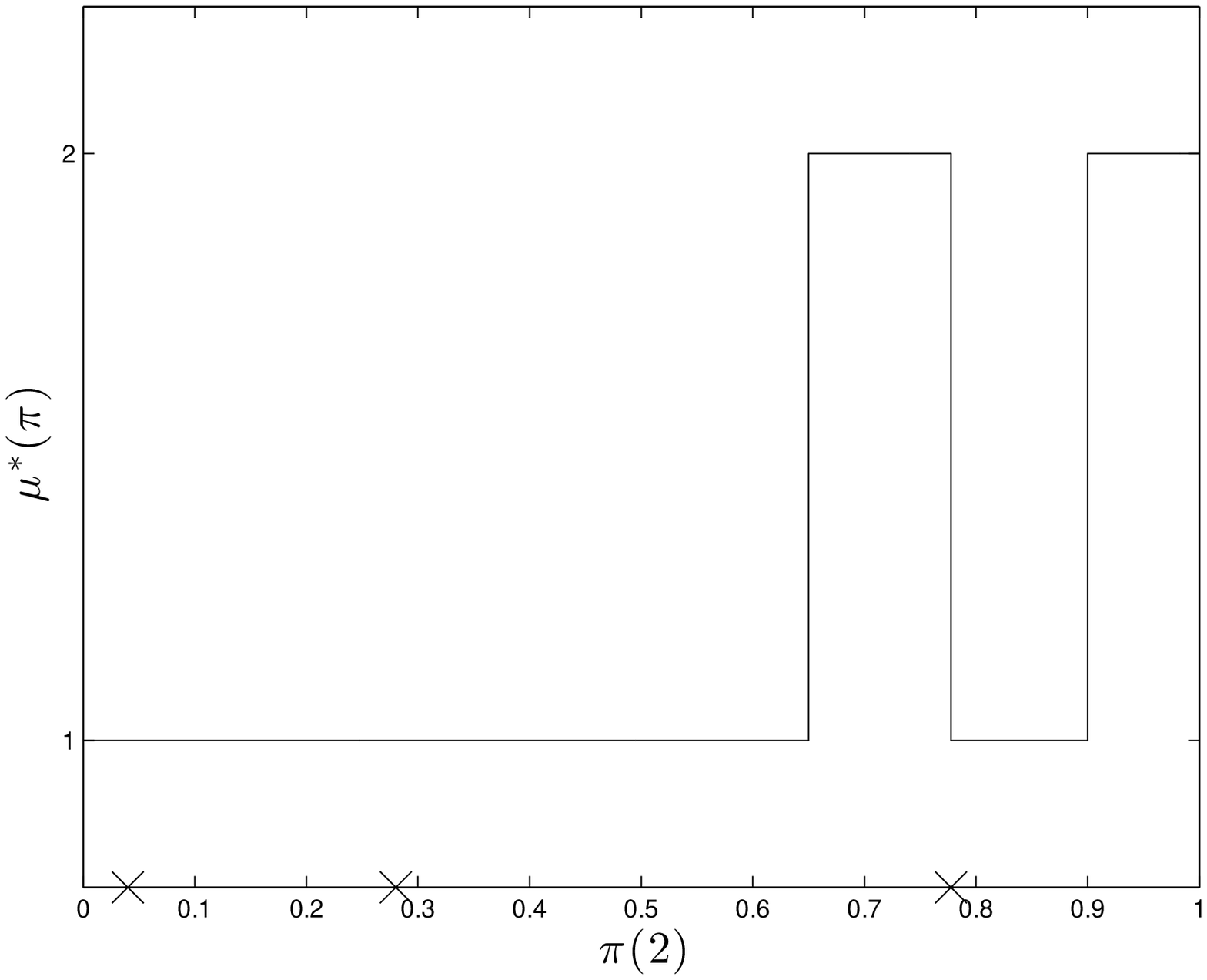,width=0.4\linewidth}} \quad
\subfigure[Value function $V(\pi)$]{\epsfig{figure=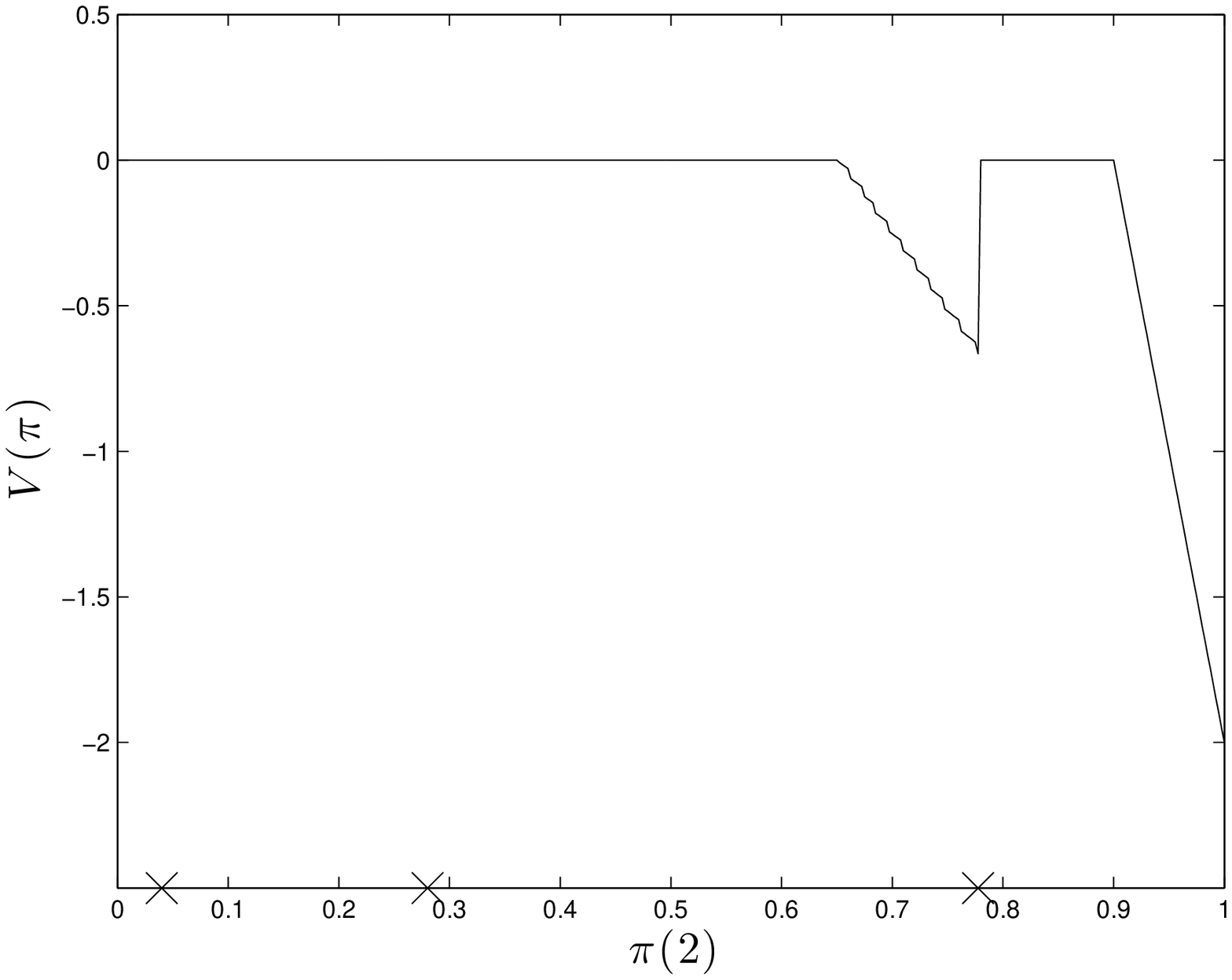,width=0.4\linewidth}}}
\mbox{\subfigure[Policy $\mu^*(\pi)$]
{\epsfig{figure=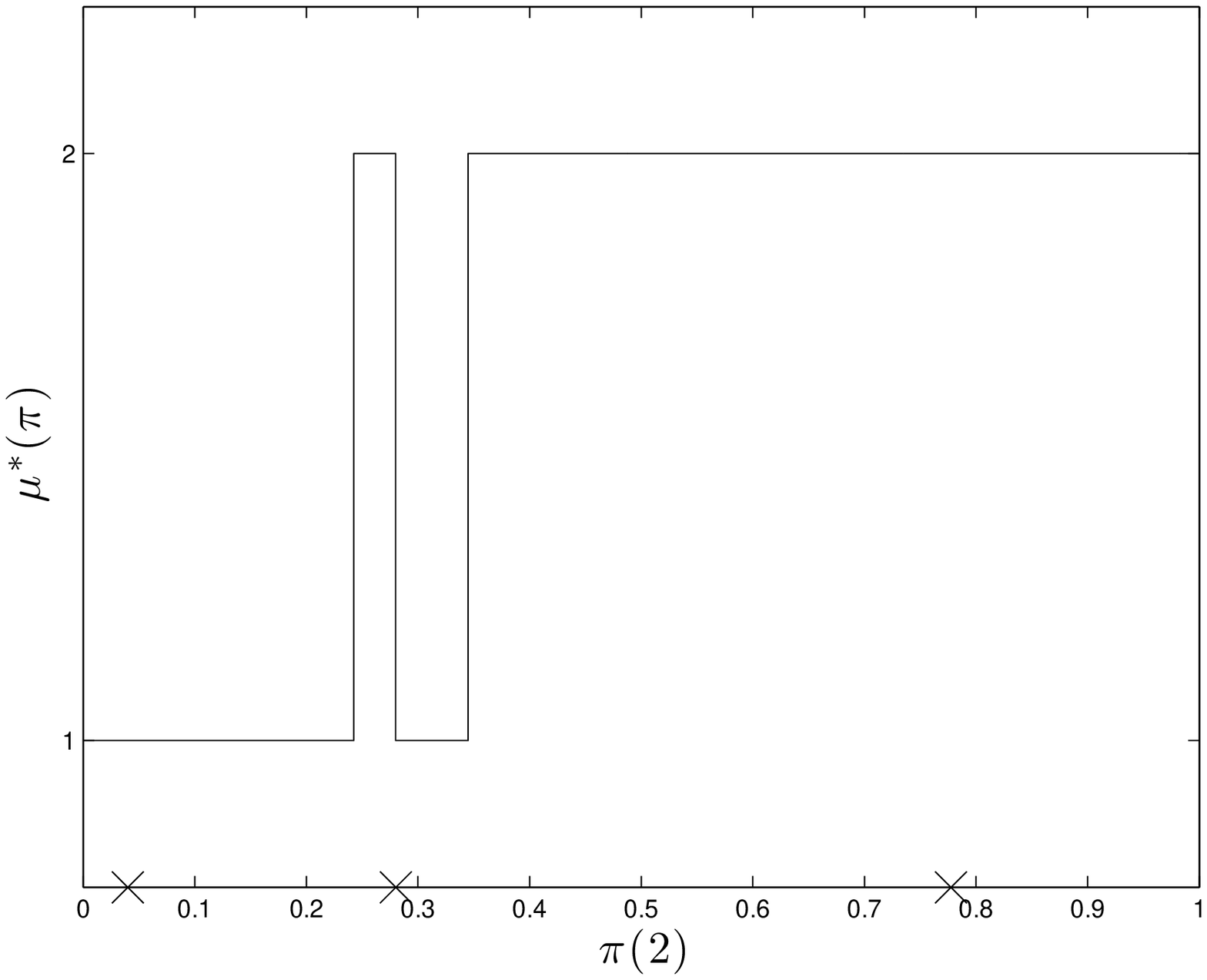,width=0.4\linewidth}} \quad
\subfigure[Value function $V(\pi)$]{\epsfig{figure=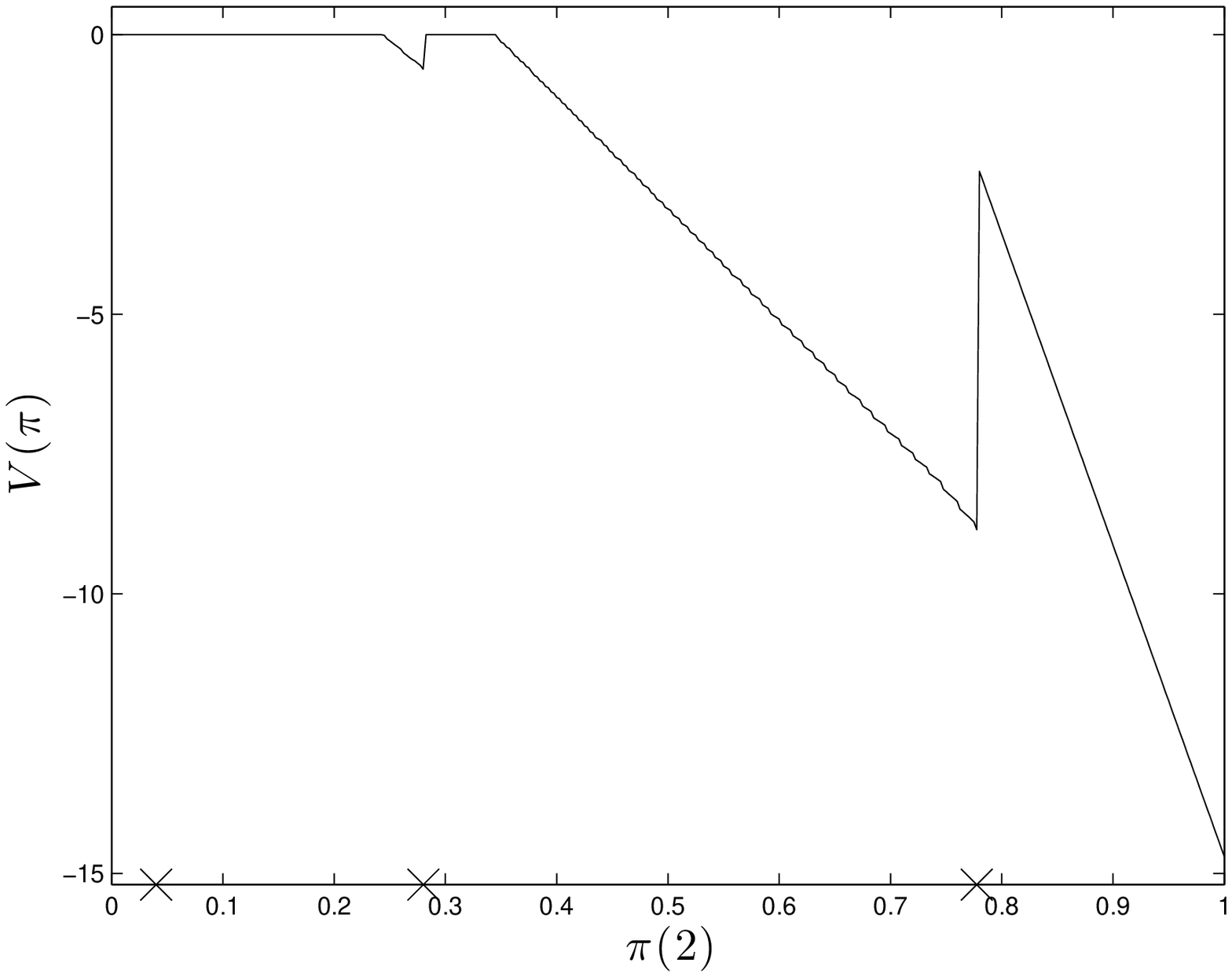,width=0.4\linewidth}}}
\caption{Double Threshold Policy in stopping time problem involving social learning. The parameters are specified
in (\ref{eq:socialexample}) and (\ref{eq:socialexample2}). Fig.\ref{fig:doublethreshold}(a) and (c) show the   optimal policy $\mu^*(\pi)$ and Fig.\ref{fig:doublethreshold} (b) and (d)   show the value
function $V(\pi)$ computed using (\ref{eq:dpsocialstop}) }
\label{fig:doublethreshold}
\end{figure}

{\em Examples}: 
(i) To illustrate the multiple  threshold structure of the above theorem, consider the 
stopping time problem (\ref{eq:stopsocial}) with the following parameters:
\beq
\rho = 0.9, \quad d = 1.8,\quad B = \begin{bmatrix} 0.9 & 0.1 \\ 0.1 & 0.9 \end{bmatrix} ,\;  c(e_i,a) = \begin{bmatrix}
4.57 & 5.57 \\ 2.57 &   0 \end{bmatrix}, \quad \beta =2.
\label{eq:socialexample}
\eeq
Fig.\ref{fig:doublethreshold}(a) and  (b) show the optimal policy and value function. These
were computed by constructing a  grid of 500 values for $\I = [0,1]$.
The double threshold behavior  of the stopping time problem when agents
perform social learning is due to the discontinuous dynamics  of the Bayesian social learning
filter (\ref{eq:piupdate}).

 (ii) Consider the following generalization of the
sequential detection problem (\ref{eq:stopsocial}). In  addition to the delay and error
probability costs,
we consider the total social learning cost incurred by all the agents. So now instead of (\ref{eq:costsocial1}) we have 
\begin{multline} \label{eq:socialext}
  J_\mu(\pi_0)  =
     \Ep\biggl\{\sum_{k=1}^{\tau-1} \discount^{k-1}  \E\left\{ \left.  \min_{a} \E\{c(x,a)|\H_k\} + 
  d I(x = e_1) \right\vert \mathcal{G}_{k-1}\right\} \\  + \rho^{\tau-1} \beta \E\{ I(x \neq e_1) | \mathcal{G}_{\tau-1}\}    
  +  \rho^{\tau-1} \E\{ \min_{a} \E\{c(x,a)|\H_{\tau}\}| \mathcal{G}_{\tau-1}\}\biggr\} . \end{multline}
Here $\mathcal{G}_k,\mathcal{H}_k$ are defined in (\ref{eq:siggn}).
The first and last terms above constitute the total social learning cost (\ref{eq:step2}) from time
1 to $\tau$.
 In terms of the public belief
$$ \E\{ \min_{a} \E\{c(x,a)|\H_{k}\}| \mathcal{G}_{k-1}\}  =
\sum_{y\in \Y} \min_a c_a^\p  T(\pi_{k-1},y) \sigma(\pi_{k-1},y) = \sum_{y \in \Y} \min_a c_a^\p B_y
 \pi_{k-1}.$$
 Then it can be shown that Theorem \ref{thm:stopsocial} holds providing $c(e_i,a)$ is decreasing in $i$.
   We chose the following parameters: Last term in (\ref{eq:socialext}) set to zero,
\beq
\rho = 0.9, \quad d = 1,\quad B = \begin{bmatrix} 0.9 & 0.1 \\ 0.1 & 0.9 \end{bmatrix} ,\;  c(e_i,a) = \begin{bmatrix}
2.1 & 3.1 \\ 3.1 &   0.53 \end{bmatrix}, \quad \beta =20.
\label{eq:socialexample2}
\eeq
Fig.\ref{fig:doublethreshold}(c) and  (d) show the optimal policy and value function. The
 optimal stopping policy is again a double threshold,
 and the value function is monotone on individual intervals.

\subsubsection*{Discussion}
The multiple threshold behavior (nonconvex stopping set $\mathcal{R}_1$) of Theorem
\ref{thm:stopsocial}   is unusual.  One would have thought
that if it was optimal to `continue'  for a particular  
belief $\pi^*(2)$, then it should be optimal to continue  for all beliefs $\pi(2)$ larger than $\pi^*(2)$.
The multiple threshold optimal policy shows that this is not true. Fig.\ref{fig:doublethreshold}(a) shows
that 
as the public belief $\pi(2) $ of state 2 decreases, the optimal decision switches from `continue' to `stop'
to `continue' and finally `stop'.
Thus 
  the global decision (stop or continue) is a non-monotone function
of public  beliefs obtained from local decisions. 

The main reason for this unusual  behavior is the  dependence of the  action likelihood  $\Bs_a$  on the belief state $\pi$.
This causes the social learning Bayesian filter to have a discontinuous update.
The value function is no longer concave on $\I$ and  the optimal policy is not necessarily monotone.
As shown in the proof of Theorem \ref{thm:stopsocial}, the value function $V(\pi) $ is concave on each of the 
intervals 
$\mathcal{P}_{l}$, $l=1,\ldots,4$.

To explain the 
 final claim of the theorem,
 let us define the intervals $\mathcal{P}_1$ and $\mathcal{P}_4$ more explicitly:
 \begin{align}
 \mathcal{P}_1 &= \{\pi : \min_a c_a^\p B_y \pi = 2, \;  \forall y \in \Y \} \\
 \mathcal{P}_2 &= \{\pi : \min_a c_a^\p B_y \pi = 1, \;  \forall y \in \Y \} \nonumber .\end{align}
For public belief $\pi \in \mathcal{P}_1$, the optimal local action
is $a=2$
 irrespective of the observation
$y$; similarly for  $\pi \in \mathcal{P}_4$, the optimal local action 
 is $a = 1$ irrespective of the observation
$y$. 
 Therefore, on intervals $\mathcal{P}_1$ and  $\mathcal{P}_4$, there is no social learning since the  local action
$a$ reveals nothing about the observation $y$ to subsequent agents. 
 Social learning only takes place when the public belief is in  $\mathcal{P}_2$ and $\mathcal{P}_3$.

Finally, we comment on the intervals $\mathcal{P}_l$, $l=1,\ldots,4$. They form a partition of
$\I$ such that if $\pi \in \mathcal{P}_l$, then $T(\pi,1) \in \mathcal{P}_{l+1}$ and $T(\pi,2) \in
\mathcal{P}_{l-1}$ (with obvious modifications for $l=1$ and $l=4$), see Lemma \ref{lem:social} in the appendix for details. In fact, $\eta_1$ and
$\eta_3$ are  fixed points of the composition  Bayesian maps: $\eta_1 = T(T(\eta_1,1),2)$ and $\eta_3 = T(T(\eta_3,2),1)$.
Given that the updates of the social Bayesian filter can be localized to specific intervals, we can
then inductively prove that the value function is concave on each such interval. This is the main
idea behind Theorem~\ref{thm:stopsocial}.

\subsection{Example 5: Constrained Social Optimum and Sequential Detection} \label{sec:cso}
In this subsection, we consider the constrained social optimum formulation in Chamley \cite[Chapter 4.5]{Cha04}.
We show in Theorem \ref{thm:socialopt} that the resulting stopping time problem has a threshold switching curve. Thus Chamley's  optimal social learning
 can be implemented efficiently in a multi-agent system.
  This is in contrast to the multi-threshold behavior
of  the stopping time problem in Sec.\ref{sec:social1} when agents were selfish in choosing their local actions.

The constrained social optimum formulation in \cite{Cha04} is motivated by the following question\footnote{The author gratefully acknowledges discussions with Dr C. Chamley who authored the book \cite{Cha04}. The
 constrained social optimum formulation presented in this section is due to Chamley and is  presented in \cite{Cha04}. 
 Our formulation considers a multi-state version together with sequential detection of state $e_1$.}:
How can agents aid social learning by acting benevolently and  choosing their action to sacrifice their local cost but optimize a social welfare cost? 
In  Sec.\ref{sec:social1},
 agents  ignore the information benefit their action provides
to others  resulting in information cascades where social
learning stops. 
 By constraining the choice by which agents pick their local action to two specific decision rules, the optimal  choice between
the two rules becomes a sequential decision problem. 

Assume $\A = \Y$. As in the social learning model above,  let $c_a$ denote the cost vector for picking local action $a$ and $\eta_k$ 
(see (\ref{eq:mudef}))   denote
the private belief of agent $k$.

Let $\tau$ denote a stopping time adapted to the sequence of sigma-algebras $\mathcal{G}_{k}$, $k \geq 1$
defined in (\ref{eq:siggn}). As in Sec.\ref{sec:social1}, 
the goal is to solve the following sequential
detection problem to detect state $e_1$: Pick the stopping time $\tau$ to minimize
\begin{align}
  J_\mu(\pizero)  &= 
   \Ep\biggl\{\sum_{k=1}^{\tau-1} \discount^{k-1}  \E\{ c(x,a_k)| \mathcal{G}_{k-1} \}
   + 
 \Ep\{\sum_{k=1}^{\tau-1} \discount^{k-1}  d I(x = e_1) | \mathcal{G}_{k-1}\}   \nonumber \\
 &  + \rho^{\tau-1} \Ep\{ \beta I(x \neq e_1) | \mathcal{G}_{\tau-1}\}
  + \frac{\rho^{\tau-1}}{1-\rho} \min_{a \in \Y}\E\{ c(x,a)   | \mathcal{G}_{\tau-1}\} \biggr\}.
 \label{eq:social0}
\end{align}
Similar to  (\ref{eq:socialext}), the second and third terms are the delay cost and error
probability in stopping and announcing state $e_1$.
The first and last terms model the total 
 social welfare cost involving all agents based on their local action. Let us explain these two terms.
The key difference compared to (\ref{eq:socialext}) is that agents now pick 
their local action 
according to the decision rule
$a(\pi,y,\mu(\pi)))$ (instead of myopically) as follows.
As in \cite{Cha04}, we constrain decision rule $a(\pi,y,\mu(\pi)))$ to two possible modes:
\beq \label{eq:acons}
a_k  = a(\pi_{k-1},y_k,u_k) = \begin{cases}  y_k  & \text{ if } u_k = \mu(\pi_{k-1}) = 2  \text{ (reveal observation)}\\
							\arg\min_a c_a^\p \pi_{k-1} & \text{ if } u_k = \mu(\pi_{k-1}) = 1 \text{ (stop) }.\end{cases}
							\eeq
Here the stationary policy $\mu:\pi_{k-1} \rightarrow u_k$ specifies which one of the two modes the benevolent agent $k$ chooses.
In  mode $u_k=2$, the  agent $k$ sacrifices is immediate cost $c(x,a_k)$ and picks action $a_k=y_k$ to reveal full information
to subsequent agents, thereby enhancing social learning.  

In mode $u_k=1$ the agent `stops and announces
state 1'.
Equivalently, using the terminology of \cite{Cha04}, the agent 
``herds" in mode $u_k=1$.  It ignores its private
observation $y_k$, and chooses its action selfishly to  minimizes its cost given the public belief $\pi_{k-1}$. So agent $k$ chooses
$ a_k = \arg\min_a c_a^\p \pi_{k-1} = a_{k-1} $. Then clearly from (\ref{eq:aprob}), $P(a|e_x,\pi)$ is functionally independent
of $x$ since $ P(y|x=e_i)$ is independent of $i$. Therefore 
 from (\ref{eq:piupdate}), if agent $k$ herds, then $\pi_k = T(\pi_{k-1},a_k) = \pi_{k-1}$, i.e.,  the public belief remains frozen. The total cost incurred in herding is then
equivalent to  final term in (\ref{eq:social0}):
$$\sum_{k=\tau}^\infty \rho^{k-1} \min_{a \in \Y}\E\{ c(x,a)   | \mathcal{G}_{k-1}\}
= \sum_{k=\tau}^\infty \rho^{k-1}\min_a c_a^\p \pi_{\tau-1} = \frac{\rho^{\tau-1}}{1-\rho} \min_a c_a^\p \pi_{\tau-1}.
$$

Define the {\em constrained social optimal  policy} $\mu^*$ such that
$J_{\mu^*}(\pizero)  = \inf_{\mu \in \Mu} J_{\mu}(\pizero)$. The sequential stopping problem
 (\ref{eq:social0}) seeks
 to determine the optimal policy $\mu^*$ to achieve the optimal tradeoff between stopping
 and announcing state 1 and the cost incurred by agents that are acting
 benevolently. 
In analogy to Theorem \ref{thm:1}, we show that $\mu^*(\pi)$ is characterized by a threshold curve.

Similar to (\ref{eq:costdef}) define the costs in terms of the belief state as
\beq
C(\pi,1) = \frac{1}{1-\rho} \min_{a\in \Y} c_a^\p \pi,\quad
C(\pi,2) = \sum_{y\in \Y} c_y^\p B_y \pi + (d + (1-\rho) \beta) e_1^\p \pi - (1-\rho) \beta \eeq
Below we list the assumptions and  main structural result which is similar to Theorem \ref{thm:1}.
These assumptions involve the social learning cost $c(e_i,a)$ and are not required if
these costs are zero.
\begin{itemize}
\item[(A1-Ex5)]  $c(e_i,a) - c(e_{i+1},a) \geq  0$ 
\item[(S-Ex5)] (i) $c(e_X,a) - c(e_i,a) \geq (1-\rho) \,\sum_y \bigl( c(e_X,y) B_{Xy} - c(e_i,y) B_{iy}\bigr) $\\
(ii)$ (1-\rho) \, \sum_y \bigl( c(e_1,y)B_{1y} - c(e_i,y) B_{iy} \bigr) \geq c(e_1,a)-c(e_i,a) $.
\end{itemize}

Similar to the discussion in  Sec.\ref{sec:assd}, (A1-Ex5) is sufficient for
$C(\pi,1)$ and $C(\pi,2)$ to be $\gr$ decreasing in $\pi \in \I$. 
This implies that the costs $C(e_i,u)$  are decreasing in $i$, i.e., state 1 is the most costly state.

(S-Ex5) is sufficient for $C(\pi,u)$ to be submodular. It implies that
$C(e_i,2) - C(e_i,1)$ is decreasing in $i$. This
gives economic incentive for agents to herd when approaching
the state $e_1$, since the differential cost between continuing and stopping is largest for $e_1$.   Intuitively,
the decision to stop (herd) should be made when the state estimate is sufficiently accurate so that revealing private observations is no longer
required.

\begin{theorem} \label{thm:socialopt}
Consider the sequential detection problem for state $e_1$ with
social welfare cost in  (\ref{eq:social0}) and
 constrained decision rule (\ref{eq:acons}). Then:\\
(i) Under (A1-Ex5), (A2), (S-Ex5),  constrained social optimal policy $\mu^*(\pi)$ satisfies the structural properties of Theorem \ref{thm:1}. 
(Thus a threshold switching curve exists).
\\
(ii) The stopping  set $\mathcal{R}_1$ is  the union
of $|\Y|$ convex sets (where $|\Y|$ denotes  cardinality of $\Y$). Note also that $\mathcal{R}_1$ is a  connected set by Statement (i). (Recall $\Y = \A$ in our formulation).
 \qed \end{theorem}

The proof of Theorem \ref{thm:socialopt} is in Appendix \ref{app:socialopt}.
The main implication of Theorem \ref{thm:socialopt}  is that the constrained optimal social learning scheme formulated in
Chamley \cite{Cha04} has a monotone structure.  This is in contrast to the multi-threshold behavior
of  the stopping time problem in Sec.\ref{sec:social1} when agents were selfish in picking their local actions.
In \cite[Chapter 4.5]{Cha04}, the above formulation is used for pricing information externalities in social learning.
From an implementation point of view, the existence of a threshold switching curve implies that
the  protocol only needs individual agents to store the optimal linear MLR
policy (computed, for example, using Algorithm \ref{alg1}).   Finally,  $\mathcal{R}_1$ is the union 
of $|\Y|$ convex sets  and  is non-convex in general. This is different to standard stopping problems
where the stopping set  is convex.

\section{Example 6: Multi-agent Scheduling in a Changing World} \label{sec:fast}

So far we have considered models where the underlying state is a constant (Sec.\ref{sec:change}),
or
a Markov chain that jumps once into an absorbing state
(change detection of Sec.\ref{sec:main}) or jumps twice (transient detection of Sec.\ref{sec:qtrans}).
In this section, we 
 consider   a more general model where the target state $x$ evolves on the same time scale
as the observation process. The target state $x$ jumps with time according
to a finite state Markov chain over the state space $\X$ with transition
probability matrix $P$. Also, unlike previous sections, decision $u=1$ does not `stop' the evolution
of the belief state.
So instead of a stopping time
problem, we have a  more general partially-observed stochastic control problem.

As mentioned in Sec.\ref{sec:intro}, this section is motivated by two questions: (i)
How can the optimal policy be bounded? (ii) How does the  optimal achievable cost vary with transition probability?
The main results of this section are two-fold. First, using Blackwell
dominance, we show that the optimal policy is lower bounded by a myopic policy (Theorem \ref{thm:compare2}).
Next, Theorem \ref{thm:tmove} shows that for the underlying Markovian state,
the larger the transition matrix (in an order defined in (\ref{eq:mor})), the cheaper the
expected optimal cost. This is useful in comparing the optimal achievable cost of quickest time detection
with different PH-distributions.

\subsection{Myopic Policy Bound to Optimal Decision Policy} \label{sec:myopic}

Consider a countable number of agents where
each agent acts once  in a predetermined sequential order indexed by $k=1,2,\ldots$ as follows: Based on the current
belief state $\pi_{k-1}$, agent $k$ chooses mode $$u_k \in \{1 \text{ (low resolution) }, 2 \text{ (high resolution)}\}.$$
Depending on its mode $u_k$, agent $k$  views the world according to this mode  -- that is, 
it  obtains  observation  
from a distribution that depends on $u_k$. 
Assume that for mode $u\in \{1,2\}$, the observation $y^{(u)} \in \Y^{(u)} = \{1,\ldots,Y^{(u)}\}$ is obtained
from  the matrix of conditional probabilities
\beq B^{(u)} = (B^{(u)}_{iy^{(u)}} , i \in \X,  y^{(u)} \in \Y^{(u)}) \text{ where }
B^{(u)}_{iy^{(u)}} = P(y^{(u)}|x=e_i,u) .\eeq
The notation $\Y^{(u)}$ allows for mode dependent observation spaces.
In  sensor scheduling \cite{Kri02}, the tradeoff is as follows: Mode
$u= 2$ yields more accurate observations of the state than mode $u=1$, but the cost of choosing mode  $u=2$ is higher
than mode $u=1$.
Thus
there is an tradeoff between the cost of acquiring information and the value of the information.
The assumption that mode $u=2$ yields more accurate observations than mode $u=1$ is modelled by
\beq \label{eq:bd}
   B^{(1)} = B^{(2)} Q . \eeq
   Here $Q$ is a $Y^{(2)} \times Y^{(1)}$ stochastic matrix. $Q$ can be viewed as a {\em confusion matrix} that maps $\Y^{(2)}$ probabilistically to $\Y^{(1)}$.
   (In a communications context, one can view $Q$ as a noisy discrete memoryless channel with input $y^{(2)}$ and output $y^{(1)}$).

When agent $k$ chooses mode $u_k \in \{1,2\}$, it incurs the expected cost
 \begin{align} \label{eq:exccost} C(\pi_{k-1},u_k) &=  \alpha_{u_k} \E\{\|(x_k - \E\{x_k|\F_{k}\})^\p g\|^2 |\F_{k-1}\} + \E\{c(x_k,u_k) | \F_{k-1} \}  \\
 &=  \alpha_{u_k} \bigl(G^\p P^\p\pi_{k-1} - (g^\p P^\p \pi_{k-1})^2 \bigr) + c^\p_{u_k} P^\p \pi_{k-1} \nonumber
 \end{align}
  where $c_u = (c(x=e_i,u),i\in \X)$, $\F_k$ is defined in (\ref{eq:sigf}), $g$ and $G$ are defined in (\ref{eq:varder}).
In  (\ref{eq:exccost}), the tradeoff between information obtained from a mode and the cost of operating in the mode is modelled as follows:
Choose  $\alpha_1 > \alpha_2$ to penalize choosing the less accurate mode 1 in terms of the variance, while
$c(e_i,1) < c(e_i,2)$ since mode 1 incurs a cheaper operating cost.

The goal is to compute the optimal policy $\mu^*(\pi) \in \{1,2\}$ to minimize the overall cost incurred by  all the agents
\beq
J_\mu(\pizero) = \Ep \{ \sum_{k=1}^\infty \rho^{k-1} C(\pi_{k-1},u_k) \} .
\eeq
The above problem is not a stopping time problem, since if mode $u=1$ is chosen,
 the problem does not terminate.  The mode $u$ chosen by each agent will affect the modes chosen by subsequent  agents, and hence  affects the total  cost. For such a  partially observed stochastic control problem, determining
 the optimal policy $\mu^*(\pi)$ is computationally intractable.
 However, using Blackwell dominance, we show below that a myopic policy forms a lower
 bound for the optimal policy.

The
value function $V(\pi)$ and optimal policy $\mu^*(\pi)$  satisfy the dynamic programming equation
\begin{align} \label{eq:dp_algmove}
V(\pi) &= \min_{u \in \U} Q(\pi,u), \quad
\mu^*(\pi)= \arg\min_{u \in \U} Q(\pi,u) ,\; J_{\mu^*}(\pi) = V(\pi) \\
 Q(\pi,u) &=  C(\pi,u) 
+ \discount \sum_{y^{(u)} \in \Y^{(u)}}  V\left( T(\pi ,y^{(u)}) \right) \sigma(\pi,y^{(u)}), \nonumber \\
T(\pi,y^{(u)}) &= \frac{B_{y^{(u)}} P^\p \pi}{\sigma(\pi,a)},
\; \sigma(\pi,y^{(u)}) = \mathbf{1}_X^\p B_{y^{(u)}} P^\p \pi .\nonumber
\end{align}

We now present the structural result.
Let $\Pi^s \subset \I$ denote the set of belief states for which
$C(\pi,2) < C(\pi,1)$. 
Define the  myopic policy
$$\bar{\mu}(\pi) = \begin{cases} 2 & \pi \in \Pi^s \\
 								1 & \text{ otherwise } \end{cases}$$

\begin{theorem}\label{thm:compare2} The myopic  policy $\bar{\mu}(\pi)$ satisfies
the following property:
For all $\pi \in \Pi^s$,  $\mu^*(\pi) = \bar{\mu}(\pi)$,
i.e., it is optimal to pick action 2.
Therefore,
 $\bar{\mu}(\pi)$ forms is a lower   bound for the optimal
policy $\mu^*(\pi)$, i.e.,  $\mu^*(\pi) \geq \bar{\mu}(\pi)$ for
all $\pi \in \I$. 
\qed
\end{theorem}

Theorem \ref{thm:compare2} is proved in Appendix \ref{app:compare2}.
The usefulness of Theorem \ref{thm:compare2} stems from the fact that $\bar{\mu}(\pi)$ is trivial to compute. It  forms a rigorous
lower bound to the computationally intractable optimal policy $\mu^*(\pi)$.
What the
theorem says is that $\bar{\mu}(\pi) $  lower bounds the optimal policy, and coincides
with the optimal policy in region $\Pi^s$.
Since $\bar{\mu}$ is sub-optimal, it incurs a higher  cost. This cost can be evaluated via simulation and forms
an upper bound to the optimal achievable cost.

Theorem \ref{thm:compare2} is  non-trivial.
Just because at some time $k$, the expected instantaneous costs
satisfy $C(\pi_k,2) < C(\pi_k,1)$,  does not necessarily imply that the myopic policy 
$\bar{\mu}(\pi)$ coincides with the optimal policy $\mu^*(\pi)$, since the optimal policy applies to
the infinite trajectory of the dynamical system.

 The proof uses Blackwell dominance of measures.
The first instance of a similar proof using Blackwell dominance for POMDPs
 was given in \cite{WD80}, see also \cite{Rie91}.
Our proof is similar and  uses concavity
of the value function $V(\pi)$ and
Blackwell dominance of observation probabilities. 
In particular, 
observation $y^{(2)}$ is 
 more informative than (Blackwell dominates) observation $y^{(1)}$, if  (\ref{eq:bd}) holds,
see \cite{Rie91}.
The proof of  Theorem \ref{thm:compare2} in the appendix
comprises of first proving concavity of $V(\pi)$.
The  proof is a non trivial extension, since $C(\pi,1)$ is nonlinear in~$\pi$.

\subsection{Effect of State Transition Matrix}
The next structural result  establishes
 how the  optimal expected  cost varies 
  with
different transition matrices.
 The model we consider applies to both stopping time problems (such as quickest time detection)
and  multi-agent scheduling considered above.

 Suppose  $\Pone$, $\Ptwo$ are two distinct
 transition probability matrices corresponding to two distinct models of Markov state evolution. 
  Let $V(\pi;\Pone)$ and 
$V(\pi;\Ptwo)$ denote the corresponding  optimal value function in (\ref{eq:dp_algmove}).
The question we pose is:  How does $V(\pi;P)$ vary with transition matrix $P$? For example, in the quickest  detection
problem, do certain phase-type distributions result in larger total optimal cost compared to other phase-type
distributions? A similar question can be posed for the stochastic control problem considered above.

We consider costs  that are linear in the belief state.
 To show the explicit dependence on the transition matrix, define the costs
 (recall $c_u = (c(e_i,u),\,i\in \X)$)
 $$
 C(\pi,u;P) = c_u^\p P^\p \pi , \quad u \in \{1,2\}.$$
The result below also applies to the case where $C(\pi,u) = c_u^\p \pi$ (i.e., the cost at each
stage is not an explicit function of transition matrix).

 Define the following ordering of transition matrices $\Pone$ and $\Ptwo$: 
\beq \Pone\succeq \Ptwo \text{ if } \Pone_{ij} \Ptwo_{m,l} \leq \Ptwo_{ij} \Pone_{m,l}\; , \quad
l > j,\quad  i,j,l,m \in \X. \label{eq:mor}\eeq

\begin{theorem} \label{thm:tmove}
Assume $\Pone\succeq \Ptwo$,
$c(e_i,u)$ is decreasing in $i$, $C(\pi,u;\Pone) \leq C(\pi,u;\Ptwo)$ 
and  (A2), (A3)  hold. Then:
\\
(i)  $V(\pi;\Pone)\leq V(\pi;\Ptwo)$. That is, the larger the transition matrix  (with respect to the partial  ordering (\ref{eq:mor})),
the lower the optimal expected  cost incurred when making optimal decisions.\\
(ii) Consider the quickest time detection problem  with  $\alpha=0$ and costs in (\ref{eq:stylized}).
The optimal expected  cost  $\Vb(\pi;P)$ (see (\ref{eq:dp_initial}))
with change time distribution $\Pone$ is less than that with $\Ptwo$.\\
\mbox{} \qed
\end{theorem}

Theorem \ref{thm:tmove} is proved in Appendix \ref{app:tmove}.
We now present three examples.

\begin{figure} \centering
\epsfig{figure=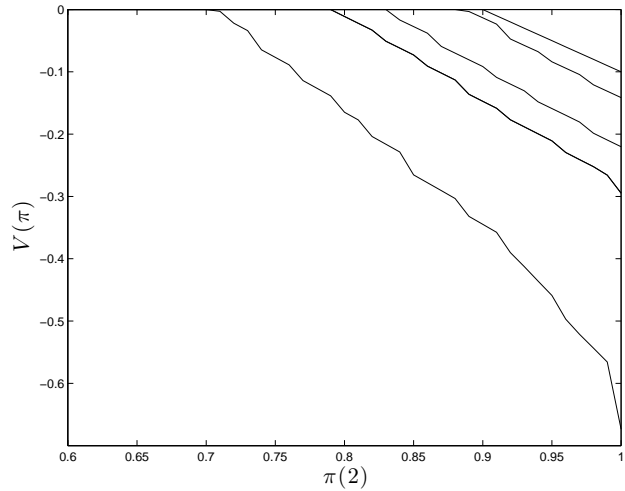,width=0.5\linewidth} 
\caption{Monotone behaviour of optimal expected  cost (value function $V(\pi;P^{(p)})$) versus probability $p$
for  quickest
detection with geometric change time, see (\ref{eq:ppd}) for notation.
The values of $p$ are 0.99 (lowest plot), 0.95, 0.9, 0.8, and 0.01 (highest plot).
} \label{fig:pdominance}\end{figure}

\noindent {\em Example (i)}:
Consider $\X=\{1,2\}$ and the dynamic decision making formulation of Sec.\ref{sec:myopic}. Then  using (\ref{eq:mor}) it can be verified that  the transition matrices
$$ \Pone = \begin{bmatrix} 0.2  & 0.8 \\ 0.1 & 0.9 \end{bmatrix} \succeq  
\Ptwo = \begin{bmatrix}  0.8 & 0.2 \\ 0.7 & 0.3 \end{bmatrix}. $$
Note that $\Pone$ and $\Ptwo$ above are TP2 (as required by (A3)).
So Theorem \ref{thm:tmove} applies.
\\
{\em Example (ii)}:
The transition matrix
$\begin{bmatrix}  p & (1-p) \\ p  & (1-p) \end{bmatrix}$ corresponding to a
two state iid process is decreasing with respect to 
the order (\ref{eq:mor}) as $p$ increases from 0 to 1.
So the theorem says that the smaller $p$ is, the cheaper the  optimal expected  cost. 
So even though the underlying process has maximum uncertainty
(entropy) when $p=0.5$, Theorem \ref{thm:tmove} says that the largest total cost incurred is when $p=1$.
\\
{\em Example (iii)}: Consider the quickest detection costs (\ref{eq:stylized}) with $\alpha=0$.
 Consider first the geometric distributed change time case ($X=2$) with transition matrix 
 $P^{(p)}$ and  parameters in (\ref{eq:stylized})
\beq P^{(p)} = \begin{bmatrix} 1 & 0 \\ 1-p & p \end{bmatrix} , \;
B = \begin{bmatrix} 0.8 & 0.2 \\ 0.2 & 0.8 \end{bmatrix}, \; \rho =0.9, \; d = 0.9.
\label{eq:ppd}
\eeq
It can be verified that $P^{(p)}$ is increasing (in terms of (\ref{eq:mor})) in $p$.
So Theorem~\ref{thm:tmove} says that the larger $p$ is, the smaller the average optimal cost 
(value function $V(\pi;P^{(p)})$) for
quickest time detection.
In Fig.\ref{fig:pdominance} we plot the value function $V(\pi;P^{(p)})$ for several values of $p$, to illustrate
this behavior.

Next, consider quickest detection with PH-distributed change times (\ref{eq:nu}) modelled by
 the following transition matrices:
 in quickest detection:
$$ \Pone = \begin{bmatrix}
1 & 0 & 0 \\ 0.5 & 0.3 & 0.2 \\
0.3 & 0.4 & 0.3 \end{bmatrix} \succeq \Ptwo = \begin{bmatrix}
1 & 0 & 0 \\ 0.9 & 0.1 & 0 \\ 0.8 & 0.15 & 0.05 \end{bmatrix} .$$
Since $\Pone$ and $\Ptwo$ are TP2 by (A3),
Theorem \ref{thm:tmove}  
implies that the optimal expected cost incurred in quickest  change detection with PH-distributed
change time $\Pone$ is less than
that of $\Ptwo$.

{\em Discussion. (i) TP2 dominance versus dominance in (\ref{eq:mor})}.
 It is shown in \cite{Lov87a,Rie91}  that if transition matrices  are ordered in the TP2 sense,
 namely  $\Pone \gtp \Ptwo$  (see Defn.\ref{def:tp2}(i) in
Appendix), then Theorem \ref{thm:tmove}
holds under the same assumptions as above. 
It is easy to prove that for $2 \times 2 $ case,
 only  matrices with identical rows, i.e., transition matrices modelling independent
 and identically distributed (iid) finite state processes,
   satisfy $\Pone \gtp \Ptwo$.
 Our conjecture is that the only examples of   transition
matrices that satisfy $\Pone \gtp \Ptwo$  are transition matrices corresponding to iid processes. So
TP2 dominance is less useful than the ordering (\ref{eq:mor}).

 (ii) {\em Kolmogorov--Shiryayev criterion}: If the 
 Kolmogorov--Shiryayev criterion (\ref{eq:ksd}) is considered then a similar proof to Theorem
 \ref{thm:tmove} shows that $V(\pi;P)$ is increasing with $P$.
The reason is that   in this case
$C(\pi,2;P) $ in (\ref{eq:costdef}) (with  $d e_1^\p P^\p \pi$ 
replaced by $d e_1^\p \pi$, see Theorem \ref{thm:modified}) has the term $- \rho(\alpha+\beta)  e_1^\p P^\p \pi$. This 
is increasing in $P$ (wrt  ordering (\ref{eq:mor})).

\section{Numerical Examples} \label{sec:numerical}
For state-space $\X=\{1,2,3\}$, i.e., $X=3$ states, the belief state space $\I$ is an equilateral triangle, and
the various results of this paper can be  illustrated visually.
There is much flexibility for choice
of parameters that satisfy the general assumptions (A1), (A2), (A3), (S) in the appendix. 

{\bf Example 1}:
We illustrate the structural
result Theorem \ref{thm:1}  for quickest time change detection with PH-distributed change time
and variance penalty.
The following  parameters were chosen in (\ref{eq:stylized}):
$$\beta=1,\;\rho=1,\; P=\begin{bmatrix} 1 &0  & 0 \\ 0.3 & 0.1 & 0.6 \\ 0 & 0.02 & 0.98 \end{bmatrix}$$
Assume Gaussian  observation noise with variance $0.01$; so  $\Y = \reals$ and the observation likelihoods are
$ B_{1y} \sim N(0,0.01)$, $B_{2y} \sim N(1,0.01)$. The operational
cost $\cop = 10^{-3} $ (see discussion below (\ref{eq:exd})).

\begin{figure} \centering
\mbox{\subfigure[$\alpha=0$, $d=1$]
{\epsfig{figure=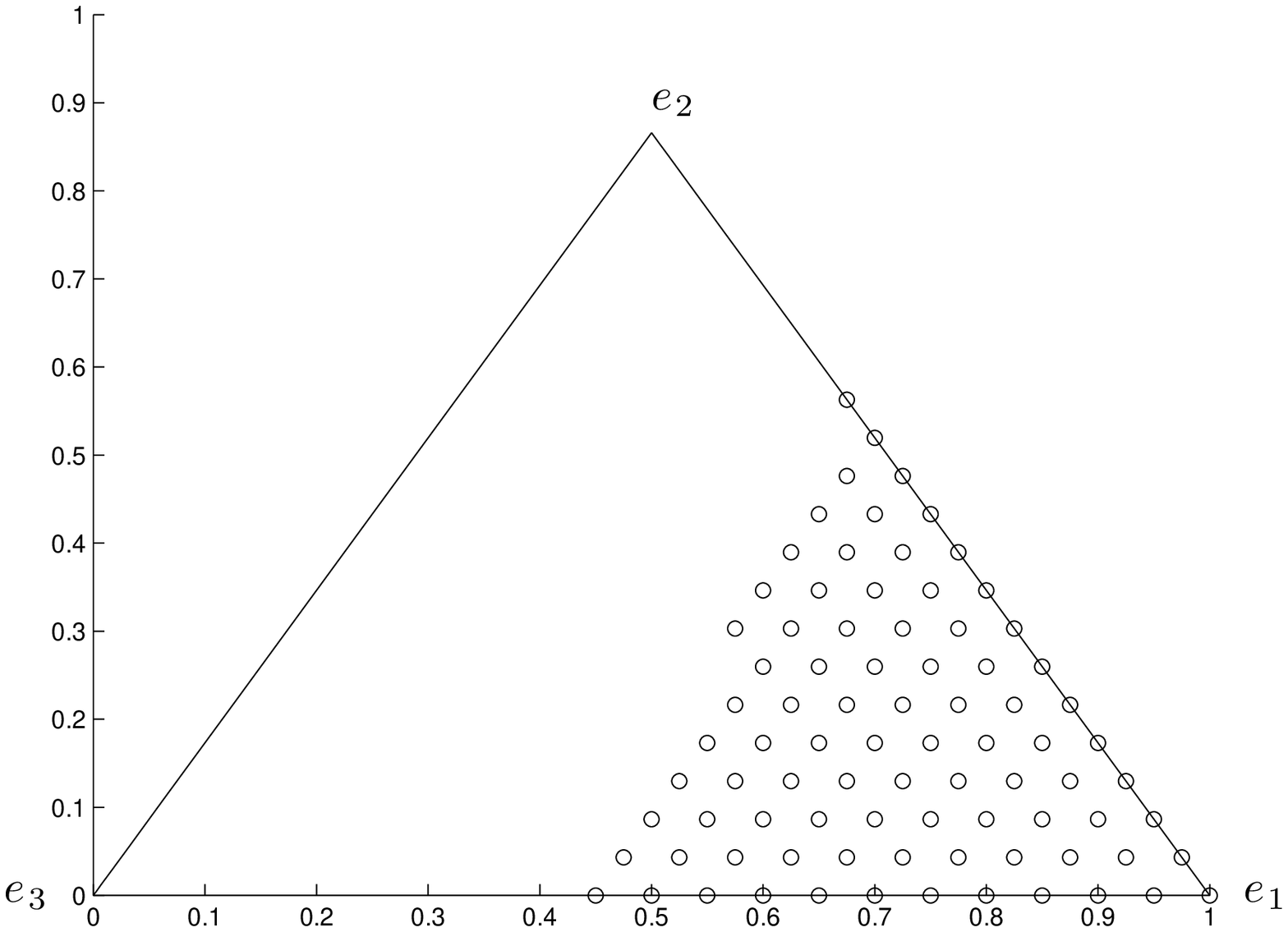,width=0.45\linewidth}} \quad
\subfigure[$\alpha=1$, $d=2$]{\epsfig{figure=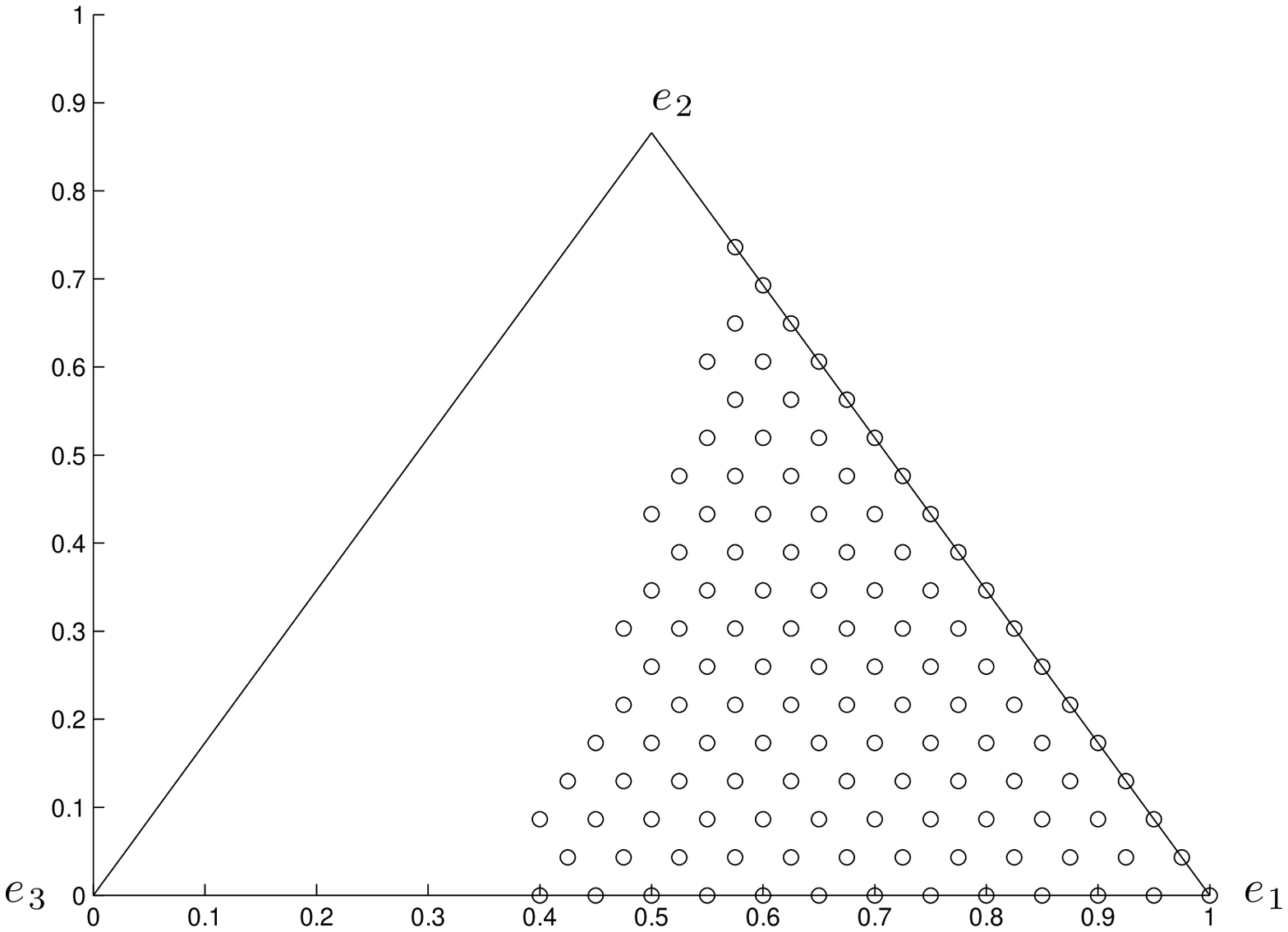,width=0.45\linewidth}}} \\
\mbox{\subfigure[$\alpha=10$, $d=11$]{\epsfig{figure=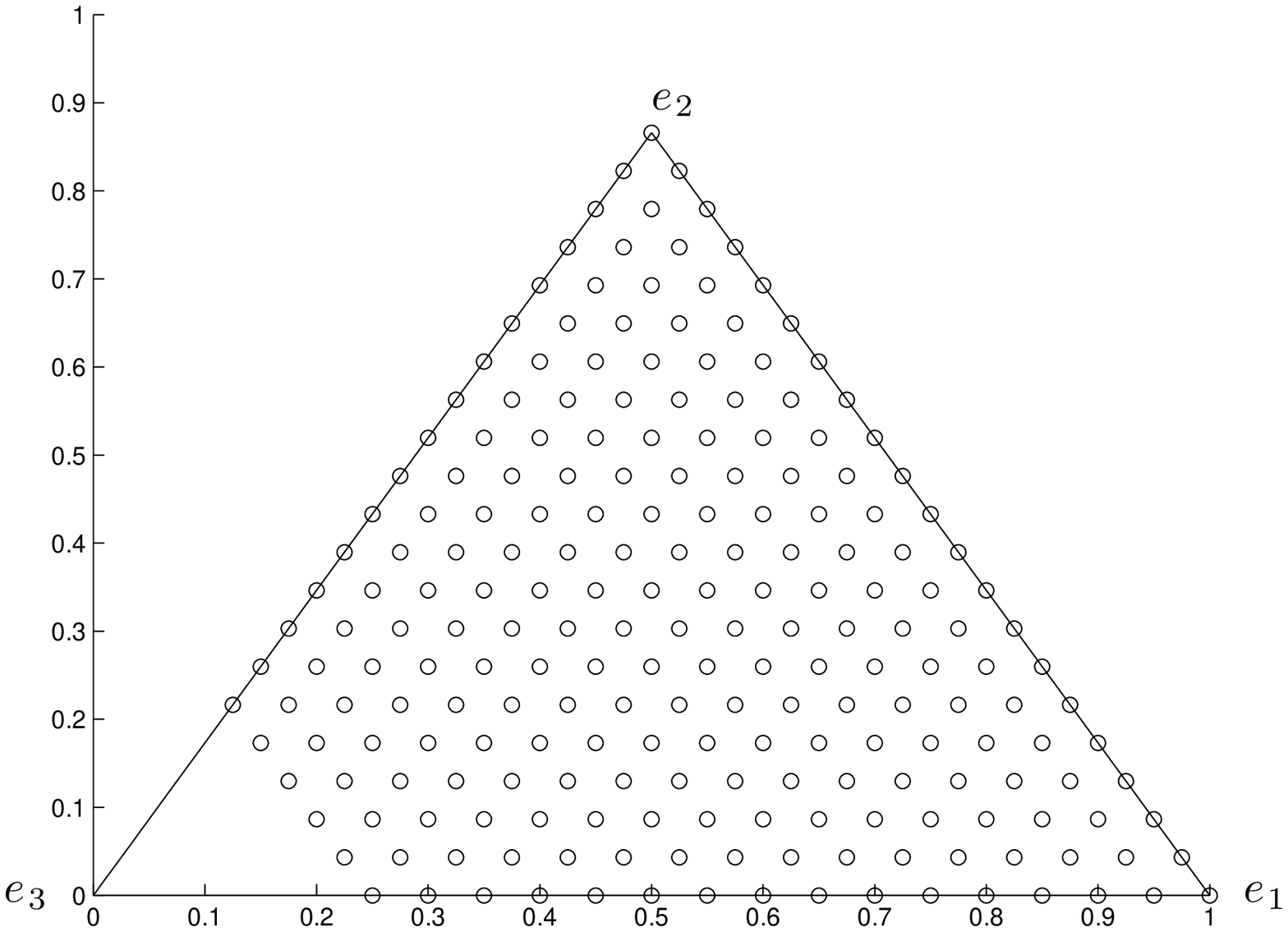,width=0.45\linewidth}} \quad
\subfigure[$\alpha=10$, $d = 5$]{\epsfig{figure=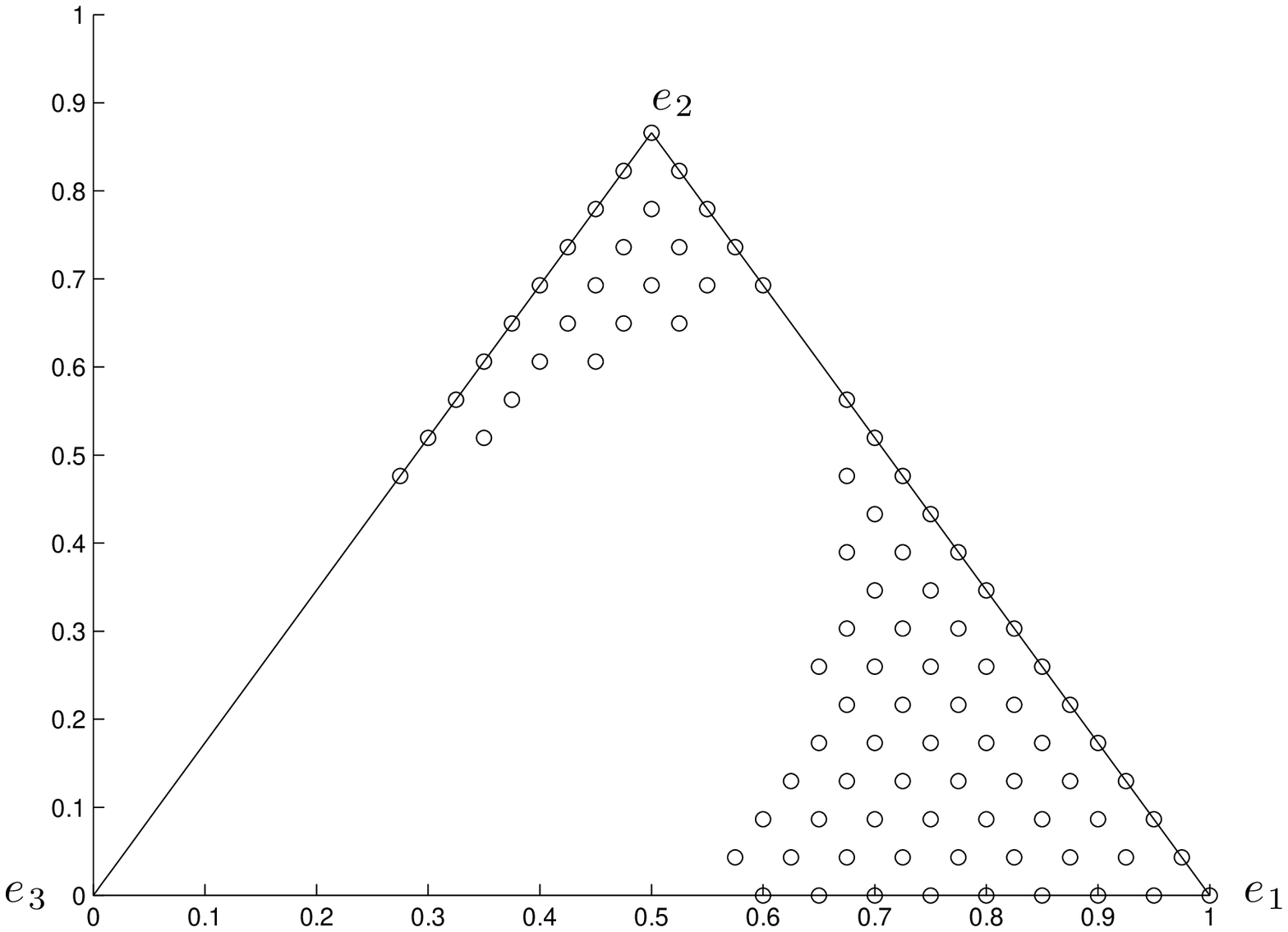,width=0.45\linewidth}}} 
\caption{Quickest Time Detection illustrating  Theorem \ref{thm:1}. The $\circ$ region
represents the stopping set $\R_1$ where decision  $u = 1 \text{ (stop)}$ is optimal.  The empty region in the simplex represents $\R_2$
where $u = 2 \text{ (continue)}$ is optimal. Cases (a), (b) and (c) satisfy the assumptions of Theorem 1.
The optimal linear threshold for these 3 cases  was estimated using Algorithm \ref{alg:spsa}.
Case (d) does not satisfy Assumptions (S-Ex1) and $\R_1$ is not a connected set.} \label{fig:qd1}
\end{figure}

The optimal policy was computed 
by forming a grid of $230$ values in the 2-dimensional unit simplex, and then solving the
value iteration algorithm (\ref{eq:vi}) over this grid and a horizon length of 200.
The optimal policy is shown in Fig.\ref{fig:qd1} for four different choices of $\alpha$ and $d$.
The first three examples, namely  $\alpha,d$ specified in Fig.\ref{fig:qd1}(a), (b) and (c),
satisfy assumptions (A1-Ex1), (A2), (A3) and (S-Ex1).
Therefore
the optimal policy $\mu^*(\pi)$ satisfies Theorem \ref{thm:1} and is characterized
by a threshold curve $\Gamma(\pi)$.
The figures clearly show the existence of a threshold curve that partitions $\I$ into two
individually connected regions. 
 Recall that for the case $\alpha = 0$, Theorem \ref{thm:1} says
that the optimal stopping set $\R_1$ is a  convex set.

Is the stopping set $\R_1$ always a connected set even when the assumptions of Theorem \ref{thm:1} do not hold? Recall the assumptions
in Theorem \ref{thm:1} are sufficient conditions. The parameters $\alpha=10$, $d=5$ in Fig.\ref{fig:qd1}(d)
does not satisfy condition (S-Ex1) in Appendix \ref{sec:applp}. As shown in Fig.\ref{fig:qd1}(d),
the optimal stopping set $\R_1$ is no longer connected.  This highlights the importance of developing useful sufficient conditions that result in monotone policies $\mu^*(\pi)$, such as the assumptions presented in this paper.

{\bf Example 2}: Here we consider the classical delay cost (\ref{eq:exd2}). We
  illustrate how the optimal stopping region $\R_1$ varies with
  transition probabilities of the PH-distribution for change time.
  Since all these constraints in $\f$  are linear,
   determining feasible choices  is straightforward using the Matlab
command
{\tt linprog}. 

We chose $B_{1y} \sim N(0,4)$, $B_{2y} = N(1,4)$ and the following parameters in (\ref{eq:modified2}):
\beq \alpha = 0.5,\; \beta = 1,\;d = 1,\; \rho = 0.75,\; \f = \begin{bmatrix}
0 \\ 1 \\ 2  
\end{bmatrix},\; P = \begin{bmatrix} 1 & 0 & 0 \\ 0.3 & 0.6 & 0.1 \\
0.1  & p & 0.9- p
  \end{bmatrix}. \label{eq:effect}
\eeq
Then (A3) holds, i.e., $P$ is TP2 for $p \in [0.2, 0.7714]$. Also it can be verified that
all the other assumptions of Theorem \ref{thm:modified} hold.

\begin{figure} \centering
\mbox{\subfigure[$p=0.2$]
{\epsfig{figure=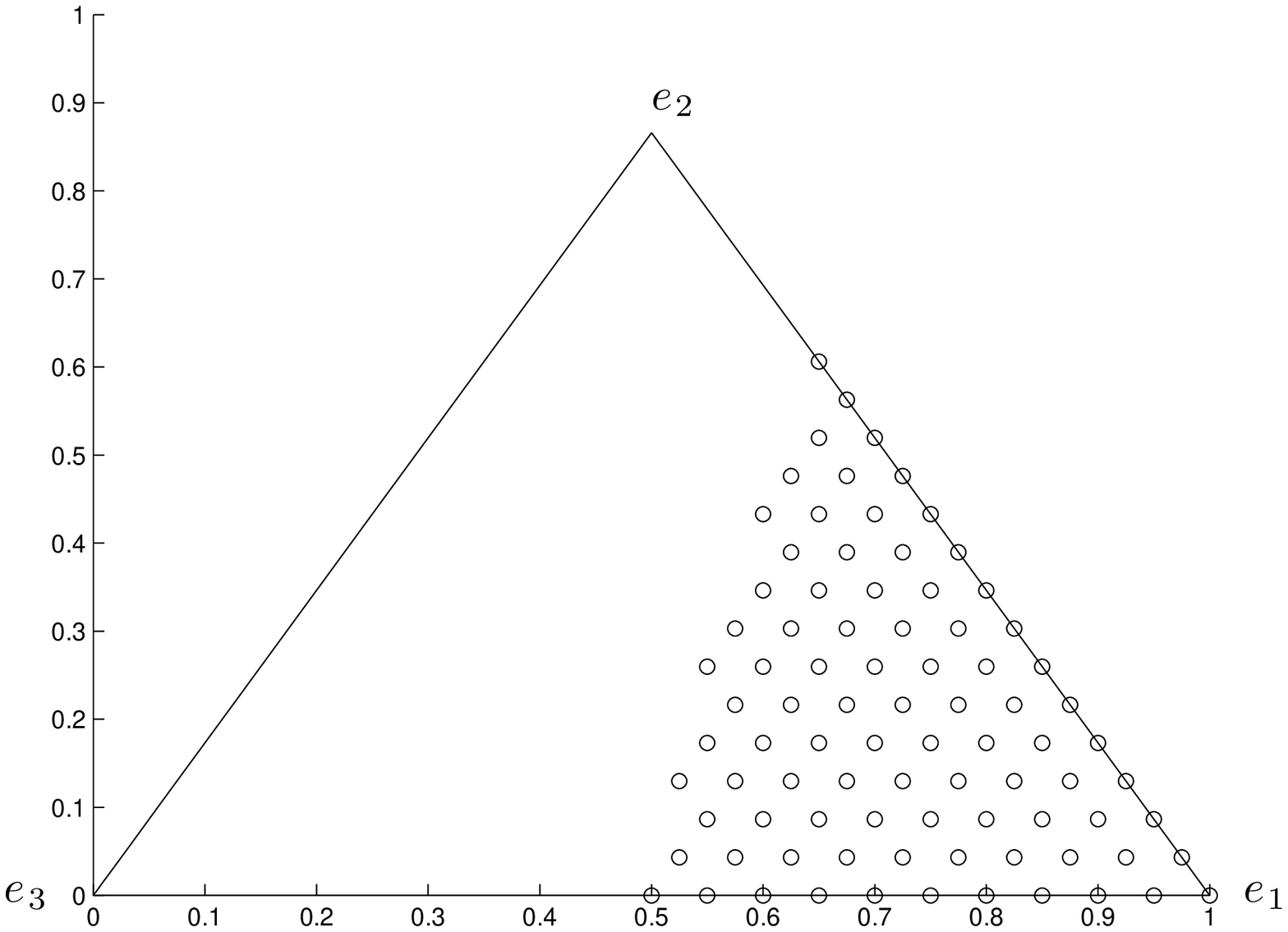,width=0.45\linewidth}} \quad
\subfigure[$p=0.77$]{\epsfig{figure=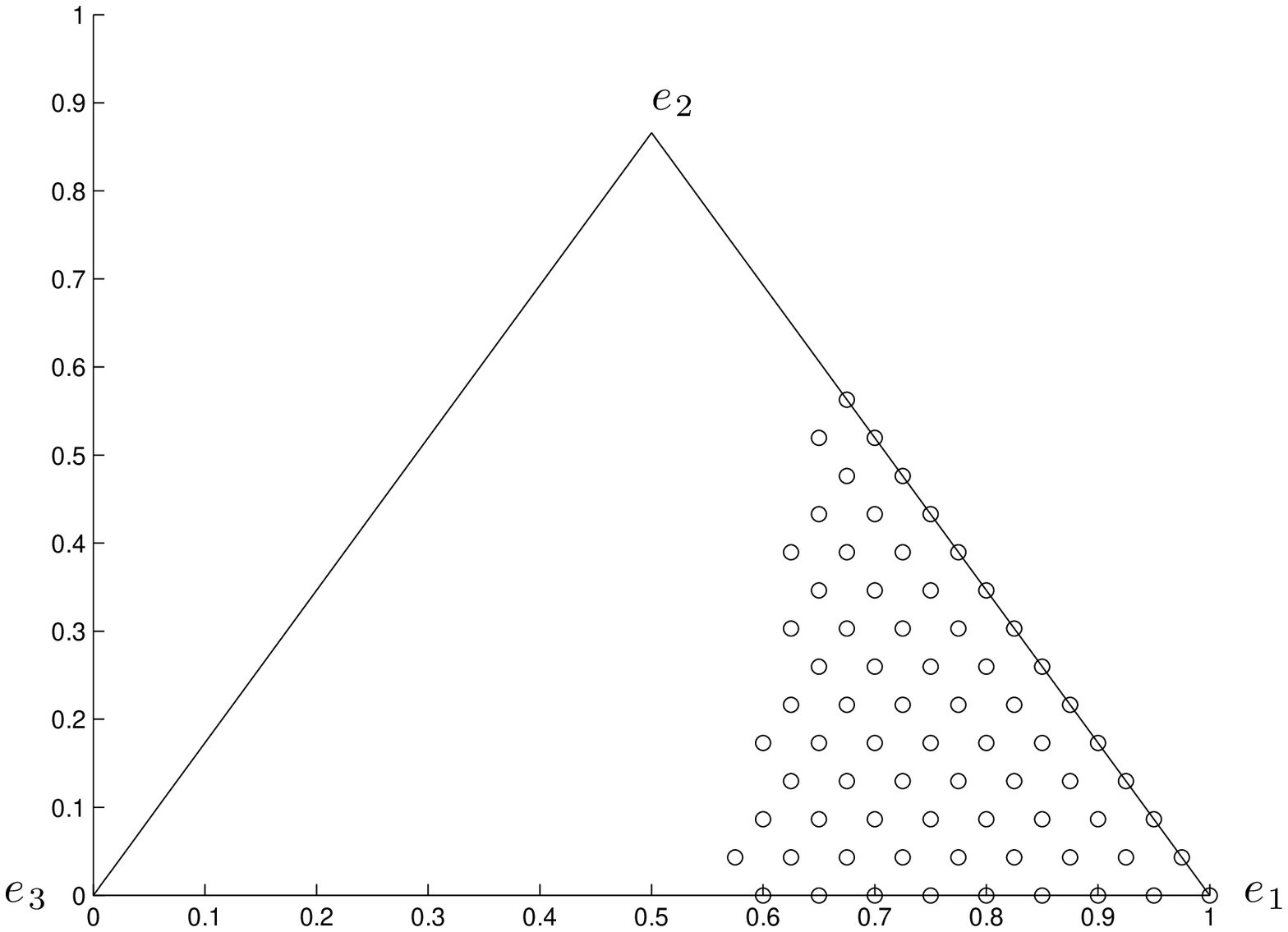,width=0.45\linewidth}}}\\
\mbox{\subfigure[Change time distribution]{\epsfig{figure=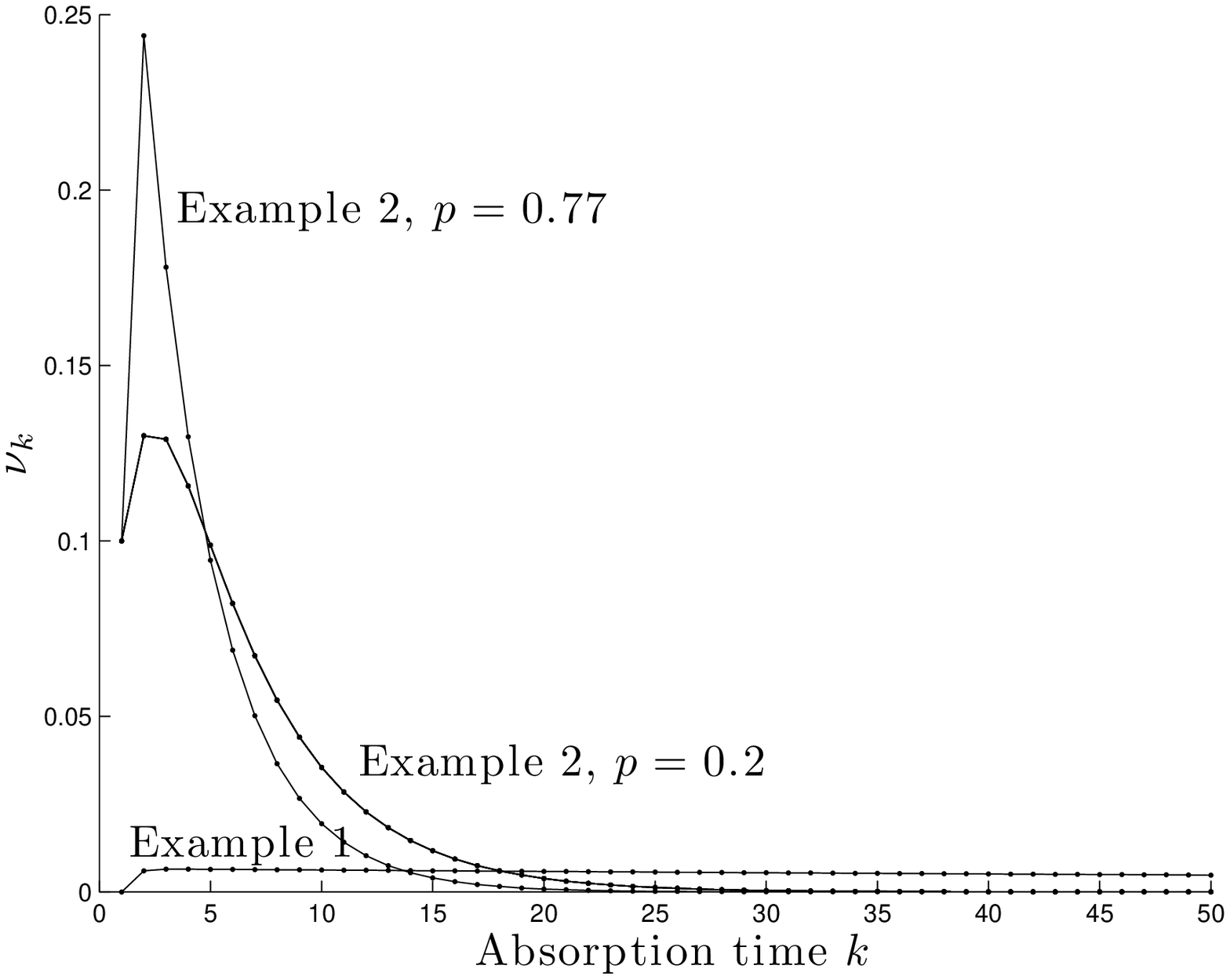,width=0.45\linewidth}}}
\caption{Effect of PH-distribution probabilities (\ref{eq:effect})  on optimal stopping region $\R_1$. The region marked
with $\circ$ denotes $\R_1$.} \label{fig:rev_transp}
\end{figure}

Fig.\ref{fig:rev_transp}(a),(b) illustrate the optimal stopping region $\R_1$ for $p = 0.2$ and  $p = 0.77$,
respectively. Fig.\ref{fig:rev_transp}(c) plots the PH-distribution probability mass function $\nu_k$
in  (\ref{eq:nu}) vs time $k$ for the transition probabilities in Example 1 and Example 2 for $p = 0.2$
and $p=0.77$.
It can be seen that even with a 3-state Markov chain the behavior is quite different
to a geometric distribution.

\section{Conclusions}
This paper has presented structural results for Bayesian quickest time detection with PH-distributed change time and variance penalty.
The main result is  Theorem \ref{thm:1}
which proves the existence of a threshold switching curve for optimal
decision making under general assumptions (A1), (A2), (A3), (S) given in the appendix.  Theorem \ref{thm:dep}
gave  necessary and sufficient conditions for a linear threshold policy to approximate
the threshold curve. Then several examples were considered, namely quickest transient detection, 
 quickest time detection with exponential penalty,  stopping time problems in social learning, constrained optimal social learning, and multi-agent scheduling
 in a changing world.
In the case of exponential penalty we used a risk sensitive stochastic control formulation.
 In all these examples, under similar assumptions
to Theorem~\ref{thm:1}, the threshold switching curve holds. The proofs of the results use lattice programming and stochastic orders on
the unit simplex.
The structural results of this paper are ÔclassÕ type results, that is, for parameters belonging to a set, the results 
hold. Hence there is an inherent robustness in these results since even if the underlying  parameters are 
not exactly specified but still belong to the appropriate sets, the  results still hold.
It would be useful to do a performance analysis of the various optimal detectors proposed in this paper --
see \cite{VV09} and references therein.

\appendix

\section{Proofs}

\subsection{Stochastic Orders and Submodularity} \label{sec:mlrdef}
Theorem \ref{thm:1} below requires  proving
that the quickest time detection policy $\mu^*(\pi)$ is monotonically increasing in belief
state $\pi$. That is, $\pi \leq \tpi $ (in a sense to be made clear below),
implies $\mu^*(\pi) \leq \mu^*(\tpi)$.
 In order to compare  belief states
$\pi$ and $\tpi$, 
we will use the monotone likelihood ratio (MLR)
stochastic ordering and a specialized version of the MLR order restricted to lines in
the simplex $\I$.
This stochastic order is  useful  since it is preserved under conditional expectations
 \cite{Rie91,KR80,Whi82,MS02}. Below we introduce
 several important definitions that will be used subsequently.
 
\begin{definition}[MLR ordering, \cite{MS02}]
Let $\pi_1, \pi_2 \in \I$ be any two belief state vectors.
Then $\pi_1$ is greater than $\pi_2$ with respect to the MLR ordering -- denoted as
$\pi_1 \gr \pi_2$,
 if 
\beq \pi_1(i) \pi_2(j) \leq \pi_2(i) \pi_1(j), \quad i < j, i,j\in \{1,\ldots,X\}. 
\label{eq:mlrorder}\eeq
\end{definition}
Similarly $\pi_1 \lr \pi_2$ if  $\leq$ in (\ref{eq:mlrorder}) is replaced
by a $\geq$.

\begin{definition}[First order stochastic dominance, \cite{MS02}]
 Let $\pi_1 ,\pi_2 \in \I$.
Then $\pi_1$ first order stochastically dominates $\pi_2$  -- denoted as
$\pi_1 \gs \pi_2$ --
 if 
$\sum_{i=j}^X \pi_1(i) \geq \sum_{i=j}^X \pi_2(i)$  for $ j=1,\ldots,X$.
\end{definition}

\begin{result}[\cite{MS02}] \label{res1}
 (i)  Let $\pi_1 ,\pi_2 \in \I$.
Then $\pi_1 \gr \pi_2$ implies $\pi_1\gs \pi_2$.\\
(ii) Let $\mathcal{V}$ denote the set of all $S$ dimensional vectors
$v$ with 
 nondecreasing components, i.e., $v_1 \leq v_2 \leq \cdots
v_X$.
Then $\pi_1 \gs \pi_2$ iff for all $v \in \mathcal{V}$,
 $v^\p \pi_1 \geq v^\p \pi_2$. 
\end{result}

For state-space dimension $X =2$, MLR is a complete order and coincides with
first order stochastic dominance.
For state-space dimension $X >2$,
MLR is a  {\em partial order}, i.e., $[\I,\gr]$ is a partially ordered set (poset) since it is not always
possible to order any two belief states $\pi \in \I$. However, on line
segments in the simplex
 defined below, MLR is a total ordering.

For $i \in \X$, define the sub simplex 
$\H_i \subset \I$  as
\beq \label{eq:hi} \H_i =  \{\pi \in \I:  \pi(i) = 0 \}. \eeq
Denote belief states that lie in $\H_i$ by $\bp$.
 For each $\bp \in \H_i$, construct the line segment $\l(e_{i},\bp)$ that connects $\bp$ to $e_{i}$. 
Thus
$\l(e_{i},\bp)$ comprises of belief states 
$\pi$ of the form:
\beq \l(e_{i},\bp) = \{\pi \in \I: \pi = (1-\epsilon) \bp + \epsilon e_{i}, \;
0 \leq \epsilon \leq 1 \} ,
 \bp \in \H_i.  \label{eq:lines}
 \eeq

\begin{definition}[MLR ordering  ${\gl}$  on  lines]  \label{def:tp2l}
 $\pi_1$ is greater than $\pi_2$ with respect to the MLR ordering on
the line $\l(e_{i},\bp)$ -- denoted as $\pi_1\gl \pi_2$, if 
$\pi_1,\pi_2 \in \l(e_i,\bp)$ for  some $\bp \in \H_i$, i.e., $\pi_1$,$\pi_2$  are on the same line connected to vertex $e_i$ of simplex $\I$, and
$\pi_1 \gr \pi_2$. 
\end{definition}

Note that $[\I,\glX]$ and $[\I,\glone]$ are chains\footnote{A chain is totally ordered subset of a partially ordered set.}, i.e., all elements
$\pi,\tpi \in \l(e_{X},\bp)$ are comparable, i.e., either $\pi\glX \tpi$ or $\tpi \glX \pi$
(and similarly for  $\l(e_{1},\bp)$) In Lemma \ref{lem:convex}, we summarize useful properties of $[\Pi(x),\gl]$ that
will be used in our proofs.

\begin{lemma} \label{lem:convex} The following properties hold on
 $[\I,\gr]$, $[\l(e_X,\bp),\glX]$.\\
(i) On  $[\I,\gr]$, $e_1$ is the least and  $e_X$ is the greatest element.
On $[\l(e_X,\bp),\gl]$, $\bp$ is the least  and $e_X$ is the greatest element.\\
(ii) Convex combinations of MLR comparable belief states form a chain. 
For any $\gamma \in [0,1]$,
$\pi \lr \tpi \implies \pi \lr \gamma\pi + (1-\gamma) \tpi \lr \tpi $.
(iii) All points on a line $\l(e_X,\bp)$ 
are MLR comparable. Consider  any two points
$\pi^{\gamma_1},\pi^{\gamma_2}\in \l(e_X,\bp)$ (\ref{eq:lines}).
Then 
$\gamma_1 \geq \gamma_2$, implies $\pi^{\gamma_1}
\gl \pi^{\gamma_2}$.  \end{lemma}

Let $\i = (i_1,\ldots,i_L)$ and $\j = (j_1,\ldots,j_L)$ denote the indices
of two $L$-variate probability mass functions
Denote the lattice operators 
\begin{align}
\i \wedge \j &= [\min(i_1,j_1),  \ldots, \min(i_L,j_L) ]^\p,\quad
\i \vee \j = [\max(i_1,j_1), \ldots, \max(i_L,j_L) ]^\p .
\end{align}

\begin{definition}[TP2 ordering and Reflexive TP2 distributions]  \label{def:tp2}
Let $P $ and $Q$ 
 denote any two $L$-variate probability mass functions.
Then: \\
 (i) $P \gtp Q$ if $P(\i) Q(\j) \leq P(\i \vee  \j) Q(\i \wedge \j)$.
 If $P$ and $Q$ are univariate, then this definition is equivalent to
 the MLR ordering  $P\gr Q$ defined above.\\
(ii)  A multivariate
distribution $P$ is said to be multivariate TP2 (MTP2) if $P \gtp P$  holds,
i.e.,  $P(\i) P(\j) \leq P(\i\vee  \j) P(\i \wedge \j)$.\\
(iii) 
If
 $\i,\j\in \{1,\ldots,X\}$ are scalar indices,
Statement (ii)  is equivalent to saying that an $X\times X$ matrix $P$ is  TP2  if all 
second order minors are non-negative.
 Equivalently, $P_{i+1,\cdot} \gr P_{i,\cdot}$, where $P_{i,\cdot}$ denotes the $i$th
 row of matrix $P$.
\end{definition}

To prove the existence of a threshold switching curve,
we will show that
 $Q(\pi,u)$  in   (\ref{eq:dp_alg}) is a submodular function on chains
 $[\I,\glX]$ and $[\I,\glone]$.

\begin{definition}[Submodular function \cite{Top98}] \label{def:supermod} Suppose $i = 1$ or $X$. Then
 $f:\l(e_i,\bp)\times \u \rightarrow \reals$  is  submodular (antitone differences)
if 
$f(\pi,u) - f(\pi,\bar{u}) \leq f(\tilde{\pi},u)-f(\tilde{\pi},\bar{u})$, for
$\bar{u} \leq u$, $\pi \gl \tpi$.
\end{definition}


The following result says that for a submodular function  $Q(\pi,u)$, 
$\mu^*(\pi)=\argmin_u Q(\pi,u)$  is increasing in its argument $\pi$. This implies $\mu^*(\pi)$ is MLR increasing  on 
the line segments $\l(e_{1},\bp)$ and $\l(e_{X},\bp)$, which in turn will be used to prove the existence
of as threshold decision curve.

\begin{theorem}[\cite{Top98}] \label{res:monotone}
\label{res:supermod} Suppose $i=1$ or $X$. If $f:\l(e_i,\bp)\times \u \rightarrow \reals$ is submodular, then
there exists a $\mu^*(\pi) = \argmin_{u\in \u} f(\pi,u)$, that  is increasing on $[\l(e_i,\bp),\gl]$,
i.e., $\tpi \gl {\pi} \implies \mu^*(\pi) \leq \mu^*(\tpi)$.
\end{theorem}


\subsection{Meta-Theorem: Lattice Programming on Simplex} 
\label{sec:applp}
We start with the following general assumptions and meta-theorem. The meta-theorem is a major step
in all our structural results.


\begin{itemize}
\item[(A1)] $C(\pi,1)$ and $C(\pi,2)$ are first order stochastic decreasing in $\pi$.
\item[(A2)] $B_{xy}$ is TP2.
\item[(A3)] $P$ is TP2.
\item[(S)] $C(\pi,2) - C(\pi,1)$ is $ [\l(e_X,\bp),\glX]$ decreasing and $[\l(e_1,\bp),\glone]$ decreasing.\end{itemize}

\begin{theorem} \label{thm:key} Consider the generic dynamic programming equation
(\ref{eq:dp_alg}) and the above assumptions. Then the following properties hold.
\begin{enumerate}
\item 
$\pi_1 \gr \pi_2$ implies 
 $\Tp(\pi_1,y)\gr \Tp(\pi_2,y)$ if (A3) holds.
 Under (A2) and (A3), 
 $\sigp(\pi_1,y) \gs \sigp(\pi_2,y)$.
\item 
For $y,\bar{y} \in \Y$,  $y > \bar{y}$ implies $\Tp(\pi_1,y)\gr \Tp(\pi_1,\bar{y})$ iff 
(A2) holds.

\item 
Assumptions  (A1-Ex1) in Sec.\ref{subsec:main}, 
(AS-Ex1)(i) and (ii) in Sec.\ref{subsec:main},
(A1-Ex2) in  Sec.\ref{sec:qtrans},
 (A1-Ex3) in  Sec.\ref{sec:risk},  and (A1-Ex5)  in Sec. \ref{sec:cso}
 are  sufficient conditions for (A1).

\item Under (A1), (A2), (A3), $Q(\pi,u)$ is MLR decreasing wrt $\glone$ and $\glX$.

\item Assumptions 
(S-Ex1) in Sec.\ref{subsec:main}, (AS-Ex1)(i) and (iii) in Sec.\ref{subsec:main},
 (S-Ex2) in Sec.\ref{sec:qtrans},  (S-Ex3) in Sec.\ref{sec:risk}
and (S-Ex5) 
in Sec.\ref{sec:cso}
 are  sufficient conditions for (S).

\item Under (A1), (A2), (A3), (S), $Q(\pi,u)$ is submodular on $[\I,\glX]$ and $[\I,\glone]$.
Thus by Theorem \ref{res:supermod}, the optimal policy $\mu^*(\pi)$ is MLR increasing on lines
$\l(e_X,\bp)$ and $\l(e_1,\bp)$.
\end{enumerate}
\end{theorem}
Part 1 and Part 2 use elementary properties of positive matrices
and are proved in \cite[Lemma 4.1 and 4.2]{Rie91}.

{\bf Proof of Part 3 for $C(\pi,1)$:} 
To give sufficient conditions for  $C(\pi,1)$ to $\gr$ decrease wrt $\pi \in \I$,
we start with the following convenient parametrization of the family of
 belief states
that first-order stochastic dominate another belief state, see \cite{MS02} for proof.

\begin{lemma} \label{lem:pie}
(i) For any $\pi,\tpi \in \I$, all belief states
  $\tpi \ls \pi$ are of the form
\beq \label{eq:pie}
\tpi = \pie \ole \pi + \epsilon_1(e_1-e_2) + \epsilon_2(e_2-e_3) + \cdots + \epsilon_{X-1} (e_{X-1} - e_X).
\eeq
where the variables $\epsilon_j$ satisfy
$ 0 \leq \epsilon_j \leq \min\{1-\pi(j),\pi(j+1)\}$,  $j=1,\ldots,X-1$.
Moreover, the $\epsilon$-parametrized belief state $\pie$ is stochastically decreasing in the elements $\epsilon_j$. That is, for any
$j=1,\ldots,X-1$,
$\epsilon_j \leq \bar{\epsilon}_j$  implies that $\pieone \gs \pietwo$. \qed
\end{lemma}

\noindent{\em Remark}:  The above constraints on $\epsilon_j$ ensure that $\pie$ is a valid belief state.
 
 In light of the above lemma, it suffices to prove that $C(\pie,1)$ is increasing
 in $\epsilon_i$, $i=1,2,\ldots,X$. We introduce the following lemma.  The proof follows straightforwardly using
 $\partial F(\pie)/ \partial \epsilon_i \geq 0$ and is omitted.
 \begin{lemma} \label{lem:qpie}
 Suppose $F(\pie) = \phi^\p \pie - \alpha (h^\p \pie)^2$, where $\phi,h \in \reals^X$. 
 Then if $h_i>0$, a sufficient condition for $F(\pie)$ to be $\gs$ increasing wrt $\epsilon$ is
 \beq  \label{eq:suffpie}  \phi_i - \phi_{i+1} \geq 2 \alpha h^\p \pi  (h_i - h_{i+1}) \qquad \forall \pi \in \I\eeq
 If  $h_i\geq 0$ is either
 monotone increasing or decreasing in $i$, then a sufficient condition for (\ref{eq:suffpie}) is
 \beq   \phi_i - \phi_{i+1} \geq 2 \alpha h_1  (h_i - h_{i+1}) .   \label{eq:suffmon}\eeq  \qed
 \end{lemma}

\begin{itemize}
\item Theorem \ref{thm:1}: Set $\phi=2 \alpha e_1$, $h  = e_1$ in  (\ref{eq:suffmon}).
This yields $2 \alpha \geq 0$ and $2 \alpha \geq 2 \alpha$ which always hold. So $C(\pi,1)$ is
$\gr$ decreasing in $\pi$ for any non-negative $\alpha$.
 
 \item Theorem \ref{thm:modified}: Set $\phi = \alpha e_1 - \alpha\f$ in (\ref{eq:suffmon}).
 This yields $f_2 \geq 1$ and $f_{i+1} \geq f_i$. Clearly (AS-Ex1)(i) and (ii) are sufficient
 conditions for this. In particular, $P$ TP2 and (AS-Ex1)(ii) implies $f_{i+1} \geq f_i$.

\item Theorem \ref{thm:qt}: 
Note that the variance
constraint $\alpha(e_3^\p \pi - (e_3^\p \pi)^2) = \alpha((\pi_1+\pi_2) - (\pi_1+\pi_2)^2)$.
Accordingly,
set  $\phi = 
2\alpha(e_1+e_2)$, $h = e_1+e_2$ in  (\ref{eq:suffmon}). This yields $2 \alpha \geq 0$ and $2\alpha \geq
2 \alpha$ which always hold for $\alpha \geq 0$.

\item In Theorem \ref{thm:risk}, $C(\pi,1)=0$, see (\ref{eq:riskcosts}), so there is nothing to prove.
\item In Theorem \ref{thm:socialopt}, to show that $C(\pi,1)$ 
 is $\gr$ decreasing in $\pi$, it suffices to show that for each $a \in \A$,
that $\ca^\p \pi $  
is $\gr$ decreasing in $\pi$.
So in (\ref{eq:suffmon}) choose $\phi= c_a$,  $h = 0$.  This yields (A1-Ex5).
\end{itemize}

{\bf Proof of Part 3 for $C(\pi,2)$:} 
Since  $C(\pi,2)$ is linear in $\pi$, to show that $C(\pi,2) \gr$ decreases wrt $\pi$,
from Result \ref{res1}(ii) (Appendix \ref{sec:mlrdef}), it suffices that 
\beq  C(e_i,2) \geq
C(e_{i+1},2), \quad i = 1,\ldots,X-1 . \label{eq:check} \eeq
Theorem \ref{thm:1}: This yields (A1-Ex1).\\
 Theorem \ref{thm:modified}: (\ref{eq:check}) is equivalent to  $f_2 \geq \rho \f^\p P^\p e_2 - \frac{d}{\alpha+\beta}$
 and (AS-Ex1)(ii). Note (AS-Ex1)(i) is sufficient for $f_2 \geq \rho \f^\p P^\p e_2 - \frac{d}{\alpha+\beta}$.
\\
Theorem \ref{thm:qt}: (\ref{eq:check}) is equivalent to $-\rho(\alpha+\beta)+d_1 + (\alpha+ \beta) \geq -\rho(\alpha+\beta)+d_2 + (\alpha+ \beta) \geq -\rho(\alpha+\beta) P_{32}$ which always holds for $d_1\geq d_2$.

{\bf Proof of Part 4:}
The proof is by mathematical induction on the value iteration recursion (\ref{eq:vi}).
Clearly  $V_0(\pi)= -\beta(1-e_1^\p \pi)$ in (\ref{eq:vi}) is a  MLR decreasing function of $\pi$.
Consider (\ref{eq:vi}) at any stage $k$. Assume
that $V_k(\pi)$ is MLR decreasing in $\pi$. From Part 1 and 2, it follows that under (A2), (A3), the term 
$\sum_{y \in \Y}  V_k\left( T(\pi,y) \right) \sigma(\pi,y)$ is  MLR decreasing in
$\pi$. From Part 3, under (A1), $C(\pi,u)$ is MLR decreasing.
Since the sum of decreasing functions is decreasing, the result follows.

{\bf  Proof of Part 5}: Here we show that general assumption (S) at the beginning of  Appendix~\ref{sec:applp} takes on the forms (S-Ex1) to (S-Ex5) for the various examples.
We start with the following characterization of belief states on lines $\l(e_X,\bp)$ and $\l(e_1,\bp)$ and submodularity on these lines. 
\begin{lemma} \label{lem:subquad}
(i) $\pi_1 \glX \pi_2$  is equivalent to  $\pi_i = (1-\epsilon_i) \bp + \epsilon_i e_X$ and $\epsilon_1 \geq \epsilon_2$ for $\bp \in \H_X$ where $\H_X$ is defined in (\ref{eq:hi}). So submodularity on $\l(e_X,\bp)$
is equivalent to showing  \beq \pi^\epsilon = (1-\epsilon) \bp + \epsilon e_X  \implies
C(\pi^\epsilon,2)-C(\pi^\epsilon,1) \text{ decreasing wrt $\epsilon$}  \label{eq:sx}\eeq
(ii)
$\pi_1 \glone \pi_2$  is equivalent to  $\pi_i = (1-\epsilon_i) \bp + \epsilon_i e_X$ and $\epsilon_1 \leq \epsilon_2$ for $\bp \in\H_1$ where $\H_1$ is defined in (\ref{eq:hi}).
So submodularity on  $\l(e_1,\bp)$ is  equivalent to showing
\beq \pi^\epsilon = (1-\epsilon) \bp + \epsilon e_1 \implies
C(\pi^\epsilon,2)-C(\pi^\epsilon,1) \text{ increasing wrt $\epsilon$. } \label{eq:s1} 
\eeq \end{lemma}
\qed

The proof of Lemma \ref{lem:subquad}  follows
from Lemma \ref{lem:convex} and is omitted.

Suppose  $C(\pi,2)-C(\pi,1)$ is of the form $\phi^\p \pi+ \alpha (h^\p \pi)^2$. Then  from (\ref{eq:sx}), (\ref{eq:s1}),  sufficient conditions
for submodularity on  $\l(e_X,\bp)$ and  $\l(e_1,\bp)$ are for $\bp \in \H_X$ and $\H_1$, respectively,
\beq \phi_X - \phi^\p \bp + 2 \alpha h^\p \pi^\epsilon (h_X - h^\p \bp) \leq 0,\quad 
\phi_1 - \phi^\p \bp + 2 \alpha h^\p \pi^\epsilon (h_1 - h^\p \bp) \geq 0  \label{eq:subg}\eeq
In particular if $h_i \geq 0$ and monotone increasing or decreasing in $i$, then (\ref{eq:subg}) is equivalent to
\beq
 \phi_X - \phi^\p \bp + 2 \alpha h_X  (h_X - h^\p \bp) \leq 0,\quad 
\phi_1 - \phi^\p \bp + 2 \alpha h_X (h_1 - h^\p \bp) \geq 0  \label{eq:subm}\eeq
where  $\bp \in \H_X$ and $\bp \in \H_1$, respectively.
\begin{itemize}
\item  Theorem \ref{thm:1}: Set    $\phi_i =  ( d- \rho(\alpha+\beta))P^\p e_i + (\beta-\alpha) e_1$, $h = e_1$  in (\ref{eq:subm}).
The first inequality is equivalent to:
(i) $ (d-\rho(\alpha+\beta)) (P_{X1}-P_{i1}) \leq 0$ for $i \geq 2$
 and (ii) $(d - \rho(\alpha+\beta))(1-P_{X1}) \geq \alpha-\beta $.
Note that (i)  holds if $d \geq \rho(\alpha+\beta)$.

The second inequality in (\ref{eq:subm}) is equivalent to
$(d-\rho(\alpha+\beta)) (1 - P_{i1}) \geq \alpha - \beta$. 
Since $P$ is TP2, from footnote \ref{foot} in Sec.\ref{sec:discussion} it follows that 
(S-Ex1) is sufficient for these inequalities to hold.

\item Theorem \ref{thm:modified}: Set $\phi_1 = d-\alpha$, $\phi_i = -\beta f_i + \rho(\alpha+\beta) \f^\p P^\p e_i$, $i=2,\ldots,X$ in (\ref{eq:subm}). The first inequality yields
(i)  $ \rho(\alpha+\beta) \f^\p P^\p e_X \leq  d - \alpha + \beta f_X$ and 
(ii)  $ \rho(\alpha+\beta) \f^\p P^\p (e_X-e_i) \leq  \beta(f_X - f_i)$, $i=2,\ldots,X-1$.
The second inequality in (\ref{eq:subm}) yields
$q_i \geq \frac{\rho}{\beta}(\alpha+\beta) \f^\p P^\p e_i + \frac{\alpha - d}{\beta}$.
These inequalities imply (AS-Ex1)(i) and (iii).

\item  Theorem \ref{thm:qt}: Recall that the variance
constraint $\alpha(e_3^\p \pi - (e_3^\p \pi)^2) = \alpha((\pi_1+\pi_2) - (\pi_1+\pi_2)^2)$.
Set $\phi_1 = d_1 + \beta - \alpha - \rho(\alpha+\beta)$, $\phi_2 = d_2 + \beta - \alpha - \rho(\alpha+\beta)$, $f_3 = - \rho(\alpha+\beta) P_{32}$ in (\ref{eq:subm}). The first inequality is equivalent to
$\rho(\alpha+\beta)(1-P_{32}) \leq \beta+ d_i - \alpha$, $i=1,2$. The second inequality is equivalent
to $d_1 \geq 0$ and $d_1+\beta-\alpha \geq \rho(\alpha+\beta)(1-P_{32})$. A sufficient condition
for these is (S-Ex2).

\item In Theorem  \ref{thm:risk}, since $C(\pi,1)=0$, $C(\pi,2)$ decreasing on lines  $\l(e_X,\bp)$ and  $\l(e_1,\bp)$ implies submodularity.
This is implied by (A1-Ex3).
\item In Theorem \ref{thm:socialopt}, $f = \sum_y B_y c_y - c_a/(1-\rho) $, $h = 0$ in (\ref{eq:subm}) yields (S-Ex5).
\end{itemize}

{\bf Proof of Part 6}:
From Definition \ref{def:supermod},
to show that $Q(\pi,u)$ is submodular, requires showing
that $Q(\pi,1) - Q(\pi,2)$ is $\gl$ on lines $\l(e_X,\bp)$ for $i=1$ and $X$. 
Part 4 shows by induction that for each $k$,  $V_k(\pi)$ is $\gr$ decreasing on $\I$
 if (A1), (A2), (A3) hold. This implies that $V_k(\pi)$ is $\gl$ decreasing on lines $\l(e_X,\bp)$ and  $\l(e_1,\bp)$.
 So to prove $Q_k(\pi,u)$ in (\ref{eq:dp_alg}) is 
 submodular, we  only need to show that 
$C(\pi,1) - C(\pi,2)$ is  $\gl$ decreasing on $\l(e_i,\bp)$, $i=1,X$. But this is implied by  (S) as shown in Part 5 above.
Since submodularity is closed under pointwise limits \cite[Lemma 2.6.1 and Corollary 2.6.1]{Top98}, it follows that
$Q(\pi,u)$ is submodular on $\gl$, $i=1,X$
Having established $Q(\pi,u)$ is submodular on $\gl$, $i=1,X$,
Theorem \ref{res:monotone} in Appendix \ref{sec:mlrdef}  implies
that the optimal policy $\mu^*(\pi)$ is $\glone$ and $\glX$ increasing on lines.

\subsection{Proof of Theorem \ref{thm:1}}  \label{sec:appth1} With the above key theorem, we can now prove Theorem \ref{thm:1}.
The statement of  Theorem \ref{thm:1} that $\mu^*(\pi)$ is $\glone$ and $\glX$ increasing is proved  above. \\
{\em Statement (i): (a) Characterization of Switching Curve $\Gamma$}.
For each $\bp \in \H_X$ (\ref{eq:lines}), construct the  line
segment  $\l(e_X,\bp)$ connecting $\H_X$ to $e_X$ as in (\ref{eq:lines}).
By Lemma~\ref{lem:convex} in Appendix \ref{sec:mlrdef}, on the line
segment connecting $(1-\epsilon) \underline{\pi} + \epsilon e_{\bS}$, all belief states are MLR orderable. 
 Since  $\mu^*(\pi)$ is monotone increasing
for $\pi \in \l(e_X,\bp)$,
moving along this line segment towards $e_{\bS}$, 
pick the largest $\epsilon$ for which  $\mu^*(\pi) = 1$.  (Since $\mu^*(e_X) = 1$, such an $\epsilon$ always exists). The belief state corresponding
to this $\epsilon$ is the threshold belief state. Denote it by $\thr(\bp) = \pi^{\epsilon^*,\bp} \in \l(e_X,\bp)\text{ where
} \epsilon^* = \sup\{\epsilon \in [0,1] :
 \mu^*(\pi^{\epsilon,\bp})=1\} $.\\
 The above construction
 implies that on $\l(e_X,\bp)$,
there is a unique threshold point $\thr(\bp)$.
Note  that the entire simplex can be covered by considering
all pairs of lines $\l(e_X,\bp)$, for $\bp \in \H_X$,
i.e.,
$ \I = \cup_{\bp \in \H}  \l(e_X,\bp)  $.
Combining all  points $\thr(\bp)$ for all pairs of lines  $\l(e_X,\bp)$,
 $\bp \in \H_X$,
yields a unique  threshold curve in $\I$ denoted
$
\thr =  \cup_{\bp \in \H} \thr(\bp)$.

{\em Statement (i): (b) Connectedness of regions $\R_1$ and $\R_2$}.\\
{\em Connectedness of $\R_1$}: 
Since  $e_1 \in \R_1$, 
call $\R_{1a}$ the subset of $\R_1$ that contains  $e_1$.
Suppose $\R_{1b}$ was a subset of  $\R_1$ that was disconnected from $R_{1a}$.
Recall that 
every point in $\I$ lies on a line
segment $\l(e_1,\bp)$ for some $\bp$.
Then such a line segment starting from $e_1 \in \R_{1a}$ would leave
the region $\R_{1a}$, pass through a region where action 2 was optimal, and then intersect the region $\R_{1b}$ where action 1 is optimal.
But this violates the requirement that $\mu(\pi)$ is increasing on $\l(e_1,\bp)$. 
Hence $\R_{1a}$ and $\R_{1b}$ have to be connected. 
(Note in the special case $\alpha = 0$, then since $\R_1$ is convex (by Theorem \ref{cor:ph}), and so is
obviously connected).

{\em Connectedness of $\R_2$}: Assume $e_X \in \R_2$, otherwise $\R_2 = \emptyset $ and there is nothing
to prove. Call the region $\R_2$ that contains $e_X$ as $\R_{2a}$.
Suppose $\R_{2b}\subset \R_2$ is disconnected from $R_{2a}$.
Since 
every point in $\I$ can be joined by the line
segment $\l(e_X,\bp)$ to $e_X$. 
Then such a line segment starting from $e_X \in \R_{2a}$ would leave
the region $\R_{2a}$, pass through a region where action 1 was optimal, and then intersect the region $\R_{2b}$ (where action 2 is optimal).
But this violates the requirement that $\mu(\pi)$ is increasing on $\l(e_X,\bp)$. 
Hence $\R_{2a}$ and $\R_{2b}$ have to be connected.

{\em Statement (ii)}: Suppose $e_i \in \R_1$. Then considering lines $\l(e_i,\bp)$
and ordering $\gl$, it follows that $e_{i-1} \in \R_1$. 
Similarly if $e_i \in \R_2$, then considering lines $\l(e_{i+1},\bp)$ and ordering $\glp$, 
it follows that $e_{i+1} \in \R_2$.

{\em Statement (iii)} follows trivially since for $X=2$, $\I$ is a one dimensional simplex.

 \subsection{Proof of Theorem \ref{cor:ph}} \label{app:thm2}

We first prove in the following lemma that $V(\pi)$ is concave in $\pi$.

\begin{lemma} $V(\pi)$ in (\ref{eq:dp_alg}) is concave in $\pi \in \I$. \label{lem:c1}
\end{lemma}
{\em Proof of Lemma \ref{lem:c1}}:
Our proof constructs an outer approximation $V_k^{(n)}(\pi)$ (defined below) to $V_k(\pi)$ and comprises of two steps:
Step 1: $V^{(n)}_k(\pi)$ is concave; Step 2: $V^{(n)}_k(\pi) \rightarrow V_k(\pi)$ uniformly as $n\rightarrow \infty$.
This establishes that $V_k(\pi)$ is concave,  and therefore $V(\pi)$ is concave.

Consider $n$ arbitrary but distinct belief states $\pi^1,\ldots,\pi^n \in \I$.
 Let 
$\gamma^i(\pi)$ denote the gradient vector of 
$Q(\pi,1)$ in (\ref{eq:vi}) at $\pi = \pi^i$, $ i=1,\ldots,n$.  That is,  
$  Q(\pi^i,1) = \pi^i \gamma^i(\pi^i) $.
Now construct a piecewise linear function out of these gradient vectors as
$Q^{(n)}(\pi,1) = \min_i {\gamma^i}^\p \pi$.
It is easily seen from (\ref{eq:costdef}) that $Q(\pi,1)$ is concave.
Therefore
 $Q^{(n)}(\pi,1)$ is piecewise linear and concave in $\pi$ since a piecewise linear function composed
of tangents to a concave function is concave.

Construct the following auxiliary value function $V^{(n)}(\pi)$ via value iteration similar
to (\ref{eq:vi}):
\begin{align}\nonumber
V^{(n)}_{k+1}(\pi) &= \min\{ Q^{(n)}(\pi,1), Q^{(n)}_{k+1}(\pi,2) \}, \quad
\mu^*_{k+1}(\pi)= \argmin\{ Q^{(n)}(\pi,1), Q^{(n)}_{k+1}(\pi,2) \}  \\
\text{ where } Q_{k+1}^{(n)}(\pi,2) &=  C(\pi,2) 
+ \discount \sum_{y \in \Y}  V_k^{(n)}\left( T(\pi,y) \right) \sigma(\pi,y),
\;
Q^{(n)}(\pi,1) = \min_i {\gamma^i}^\p \pi \quad
\pi \in \I.  \label{eq:vin}
\end{align}

{\bf Step 1: Proof of concavity of $V_k^{(n)}(\pi)$}:
We prove this by induction on the
value iteration algorithm (\ref{eq:vin}) for $k=1,2,\ldots$ and fixed $n$.
Start with arbitrary concave $V_0^{(n)}(\pi)$. As mentioned below (\ref{eq:vi}),
the VI algorithm converges  for any choice of initialization.

Since both $C(\pi,2)$ (see (\ref{eq:costdef})) and $Q^{(n)}(\pi,1)$ are piecewise linear  in $\pi$, it is easily seen from (\ref{eq:vin}) that at
each iteration  $k$,
 $V_k^{(n)}(\pi)$ is  positively homogeneous, i.e., 
$V_k^{(n)}(\lambda \pi) = \lambda V_k^{(n)}(\pi)$ 
for any $\lambda \geq 0$. As a result (\ref{eq:vin}) yields,
$Q^{(n)}_{k+1}(\pi,2) = C(\pi,2) + \discount \sum_{y \in\Y} V^{(n)}_k(B_y \pi) $.
Now use mathematical induction. Assume $V_k^{(n)}(\pi)$ is concave.
Since  $B_y \pi$ is concave, and the composition of concave
functions is concave, $V^{(n)}_k(B_y \pi)$ is concave. Since $C(\pi,2)$ is piecewise linear
and concave, and the sum of concave functions
is concave, it follows that $Q_{k+1}^{(n)}(\pi,2)$ is concave. Finally since minimization
preserves concavity, it follows that  $V^{(n)}_{k+1}(\pi) = \min\{ Q^{(n)}(\pi,1), Q^{(n)}_{k+1}(\pi,2) \}$ is concave. This completes the inductive proof.

{\bf Step 2: Concavity of $V_k(\pi)$}:
Next, we show that $V_{k}^{(n)}(\pi) \rightarrow V_{k}(\pi)$ uniformly in $\I$ implying
that $V_k(\pi)$ is concave.
Since $Q(\pi,1)$ is concave,
it follows that $Q^{(n)}(\pi,1) > Q(\pi,1)$ and also $Q^{(n)}(\pi,1)$ is a monotone sequence of decreasing functions
in $n$. (Intuitively, the piecewise linear function composed of tangents always upper bound a concave function and they become tighter as more piecewise linear segments are considered).
From (\ref{eq:dp_alg}) 
and (\ref{eq:vin}) this
implies that  $V_k^{(n)}(\pi) \geq V_k(\pi)$ for all $k$ and also $V_k^{(n+1)}(\pi)<V_k^{(n)}(\pi) $.
Finally, since (i) $V_k^{(n)}(\pi)$ converges to $V_k(\pi)$ pointwise,
(ii) $\I$ is compact,  (iii) $V_k^{(n)}(\pi)$ is continuous
and $V_k^{(n)}(\pi)$ is monotone
decreasing sequence in $n$, it follows from
 \cite[Theorem 7.13]{Rud76} that $V_k^{(n)}(\pi) \rightarrow V_k(\pi)$ 
uniformly on $\I$. Therefore, $V_k(\pi)$ is concave for $k=1,2,\ldots$. Finally as
discussed below (\ref{eq:vi}), $V_k(\pi)$ converges
uniformly to $V(\pi)$; so $V(\pi)$ is concave. Thus Lemma \ref{lem:c1} is proved.  \qed

The rest of the proof of Theorem \ref{cor:ph} follows from arguments in \cite{Lov87a}. We repeat this for completeness here.
 Our goal is to show that 
$\R_{1}$ is convex.
Pick any two belief states $\pi_1,\pi_2 \in \R_{1}$. To demonstrate convexity of $\R_{1}$,
we need to show for any $\lambda \in [0,1]$,  $\lambda \pi_1 + (1-\lambda) \pi_2 \in \R_{1}$.
Since $V(\pi)$ is concave and $\alpha = 0$,
\begin{align}
V(\lambda \pi_1 + (1-\lambda) \pi_2) &\geq \lambda V(\pi_1) + (1-\lambda) V(\pi_2) \nonumber\\
&= \lambda Q(\pi_1,1) + (1-\lambda) Q(\pi_2,1)  \text{ (since $\pi_1,\pi_2 \in \R_{1}$) } \nonumber\\
&= Q(\lambda \pi_1 + (1-\lambda) \pi_2,1 ) \text{ (since $Q_{1}(\pi,1)$ is linear in $\pi$) }\nonumber \\
& \geq V(\lambda \pi_1 + (1-\lambda) \pi_2) \text{ (since $V(\pi)$ is the optimal value function) } \label{eq:convexregion}
\end{align}
Thus all the inequalities above are equalities, and $\lambda \pi_1 + (1-\lambda) \pi_2 \in 
\R_{1}$.

\subsection{Proof of Theorem \ref{thm:dep}}  \label{app:dep}
Given any $\pi_1,\pi_2 \in \l(e_X,\bp)$ with $\pi_2\glX \pi_1$, we need to prove:
$\mu_\theta(\pi_1) \leq \mu_\theta(\pi_2)$ iff  $\theta(X-2) \geq 1 $, $\theta(i) \leq \theta(X-2)$ for $i< X-2$.
But from the structure of (\ref{eq:linear}), obviously $\mu_\theta(\pi_1) \leq \mu_\theta(\pi_2)$ 
is equivalent to 
$ \begin{bmatrix} 0 & 1 & \theta^\p\end{bmatrix}^\p \begin{bmatrix}\pi_1 \\ -1 \end{bmatrix} 
\leq \begin{bmatrix} 0 & 1 & \theta^\p\end{bmatrix}^\p \begin{bmatrix}\pi_2 \\ -1 \end{bmatrix}$,
or equivalently, 
$\begin{bmatrix} 0 & 1 & \theta(1) & \cdots & \theta(X-2)\end{bmatrix} (\pi_1 - \pi_2 ) \leq 0$.

Now from Lemma \ref{lem:convex}(iii),  $\pi_2 \glX\pi_1 $ implies that
 $\pi_1 = \epsilon_1 e_{X} + (1-\epsilon_1) \bp$,
$\pi_2 = \epsilon_2 e_{X} + (1-\epsilon_2) \bp$ and $\epsilon_1 \leq \epsilon_2$.
Substituting these into the above expression, we need to prove
$$
( \epsilon_1 -  \epsilon_2)\bigl(\theta(X-2) -  \begin{bmatrix} 0 & 1 & \theta(1) & \cdots & \theta(X-2)\end{bmatrix}^\p \bp \bigr) \leq 0, \quad \forall \bp \in \H_X$$  iff  $\theta(X-2) \geq 1 $, $\theta(i) \leq \theta(X-2)$, $i < X-2$. This is obviously true.

A similar proof shows that  on lines $\l(e_1,\bp)$ the linear
threshold policy satisfies
$\mu_\theta(\pi_1) \leq \mu_\theta(\pi_2)$ iff  $\theta(i) \geq 0$ for $i < X-2$.

\subsection{Proof of Theorem \ref{thm:risk}}  \label{app:risk}
The only difference compared to the meta-theorem is the update of the belief state (\ref{eq:riskpi}) which
now includes the term $\text{diag}(R_u)$. The elements of $R_u$ are non-negative and 
functionally independent
of the observation $y$. Therefore the three main requirements
that $T(\pi,y)$ is MLR increasing in $\pi$, $T(\pi,y)$  is MLR increasing in $Y$,
and $\sigma(\pi,:)$ is $\gr$ increasing in $\pi$ continue to hold. Then the rest of the proof is identical
to Theorem \ref{thm:1}.

\subsection{Proof of Theorem \ref{thm:stopsocial}} \label{app:stopsocial}
The proof  is more complex than that of Theorem \ref{cor:ph} since now $V(\pi)$ in 
is not necessarily
concave over $\I$, since $\Tp(\cdot)$ and $\sigp(\cdot)$ are functions
of $\Bs_a$ (\ref{eq:aprob}) which itself is an explicit (and in general non-concave)
function of $\pi$. 

Define the matrix $\Bs = (\Bs(i,a), i=\{1,2\}, a \in \{1,2\})$, where $\Bs(i,a) = P(a|x=e_i,\pi)$.
It can be verified from (\ref{eq:aprob}) that  there are only 3 possible values for $\Bs$, namely,
\beq  \label{eq:bsvalues}\Bs=\begin{bmatrix} 0 & 1 \\ 0 & 1 \end{bmatrix},
\pi \in \mathcal{P}_1, \quad
\Bs =  B, \pi \in
 \mathcal{P}_2 \cup \mathcal{P}_3, \quad \Bs = \begin{bmatrix} 1 & 0 \\ 1 & 0 \end{bmatrix},
\pi \in \mathcal{P}_4 \eeq
Thus based on the dynamic
programming equation (\ref{eq:dpsocialstop}),
the value iteration algorithm reads
\begin{multline}
V_{n+1}(\pi) = \min\{ C(\pi,2) + \rho V_n(\pi) I(\pi \in \mathcal{P}_1) + \rho \sum_{a\in \A} V_n(T(\pi,a)) \sigma(\pi,a) [I(\pi \in \mathcal{P}_2) 
+ I(\pi \in \mathcal{P}_3) ] \\   +   \rho V_n(\pi) I(\pi \in \mathcal{P}_4)  , 0 \}  \label{eq:visocialstop}\end{multline}
Assuming $V_n(\pi)$ is  MLR decreasing on  $\mathcal{P}_1 \cup \mathcal{P}_4$ straightforwardly  implies $V_{n+1}(\pi)$ is 
MLR  decreasing on  $\mathcal{P}_1 \cup \mathcal{P}_4$ since $C(\pi,2)$ is MLR decreasing.  This proves claim (i).

We now prove inductively that $V_n(\pi)$ is piecewise linear concave on each interval  $\mathcal{P}_l$, $l=1,\ldots,4$.
The proof of concavity on  $\mathcal{P}_1$ and $\mathcal{P}_4$ follows straightforwardly since $C(\pi,2)$ is piecewise linear
and concave.
The proof for intervals $\mathcal{P}_2$
and $ \mathcal{P}_3$ is more delicate.

We need the following property of the social learning Bayesian filter. 
We use the following slight abuse in notation. Define the two dimensional vector 
$\eta_i = (1 - \eta_i, \quad \eta_i)^\p$.

\begin{lemma}  \label{lem:social}Consider the  social learning Bayesian filter  (\ref{eq:piupdate}). 
Then
$T(\eta_1,1) = \eta_2$,  $T(\eta_3,2) = \eta_2$.  Furthermore if $B$ is symmetric TP2, then 
$T(\eta_2,2) = \eta_1$,  $ T(\eta_2,1) = \eta_3$ and $\eta_3 \leq \eta_2 \leq \eta_1$.
So \\ (i)  $\pi \in \mathcal{P}_2$  implies   $T(\pi,2) \in \mathcal{P}_1$ and 
   $T(\pi,1)  \in \mathcal{P}_3$. \\
(ii)  $\pi \in \mathcal{P}_3$  implies   $T(\pi,2)  \in \mathcal{P}_2$ and
  $T(\pi,1)  \in \mathcal{P}_4$. \qed
\end{lemma}
The proof of Lemma \ref{lem:social} is as follows. Recall from (\ref{eq:bsvalues}) that
on intervals $\mathcal{P}_2$ and $\mathcal{P}_3$, $\Bs = B$.
Then it is straightforwardly verified from 
(\ref{eq:piupdate}) that $T(\eta_1,1) = T(\eta_3,2) = \eta_2$.
Next, using (\ref{eq:piupdate}) it follows that $B_{12}B_{11} = B_{22} B_{21}$ is a sufficient
condition for $T(\eta_2,2) = \eta_1$ and $T(\eta_2,1) = \eta_3$. 
Also,  applying Theorem \ref{thm:key}(2), $B$ TP2 implies $\eta_3 \leq \eta_2 \leq \eta_1$.
So $B$ symmetric TP2 is sufficient for the claims of the lemma to hold.
Statements (i) and (ii) then follow straightforwardly. In particular, from Theorem \ref{thm:key}(1),
$\eta_1 \gr \pi \gr \eta_2$ implies $T(\eta_1,1) = \eta_2 \gr T(\pi,1) \gr T(\eta_2,1) = \eta_3$,
which implies Statement (i) of the Lemma. Statement (ii) follows similarly.

Returning to the proof of Theorem \ref{thm:stopsocial}.
Assume now that $V_n(\pi)$ is  piecewise linear and concave on each interval $\mathcal{P}_l$, $l=1,\ldots,4$.
That is, for two dimensional vectors $\gamma_{m_l}$ in the set $\Gamma_l$,
$$
V_n(\pi) = \sum_l \min_{m_l \in \Gamma_l}  \gamma_{m_l}^\p \pi\, I(\pi \in \mathcal{P}_l) $$
Consider $\pi \in \mathcal{P}_2$. 
From (\ref{eq:bsvalues}), since 
 $\Bs_a = B_a$, $a = 1,2$,
Lemma \ref{lem:social} (i) together with the value iteration algorithm (\ref{eq:visocialstop}) yields
$$ V_{n+1}(\pi) =  \min\{ C(\pi,2) + \rho \left[\min_{m_3 \in \Gamma_3} \gamma_{m_3}^\p B_1 \pi + \min_{m_1 \in \Gamma_1}
\gamma_{m_1}^\p B_2 \pi\right], 0 \}. $$
Since each of the terms in the above equation  are piecewise linear and concave, it follows that 
 $V_{n+1}(\pi)$ is piecewise linear and concave on $\mathcal{P}_2$. A similar proof holds for $\mathcal{P}_3$ and this involves using Lemma \ref{lem:social}(ii).
As a result the stopping set on each interval $\mathcal{P}_l$ is a convex region, i.e., an interval. This proves claim (ii).

\subsection{Proof of Theorem \ref{thm:socialopt}} \label{app:socialopt}
Part (i)  follows directly from the proof of Theorems \ref{thm:key} and \ref{thm:1}. 
For Part (ii), define the convex polytopes $\mathcal{P}_a = \{\pi: c_a^\p \pi < c_{\bar{a}}^\p \pi, \quad \bar{a} \neq a \}$. Then on each convex
polytope $\mathcal{P}_a$, since $C(\pi,1) = c_a ^\p\pi$ (recall $\alpha = 0$), we can apply the argument of (\ref{eq:convexregion}) which
yields that $\R_1 \cap \mathcal{P}_a$ is a convex region. Thus $\R_1$ is the union of $A$ convex regions and is in general non-convex.
However, it is still  a connected set by part (i) of the theorem.

\subsection{Proof of  Theorem \ref{thm:compare2}}  \label{app:compare2}

A similar proof to Theorem \ref{cor:ph} then establishes $V(\pi)$ is concave on $\I$. 
We then use the Blackwell dominance condition (\ref{eq:bd}). In particular,
$$\Ts(\pi,\yi) =   \sum_{\yii \in \Y^{(2)}} \Tp(\pi,\yii) \frac{\sigp(\pi,\yii)}{\sigs(\pi,\yi)} P(\yi|\yii) 
\quad \text{ and } \sigs(\pi,\yi) = \sum_{\yii \in \Y^{(2)}} \sigp(\pi,\yii) P(\yi|\yii).$$
Therefore $\frac{\sigp(\pi,\yii)}{\sigs(\pi,\yi)} P(\yi|\yii) $ is a probability measure wrt $\yii$.
Since $V(\cdot)$ is concave, using Jensen's inequality it follows that
\begin{align*}
V(\Ts(\pi,\yi) ) & = V \left(\sum_{\yii \in \Y^{(2)}} \Tp(\pi,\yii) \frac{\sigp(\pi,\yii)}{\sigs(\pi,\yi)} P(\yi|\yii) \right)
\geq \sum_{\yii \in \Y^{(2)}}  V (\Tp(\pi,\yii)) \frac{\sigp(\pi,\yii)}{\sigs(\pi,\yi)} P(\yi|\yii)\\
\text{ implying }&  \sum_{\yi}  V(\Ts(\pi,\yi) ) \sigs(\pi,\yi) \geq
\sum_{\yii} V(\Tp(\pi,\yii)\sigp(\pi,\yii).
\end{align*}
Therefore for $\pi \in \Pi^s$, 
$$ C(\pi,2) + \sum_{\yii} V(\Tp(\pi,\yii)\sigp(\pi,\yii) \leq 
C(\pi,1) +  \sum_{\yi}  V(\Ts(\pi,\yi) ) \sigs(\pi,\yi)  $$
So for $\pi \in \Pi^s$, the optimal policy $\mu^*(\pi) = \arg\min_{u \in \U}Q(\pi,u) = 2$.
So $\bar{\mu}(\pi) = \mu^*(\pi)$ for $\pi \in \Pi^s$ and $\bar{\mu}(\pi)=1$ otherwise, implying that
$\bar{\mu}(\pi)$ is a lower  bound for $\mu^*(\pi)$.

\subsection{Proof of Theorem \ref{thm:tmove}} \label{app:tmove}
Identical  to the proof of meta Theorem \ref{thm:key} in Appendix \ref{sec:applp},
under the assumptions $c(e_i,u)$ decreasing in $i$, (A2), (A3), it follows that $V(\pi;\Pone)$ and
$V(\pi;\Ptwo)$ are  MLR decreasing
for $\pi \in \I$. 
We next introduce the following lemma.

\begin{lemma}\cite[Theorem 2.4]{KR80} \label{lem:mor}
$\Pone \succeq \Ptwo$ implies ${\Pone}^\p \pi \gr {\Ptwo}^\p \pi$ where $\succeq$ is defined
in (\ref{eq:mor}). \end{lemma}
The proof of the lemma is as follows: By definition
${\Pone}^\p \pi \gr {\Ptwo}^\p \pi$ is equivalent to
$$\sum_{i\in \X} \sum_{m\in\X} \left(\Pone_{ij} \Ptwo_{m,j+1} - \Ptwo_{ij}\Pone_{m,j+1}\right) \pi_i \pi_m \leq 0. $$
Thus clearly (\ref{eq:mor}) is a sufficient condition for ${\Pone}^\p \pi \gr {\Ptwo}^\p \pi$.

Returning to the proof of the theorem,  if (A2), (A3) hold, it follows from Lemma \ref{lem:mor} 
and meta Theorem \ref{thm:key} (Statements (1) and (2)) that for actions $u \in\{1,2\}$,
\beq T(\pi,y,u;\Pone) \gr T(\pi,y,u;\Ptwo), \quad  \sigp(\pi,\cdot,u;\Pone) \gs
\sigp(\pi,\cdot,u;\Ptwo). \label{eq:ssd}\eeq
The rest of the proof is by induction on the value iteration algorithm (\ref{eq:dp_algmove}).
Assume $V_k(\pi;\Pone) \leq V_k(\pi;\Ptwo)$ for $\pi \in \I$. Then from (\ref{eq:ssd}),
$$V_k(T(\pi,y,u;\Pone);\Pone) \leq V_k(T(\pi,y,u;\Pone);\Ptwo) \leq V_k(T(\pi,y,u;\Ptwo);\Ptwo) $$ Therefore, 
$$ \sum_y V_k(T(\pi,y,u;\Pone);\Pone) \sigp(\pi,y,u;\Pone) \leq
\sum_y V_k(T(\pi,y,u;\Ptwo);\Ptwo) \sigp(\pi,y,u;\Ptwo) .$$
Next since $C(\pi,u;\Pone) \leq C(\pi,u;\Ptwo)$, it follows that 
\begin{multline*}
C(\pi,u;\Pone)+\sum_y V_k(T(\pi,y,u;\Pone);\Pone) \sigp(\pi,y,u;\Pone)  \\ \leq
C(\pi,u;\Ptwo)+ \sum_y V_k(T(\pi,y,u;\Ptwo);\Ptwo) \sigp(\pi,y,u;\Ptwo) .\end{multline*}
Taking the minimum with respect to $u$ yields $V_{k+1}(\pi;\Pone) \leq V_{k+1}(\pi;\Ptwo)$.

To prove the second claim of the theorem;  first
recall from (\ref{eq:dp_initial}) that $\Vb(\pi;P)$ is the actual optimal expected  cost. 
Recall that the transformation
from $\Vb(\pi;P)$ to $V(\pi;P)$ was made to prove that the optimal policy is monotone.
It is readily verified that the quickest detection problem
 (\ref{eq:stylized}) satisfies all the assumptions of the theorem.
 So  $V(\pi;\Pone) \leq V(\pi;\Ptwo)$.
The actual optimal cost is $\Vb(\pi;P) = V(\pi;P) + (\alpha+\beta)(1 - e_1^\p \pi)$ (see (\ref{eq:stylized})).
Since $(\alpha+\beta)(1 - e_1^\p \pi)$ is functionally independent of $P$, it then follows that 
$\Vb(\pi;\Pone) \leq\Vb(\pi;\Ptwo)$.



\begin{biography}[]{Vikram Krishnamurthy}
(S'90-M'91-SM'99-F'05) was born in 1966.  He received his bachelor's
degree from the University of Auckland, New Zealand in 1988, and
Ph.D.\ from the Australian National University in 1992.
He currently is  a professor and Canada Research Chair at the
Department of Electrical Engineering, University of British Columbia,
Vancouver, Canada. 

Dr Krishnamurthy's  current research interests include computational game theory,
stochastic control in sensor networks, and stochastic
dynamical systems for
modeling of biological ion channels and biosensors.
Dr. Krishnamurthy currently serves as Editor in Chief of IEEE Journal Selected Topics
in Signal Processing. He has served as associate editor for several journals
including IEEE Transactions Automatic Control and 
IEEE Transactions on Signal Processing. In 2009-2010, he served as Distinguished lecturer for the IEEE signal
processing society.
\end{biography}

\end{document}